\documentclass[prd,onecolumn,amsmath,amssymb,superscriptaddress,nofootinbib,10pt]{revtex4-2}
\usepackage[T1]{fontenc}
\usepackage{lmodern}
\usepackage{customcommands}
\usepackage{amsmath}
\usepackage{amsthm}
\usepackage{amsfonts}
\usepackage{xcolor}
\usepackage{slashed}
\usepackage{mathtools}
\usepackage{epsfig}
\usepackage[colorlinks=true,citecolor=magenta]{hyperref}

\def\ba{\begin{eqnarray}}
\def\ea{\end{eqnarray}}

\def\mpl{M_{\rm Pl}}
\def\p{\partial}
\def\d{\mathrm{d}}
\def\mn{_{\mu \nu}}

\def\mupn{^\mu_{~\, \nu}}
\def\({\left(}
\def\){\right)}
\def\stu{St\"uckelberg }

\def\L{\mathcal{L}}
\def\K{\mathcal{K}}
\def\Lc{\Lambda_{\rm cutoff}}

\begin{document}

\title{A dynamical formulation of ghost-free massive gravity}

\author{Claudia de Rham}
\email{c.de-rham@imperial.ac.uk}

\author{Jan Ko$\dot{\mathrm{z}}$uszek}
\email{j.kozuszek21@imperial.ac.uk}

\author{Andrew J. Tolley}
\email{a.tolley@imperial.ac.uk}

\author{Toby Wiseman}
\email{t.wiseman@imperial.ac.uk}

\affiliation{Theoretical Physics Group, Blackett Laboratory, Imperial College, London SW7 2AZ, United Kingdom}

%------------------------

\begin{abstract}

We present a formulation of ghost-free massive gravity with flat reference metric that exhibits the full non-linear constraint algebraically, in a way that can be directly implemented for numerical simulations. Motivated by the presence of higher order operators in the low-energy effective description of massive gravity, we show how the inclusion of higher-order gradient (dissipative) terms leads to a well-posed formulation of its dynamics. While the formulation is presented for a generic combination of the minimal and quadratic mass terms on any background, for concreteness, we then focus on the numerical evolution of the minimal model for spherically symmetric gravitational collapse of scalar field matter.
 This minimal model does not carry the relevant interactions to switch on an active Vainshtein mechanism, at least in spherical symmetry, thus we do not expect to recover usual GR behaviour even for small graviton mass. Nonetheless we may ask what the outcome of matter collapse is  for this gravitational theory.
Starting with small initial data far away from the centre, we follow the matter
through a non-linear regime as it falls towards the origin. For sufficiently weak data the matter disperses. However for larger data
we generally find that the classical evolution breaks down due to the theory becoming infinitely strongly coupled without the presence of an apparent horizon shielding this behaviour from an asymptotic observer.

\end{abstract}

\maketitle

%------------------------

\section{Introduction}

Current and upcoming cosmological observations, event horizon mapping, and gravitational wave detections offer a unique opportunity to test the laws of gravity in unprecedented situations. While Einstein's theory of General Relativity (GR) has proven to be in outstanding agreement with all observations to date, existing cosmological challenges and the need for an ultimate high-energy completion of GR have motivated the search for alternative frameworks. Even if GR provides the ultimate description of gravity on low-energy scales, the measure of success requires comparison with alternatives against which GR can be meaningfully tested. This is particularly important when observations and detections require the use of templates or priors through which assumptions about
the underlying framework have to be made. With this aim in mind, and driven by the potential of tackling the Cosmological Constant Problem and the physics underlying the nature of the dark sector, a plethora of alternatives to GR have been formulated in the past two decades. While most of these models propose a modification of gravity through the introduction of additional modes (typically scalar fields), non-minimally coupled either to gravity or matter, a genuine modification of the graviton at low-energy (the IR)
has proven more challenging. Large extra-dimensional models of gravity provided a first class of explicit realizations, where the structure of the graviton was genuinely modified in the IR or at
 large (cosmological) distances. In particular the Dvali-Gabadadze-Porrati (DGP) model of gravity introduced in 2000 proposes a model where the graviton appears as a broad resonance of light massive modes from a four-dimensional perspective, \cite{Dvali:2000hr,Dvali:2000rv,Dvali:2000xg,ArkaniHamed:2002fu,Gabadadze:2003ii,Gabadadze:2007dv}, dubbed `soft massive gravity'. The centre and sharpness of this resonance was then further controlled by adjusting the size, scale, topology and the number of extra dimensions \cite{Gabadadze:2003ck,deRham:2007xp,deRham:2007rw,Kobayashi:2008bm,deRham:2009wb,deRham:2010rw}. Attempts to define a theory with zero width - `a hard massive gravity' - have a long history and proposals motivated by extra dimensions were given in    \cite{Gabadadze:2009ja,deRham:2009rm,deRham:2010gu,Berezhiani:2011nc}. \\

The first explicit attempts to construct a four-dimensional formulation of (hard) massive gravity were proposed in the 1970's, but the presence of a ghost at a low-energy scale, highlighted in \cite{Boulware:1973my,Aragone:1979bm,Creminelli:2005qk}, appeared to plague every explicit realization. Formulating massive gravity with the use of \stu fields, as first introduced by Delbourgo and Salam in 1975, \cite{Delbourgo:1975aj} proved particularly insightful in understanding the origin of this ghost \cite{Deffayet:2005ys} and ultimately led to a framework where it could be eradicated all-together, leading to the development of ``ghost-free massive gravity'' (sometimes refereed to as dRGT massive gravity)  \cite{deRham:2010ik,deRham:2010kj}. The absence of ghosts has not only been proven using the \stu fields, but generalized to a multitude of different formalisms
\cite{deRham:2011qq,deRham:2011rn,Hassan:2011hr,Hassan:2012qv,Hinterbichler:2012cn,Kluson:2011aq,
Kluson:2011qe,Kluson:2011rt,Golovnev:2011aa,Comelli:2012vz,Kluson:2012gz,Deffayet:2012nr,Deffayet:2012zc,Kluson:2012cqq,Comelli:2013txa,Deffayet:2015rva},  confirming the existence of  secondary constraints \cite{Hassan:2011ea,Kluson:2012wf,deRham:2015cha}, (see also \cite{deRham:2014zqa} for a review). The form of the constraint was derived on arbitrary backgrounds \cite{Mirbabayi:2011aa,Bernard:2014bfa,Bernard:2015uic,Bernard:2015mkk,Mazuet:2017hey,Mazuet:2018ysa}, including on spherically symmetric ones as will be relevant 
for the explicit numerical
study presented here \cite{Comelli:2011wq,Volkov:2013roa,Volkov:2012wp,
Volkov:2014ooa,Volkov:2014ida,Volkov:2017gox}.\\

In what follows, we shall use the ``vielbein-inspired'' or symmetric vielbein formulation of massive gravity \cite{Hinterbichler:2012cn} which utilizes a 10 component vierbein to describe the geometry. This formalism exhibits the full non-linear scalar constraint as presented in \cite{Deffayet:2012nr,Bernard:2014bfa,Bernard:2015uic} in a way which can be directly applied to numerical evolution.
In \cite{Deffayet:2012nr} the scalar constraint was explicitly identified for the minimal and quadratic models of massive gravity as a first order derivative scalar equation, derived from the Einstein equations with mass terms. It was shown to be more subtle for the cubic mass term, which cannot be expressed in a covariant way in the vielbein language. For perturbations about a general background it was shown in~\cite{Bernard:2014bfa,Bernard:2015uic} that this scalar constraint explicitly removes the unwanted Boulware-Deser ghost, leaving only the five expected dynamical degrees of freedom. In this work, one of our aims is to formulate this constraint locally and use it to explicitly eliminate the unwanted variables,
 rather than working with a first order differential equation to be solved on every timeslice. We provide this algebraic phrasing of the scalar constraint by performing a $(3+1)-$decomposition and then identifying appropriate momenta. In these variables the constraint will simply become algebraic in the time-time component of the vierbein, and furthermore for the simple scalar field matter we employ, will be either a quadratic or cubic equation in that vierbein component, depending on which mass terms one takes. \\

For concreteness,
the numerical results derived in this work will be for the minimal model, for which a Vainshtein mechanism \cite{Vainshtein:1972sx} is not expected to occur (unless one relies on the helicity-one interactions \cite{Renaux-Petel:2014pja}, which are absent in the spherically symmetric case we shall consider).
Nonetheless it is a model of gravity with a dynamical spacetime, and thus a natural question is what its behaviour is for collapse of matter.
Does it resemble GR in the sense that it forms black holes for sufficiently non-linear collapse? Or is its behaviour unlike GR, with naked singularity formation? In what follows we shall provide answers to these basic questions.
Applications to the quadratic model,  which for cosmological asymptotic conditions, is expected to have a working Vainshtein mechanism, and hence yield behaviour similar to GR for low graviton masses, will be explored in further studies. \\

Current tests of GR, direct and indirect detections of gravitational waves and astrophysical/cosmological observations already provide interesting bounds on the graviton mass, \cite{deRham:2016nuf}, however the strongest constraints remain very model-dependent. Model-independent bounds typically rely on the propagation of gravitational waves or modification of the dispersion relation, leading to a bound of the graviton mass which remains many orders of magnitude away from the phenomenologically interesting region (tackling the cosmological constant problem or the origin of dark energy requires a graviton mass of order of the Hubble parameter today, $m\sim H_0\sim 10^{-32}$eV, while model-independent constraints on the graviton mass bound it to be $\lesssim 10^{-22}$eV).
To better improve these bounds, an outstanding open question is what is the precise behaviour of 
black holes in massive gravity, and in particular, what is the effect of the graviton mass on the production of gravitational waves and the resulting waveform?\\

Since the curvature invariant related to any realistic astrophysical 
 black hole is dozens of orders of magnitude above the graviton mass\footnote{For $m\sim H_0$, only a 
  black hole of the size of the Universe would carry a curvature invariant of order of the graviton mass.}, we would expect the graviton mass to be utterly irrelevant to the dynamics of 
 black holes and to the production of gravitational waves during inspiral and 
  black hole
 mergers. However this argument relies on the existence of a smooth decoupling of scales. Such a decoupling would only occur if an efficient Vainshtein  mechanism is in place to screen out the effect of the additional graviton polarizations. In practice, the presence of such a screening mechanism has been challenging to prove formally other than in specific  configurations \cite{Deffayet:2001uk,Deffayet:2008zz,Babichev:2009us,Babichev:2010jd,deRham:2010tw,deRham:2011by,
Chkareuli:2011te,Koyama:2011yg,Kaloper:2011qc,
Belikov:2012xp,Hiramatsu:2012xj,Sbisa:2012zk,Kimura:2011dc,Gannouji:2011qz,Babichev:2011iz,deRham:2012fw,deRham:2012fg,Padilla:2012ry,
DeFelice:2011th,Chu:2012kz,Andrews:2013qva,Berezhiani:2013dw,Koyama:2013paa,Babichev:2013usa,Li:2013nua,Dar:2018dra,Brax:2020ujo,Lara:2022gof,Bezares:2021dma,Shibata:2022gec,
TerHaar:2022wus,Dima:2021mdg}.
Another major challenge to making further progress is the fact that the constraint that prevents the presence of a ghost in massive gravity also prevents the existence of highly symmetric exact solutions. This feature has inhibited the existence of exact homogeneous and isotropic (cosmological) solutions on all scales \cite{DAmico:2011jj}, where solutions can appear arbitrarily close to FLRW on scales of the order of the observable Universe but departure from homogeneity or isotropy must emerge on large distance scales beyond the current cosmological horizon. A similar feature plagues the search for 
 black hole solutions, where Birkhoff's theorem is broken and the constraint prevents the existence of perfectly static and spherically symmetric solutions \cite{Deffayet:2011rh,Berezhiani:2011mt}, (aside from solutions that exhibit a physical singularity at the horizon) see also \cite{Gruzinov:2011mm,Koyama:2011xz,Volkov:2012wp,
Cai:2012db,
Volkov:2013roa,Tasinato:2013rza,Brito:2013wya,Arraut:2013bqa,
Berezhiani:2013dca,Kodama:2013rea,Volkov:2014ooa,
Gervalle:2020mfr,Berens:2021tzd} for other 
black hole solutions in massive (bi-)gravity. As pointed out in \cite{Rosen:2017dvn}, a spherically symmetric non-singular 
black hole solution can nonetheless accommodate an asymptotic  Yukawa-like behaviour if a small time dependence (scaling as the graviton mass) is included. For a graviton mass of the order of the Hubble parameter today, this would correspond to a time dependence that only manifests itself on time scales of the order of the age of the Universe and in practice the solutions are locally indistinguishable from standard Schwarzschild solutions. These solutions were obtained perturbatively about the 
 black hole horizon. Exploring more precisely some of the features of these solutions was pioneered in \cite{Rosen:2018lki} but deriving an explicit exact solutions has remained challenging. \\

It is worth emphasizing that the physical relevance of the Kerr black hole in GR stems firstly from the fact that black holes form in generic matter collapse, as encapsulated in the Singularity Theorems, and secondly, that  due to the Uniqueness theorems, it is the only vacuum black hole solution. In massive gravity no such theorems are currently known, and thus even if candidate black hole end-states for collapse are found, their physical relevance hinges on whether it is indeed these solutions that form dynamically from gravitating matter.
In this work, we therefore initiate some of the first steps towards
understanding what the end state of dynamical collapse is in the dRGT theory of gravity. We do so by
solving numerically for spherically symmetric solutions in non-linear massive gravity by considering the spherical gravitational collapse of a `lump' of matter (described by a massless scalar field) in minimal massive gravity.\\

Before going to the core of the formulation and the numerical framework, it is worth clarifying that just like GR should be seen as the leading order  term in an infinite Effective Field Theory (EFT) expansion \cite{Weinberg:2021exr}, the same applies to massive gravity. At best, massive gravity is only ever expected to represent a low-energy description of gravity, and should always be seen as the leading order contribution in an infinite EFT expansion  where the inclusion of higher order operators is unavoidable \cite{deRham:2017xox,deRham:2018qqo,deRham:2018bgz}
\ba
\label{eq:mGREFT}
S_{\rm mGR}=
\int \d^4 x\sqrt{-g}\left[\frac{1}{16 \pi G_N}\(R[g]+\frac{m^2}{2} \sum_{n}\alpha_n\L_n[\K]\)+\Lc^4 \L^{\rm EFT}\(\frac{\nabla_\mu}{\Lc},\frac{\Lambda_3^3}{\Lc^3}\K\mupn,
\frac{R^\mu{}_{\alpha\nu\beta}}{\Lc^2}\)
\right]\,,  
\ea
where $\Lc$ is the cutoff of the EFT and $\Lambda_3=(\mpl m^2)^{1/3}\sim (m^4/8 \pi G_N)^{1/6}$. 
The Lagrangian terms $\L_n$ are the ``total derivative polynomials''  (dRGT mass terms) introduced in \cite{deRham:2010kj} and $\mathcal{K}\sim \delta-\sqrt{g^{-1}f}$ are the building blocks of the ghost-free mass term (reminiscent of the extrinsic curvature in models of massive gravity arising from extra dimensions \cite{deRham:2013awa}). 
An example of a higher order operator is a term $\K^2$ which is not of the dRGT mass term form. From the EFT point of view it is natural to include such a contribution, but it will come suppressed by the cutoff,
\be
\Delta \L \sim \frac{\Lambda_3^6}{\Lc^2} \K^2 \, .
\ee
Such a term appears to induce a ghost, but one whose mass is at the scale $\Lc $ which renders it harmless \footnote{The parametric scaling is
reflecting the fact that in the decoupling limit $\mpl \rightarrow \infty$, $m \rightarrow 0$,  $\K_{\mu\nu} \sim \partial_{\mu} \partial_{\nu} \pi/\Lambda_3^3$ with $\pi$ the helicity zero mode of the graviton, hence $\Delta \L \sim (\Box \pi)^2/\Lambda_c^2$.}.
\\

For the theory of massive gravity to make sense, it should be ghost-free up to the cutoff $\Lc$ which should be parametrically larger than the graviton mass $m$. In principle the cutoff can be separate from the strong coupling scale $\Lambda_3$ at which 
naive perturbative unitarity of the truncation to the leading $\L_n$ operators breaks down \cite{Aydemir:2012nz,deRham:2014wfa}. For the theory to enjoy a standard (Wilsonian-like) and weakly coupled high-energy completion one would need the cutoff to be parametrically smaller than the scale $\Lambda_3$ \cite{deRham:2017xox}, however it is typically expected that theories that
have a Vainshtein mechanism be embedded in alternative completions \cite{deRham:2014wfa,Dvali:2010jz,deRham:2013hsa,deRham:2014wnv,Kampf:2014rka,Keltner:2015xda}. In particular, owing to the existence of a non-renormalization theorem \cite{deRham:2012ew,deRham:2013qqa} that protects the ghost-free structure of the Lagrangian, situations where the higher order operators enter at a scale $\Lc$ parametrically larger than the strong coupling scale $\Lambda_3$ can be considered, with  $\Lc$ potentially close to the Planck scale.
Ultimately, the scale at which new physics enters, and how it manifests itself through the higher order operators 
encapsulated in $\L^{\rm EFT}$, depends on the precise details of the UV completion (Wilsonian or not, weakly coupled or not), but for the low-energy theory of massive gravity to make sense, low-energy observables should be independent of these details.\\

At the level of the classical continuum partial differential equations (p.d.e.s) describing the truncation to the leading low energy theory, short distance modes will explore this region that is out of control of the EFT. Unlike in the case of GR, as we discuss later, we believe it is unlikely that this truncation to the leading dRGT terms makes sense as a theory with a well-posed initial-value formulation in its own right. In order to perform numerical calculations of this continuum leading low energy theory, we wish to accommodate such short distance excursions while remaining as agnostic as possible to the precise operators entering in $\L^{\rm EFT}$. We cannot expect to simply ignore the issue, as if the leading low energy theory is ill-posed without high order terms, then one would not expect to have a good continuum limit when refining a numerical discretization for it. If one is unable to refine a numerical approximation to a continuum limit it is then unclear what the status of that numerical calculation is. Thus here we take the conservative view that we should start with a well-posed theory \emph{before} considering a numerical discretization, so that a good continuum limit does exist \footnote{An alternative perspective is to regard the numerical discretization itself as a cut-off, which cannot be completely removed. While perhaps a valid viewpoint, it is then difficult to assess the accuracy of the numerical calculations, and hence this is not the approach taken here.\label{foot:cutoff}}. In order to achieve this for dRGT we introduce specific higher derivative terms that are convenient for numerical simulation, and natural in our $(3+1)$-decomposition. We are able to prove the resulting continuum p.d.e. system is well-posed.
Since we include these terms within the $(3+1)$ split, they are not of the form expected from a Lorentz invariant UV completion, but nonetheless mimic the dissipative effect of the operators expected to enter in $\L^{\rm EFT}$ at the level of the dynamical equations, while ensuring that the resulting low-energy physics is insensitive to the scale at which they enter (so long as the scale is sufficiently large). Provided gradients remain below the scale of these new operators, they will be irrelevant for the dynamics, and this may be explicitly checked. This strategy complements that formulated in \cite{Allwright:2018rut,Bernard:2019fjb,Cayuso:2020lca,Lara:2021piy,Figueras:2021abd,Gerhardinger:2022bcw,Franchini:2022ukz,Barausse:2022rvg,Franchini:2022ukz} for other theories of modified gravity or scalar/vector(-tensor) EFTs involving screening. An important  outcome of this work is the existence of a manifestly well-posed initial-value formulation of the dynamics of massive gravity.\\

The rest of this work is organized as follows: We start by providing a brief review of ghost-free massive gravity in section~\ref{sec:setup}, formulating the constraint algebraically in a symmetric ``vielbein-inspired'' language. Then in section~\ref{sec:formulation}, as a first step towards formulating the dynamics of massive gravity in a way that is amenable to numerical simulations, we perform a 3+1 space and time decomposition of the dynamical variables and their associated dynamical equations and constraints.
Secondly, we are able to identify specific momenta such that the second class constraints in the system (the vector and scalar constraints) can be solved algebraically for components of the vierbein and its momenta. Specifically the scalar constraint becomes an algebraic relation for the time-time component of our symmetric vierbein. The remaining second order degrees of freedom precisely account for the five degrees of freedom in the theory.
This formulation does not require any spacetime symmetry and is non-perturbative (ie. not dependent on an expansion about a background).
From this point on we focus on the simplest theory, that with minimal mass term.
In section~\ref{sec:Harmonic} we briefly outline an alternative harmonic formulation of the theory where the vector constraint is automatically satisfied if it is imposed on the initial Cauchy surface. We then introduce higher derivative dissipative terms motivated by the higher-order EFT operators present in \eqref{eq:mGREFT} in section~\ref{sec:wellposed}.
These terms are convenient in our $(3+1)$-formulation, and we argue the resulting p.d.e.s then have a well-posed initial value formulation.
In section~\ref{sec:sphericalcollapse} we demonstrate that our formulation may be implemented numerically in a straightforward manner by studying spherical collapse.
We diagnose under which conditions the system evolves smoothly and when one hits regions of strong coupling, at which point the EFT loses predictability. We also diagnose under which conditions our results remain insensitive to the scale at which higher order EFT (or dissipative effects) kick in.  A summary of our results and outlooks for  further work are discussed  in section~\ref{sec:Conclusion}. Further details on the numerical implementation, diffusion effects and convergence are provided in the appendix~\ref{app:numdetails}. \\

While massive gravity can be formulated in any number of dimensions, throughout this work we focus for concreteness on four spacetime dimensions and use mainly $+$ signature. The relation between the reduced Planck scale and Newton's constant is given by $8 \pi G_N=\mpl^{-2}$ and we work with one or the other depend on context. For most of this work we will 
choose units where $8 \pi G_N=\mpl^{-2}=1$ but reintroduce dimensions whenever needed.

\section{Brief review of \lowercase{d}RGT massive gravity}
\label{sec:setup}

As for any massive field, the notion of mass requires a reference, which we shall denote as the reference metric $f_{\mu\nu}$.
In principle massive gravity can be formulated for any reference metric $f\mn$ \cite{Hassan:2011tf}, however the notion of mass is only unambiguous when dealing with representations of the Lorentz group or (anti)-de Sitter group, and so it makes sense to restrict to maximally symmetric spacetimes. For the remainder we make the standard choice that the reference metric is Minkowski. In that case, massive gravity in unitary gauge respects a global version of Lorentz invariance and admits Minkowski spacetime as a vacuum solution. \\

The usual dynamical metric is denoted as $g_{\mu\nu}$, and it is this metric that matter is chosen to couple  to, preserving the weak equivalence principle. The global structure of the dynamical metric can be very different than the reference metric, but for comparison with GR in this work we shall require it to be asymptotically flat. As indicated in the introduction, restricting to four spacetime dimensions the  ghost-free massive gravity action \eqref{eq:mGREFT} can be formulated  in terms of ``total derivative polynomials'' $\L_n[\K]$ given in terms of the Levi-Civita symbols $\varepsilon$ by \cite{deRham:2010kj}
\ba
\L_n[\K]=\varepsilon^{a_1 \cdots a_4}\varepsilon_{b_1 \cdots b_4} \K^{b_1}{}_{a_1}\cdots \K^{b_n}{}_{a_n}\delta^{b_{n+1}}{}_{a_{n+1}}\cdots \delta^{b_4}{}_{a_4}\,.
\ea
In particular we have
\ba
\label{L0}
\L_0[\K]&=&4!\\
\label{L1}
\L_1[\K]&=&3!\, [\K]\\
\label{L2}
\L_2[\K]&=&2!\([\K]^2-[\K^2]\)\\
\label{L3}
\L_3[\K]&=&\([\K]^3-3[\K][\K^2]+2[\K^3]\)\\
\label{L4}
\L_4[\K]&=&\([\K]^4-6[\K]^2[\K^2]+3[\K^2]^2+8[\K][\K^3]-6[\K^4]\)\,,
\ea
where square brackets represent the trace of tensors (taken with respect to the dynamical metric).
We see that $\L_0$ is a cosmological constant and $\L_1$  includes a tadpole, so there are only three linearly independent terms that will lead to the graviton gaining a mass.
The building block $\K\mupn$ defined as
\ba
\K\mupn=\delta\mupn-E\mupn=g^{\mu\alpha}\(g_{\alpha\nu}-E_{\alpha\nu}\)\,,
\ea
is constructed out of the symmetric vielbein $E_{\mu\nu} = E_{(\mu\nu)}$
\cite{Hinterbichler:2012cn} which is defined from the metric and reference metric as,
\be
g_{\mu\nu} = (f^{-1})^{\alpha\beta} E_{\alpha\mu}  E_{\beta\nu}\,,
\ee
where $(f^{-1})^{\mu\nu}$ is the inverse to the Minkowski reference metric.
We may equivalently write the relation as,
\be
f^{\mu}_{~~\nu} = (g^{-1})^{\mu\alpha} f_{\alpha\nu} = E^{\mu}_{~~\alpha} E^{\alpha}_{~~\nu}\,,
\ee
so that symbolically, we may write $E^\mu_{~~\nu} = \sqrt{ f^{\mu}_{~~\nu} }$. We also define $(E^{-1})^{\mu\nu}$ as the inverse to $E_{\mu\nu}$ in the sense that $(E^{-1})^{\mu\alpha} E_{\alpha\nu} = \delta^\mu_\nu$, so that
\be
(E^{-1})_\mu{}^\nu=\sqrt{g_{\mu\alpha}(f^{-1})^{\alpha\nu}}=\sqrt{(f^{-1})_\mu{}^{\nu}}\,.
\ee

Omitting for now the EFT contributions that enter at the cutoff scale, and including coupling to matter, the formulation of massive gravity we shall be interested in is given by
\ba
S_{\rm mGR}=
\frac{1}{16 \pi G_N}\int \d^4 x\sqrt{-g}\(R[g]+\frac{m^2}{2} \sum_{n}\alpha_n\L_n[\K]\)+ S^{\rm (matter)}[g, \psi_i]\,,
\ea
where $R$ is the standard scalar curvature of the dynamical metric $g\mn$ and matter fields $\psi_i$ only couple to the physical metric $g$.
When perturbing about flat spacetime, each Lagrangian $\L_n[\K]$ is order $n$ in fluctuations, which allows us to establish the order at which each interaction enters.
The minimal model corresponds to the special case where all coefficients $\alpha_n$ vanish aside from $\alpha_0$ and $\alpha_1$, and the cosmological constant $\alpha_0$ is tuned so as to remove the tadpole associated with $\L_1[\K]$, $\alpha_0=- \alpha_1/4$, so that $g\mn=f\mn$ is a vacuum solution. Focusing on minimal and quadratic mass terms, we may set 
$\alpha_3=\alpha_4=0$, $\alpha_0=-\alpha_1/4=-(1-\alpha_2)/6$, so the action for massive gravity hence takes the form (still omitting the higher order operators for now),
\be
S = \frac{1}{16 \pi G_N} \int \d^4x \sqrt{-g} \left(R[g] - m_1^2 (2[E]-6) - \frac{m_2^2}{2} \([E]^2-[E^2]-6\) \right) + S^{\rm (matter)}[g, \psi_i]\,,
\ee
with the two following mass terms
\ba
m_1^2=m^2(1+2\alpha_2)\quad {\rm and}\quad
m_2^2=-2 m^2 \alpha_2\,,
\ea
so that the minimal model corresponds to $\alpha_2=0$, while the quadratic model corresponds to $\alpha_2=-1/2$. In both cases, the graviton mass in the vacuum is $m$.

The resulting Einstein equation is,
\be
\label{eq:EinsteinEq}
\mathcal{E}_{\mu\nu} \equiv G_{\mu\nu} + m_1^2 M^{(1)}_{\mu\nu} + m_1^2 M^{(2)}_{\mu\nu} - 8\pi G_N T_{\mu\nu} = 0\,,
\ee
where the mass terms contributions are given by,
\ba
M^{(1)}_{\mu\nu}  & = & - E_{\mu\nu} + [E] g_{\mu\nu} - 3 g_{\mu\nu} \\
M^{(2)}_{\mu\nu} & = & \frac{1}{2} E_{\mu\alpha} E^{\alpha}_{~~\nu} - \frac{1}{2} [E] E_{\mu\nu} - \frac{1}{4} \left(
[E^2] - [E]^2 \right) g_{\mu\nu} - \frac{3}{2} g_{\mu\nu}\,.
\ea
Using the contracted Bianchi identity and matter-stress energy conservation we may take the divergence of the Einstein equation to derive
\be
\label{eq:vectordef}
V_\mu = \nabla^\nu \left(   m_1^2 M^{(1)}_{\mu\nu} + m_2^2 M^{(2)}_{\mu\nu}  \right)=0\,,
\ee
which we refer to as the `vector equation'  \footnote{Had we not set unitary gauge from the outset and kept the \stu fields arbitrary, the following vector equation would simply arise as the \stu equation of motion.}. \\

Diffeomorphism invariance can be made explicit through the use of \stu fields, however for now it will prove more convenient to
commit to either Cartesian coordinates $x^\mu = (t, x, y, z)$, or later in section~\ref{sec:sphericalcollapse} when we consider spherical symmetry, spherical coordinates $x^\mu = (t, r, \theta, \phi)$, and
formulate the theory in unitary gauge where the reference metric is given by $f_{\mu\nu} = (f^{-1})^{\mu\nu} = \eta_{\mu\nu} = \mathrm{diag}\left( -1, +1, +1, +1 \right)$ in the Cartesian case or $f_{\mu\nu} = \mathrm{diag}\left( -1, +1, +r^2, +r^2 \sin^2(\theta) \right)$ in the spherical case. Thus until 
 section~\ref{sec:sphericalcollapse} we will take $f_{\mu\nu} = \eta_{\mu\nu}$, and from now we also adopt units such that $8\pi G_N = 1$.

\subsection{Linear perturbations}
\label{sec:lintheory}
Considering linear fluctuations about flat space, by expanding the dynamical metric as
\be
g_{\mu\nu} = \eta_{\mu\nu} + h_{\mu\nu} \, ,
\ee
we recover the standard Fierz-Pauli mass term in Minkowski,
\be
m_1^2 M^{(1)}_{\mu\nu} + m_2^2 M^{(2)}_{\mu\nu} = \frac{1}{2}m^2(h_{\mu\nu} - h \eta_{\mu\nu})\,,
\ee
where indices here, and in what follows in this perturbative discussion, are raised and lowered with the Minkowski metric $\eta$, and the graviton mass (squared) is given by $m^2 = m_1^2 + m_2^2$.
Assuming the matter fields are all covariantly coupled to the dynamical metric, the resulting matter stress-energy tensor $T\mn$ is then conserved,  $\partial^\mu T_{\mu\nu} = 0$. At the linear level, the Bianchi identity~\eqref{eq:vectordef} hence imposes the condition
\ba
\label{eq:linearizedBianchi}
\p_\mu h\mupn = \p_\nu h\,.
\ea
Note that unlike in GR,  we have already set unitary gauge at this stage and there is therefore no additional gauge choice available. In particular the condition \eqref{eq:linearizedBianchi} appears as a constraint and not as a gauge choice. At the linearized level, this constraint forces the Ricci scalar to vanish, irrespectively of the trace of the stress-energy tensor. This indicates that the linearized theory is unable to properly capture coupling with external matter sources. This is at the origin of the infamous van-Dam-Veltman-Zakharov discontinuity \cite{vanDam:1970vg,Zakharov:1970cC}, whose resolution lies in the contribution from the non-linear interactions \cite{Vainshtein:1972sx} we shall discuss in section~\ref{sec:Vainshtein}. For now, carrying on with the linear analysis, the vanishing of the linearized Ricci scalar implies the following constraints
\be
\label{eq:linconstraint}
h = - \frac{2}{3 m^2} T \; , \quad  \partial^\rho h_{\rho \mu} = - \frac{2}{3 m^2} \partial_\mu T\,.
\ee
As for the dynamical equations, they are given by the linearized Einstein field equations \eqref{eq:EinsteinEq}, after substituting the linearized condition \eqref{eq:linearizedBianchi}, leading to the massive wave equation
\be
\label{eq:lineq}
- \frac{1}{2} (\nabla^2 - m^2) h_{\mu\nu} + \partial_{(\mu} \chi_{\nu)} =  T_{\mu\nu} - \frac{1}{3} \eta_{\mu\nu} T\,,
\ee
with $\nabla^2 = \partial^\rho \partial_\rho$ and $ \chi_\mu = \partial^\rho h_{\rho \mu} - \frac{1}{2} \partial_\mu h=\frac 12 \partial_\mu h$.\\

The $5$ linear constraints \eqref{eq:linconstraint} on $\partial^\rho h_{\rho \mu}$ and $h$ reduce the number of dynamical degrees of freedom from $10$ down to $5$. The special structure of the Fierz-Pauli term ensures the existence of an algebraic constraint on $h$ rather than it obeying a wavelike equation, which would yield an additional ghostly degree of freedom.
The existence of an analogous algebraic constraint in the full non-linear massive gravity was indicated in \cite{deRham:2010kj,deRham:2011qq,deRham:2011rn} and proven more generically in various languages in \cite{Hassan:2011hr,Hassan:2012qv,Hinterbichler:2012cn,Kluson:2011aq,
Kluson:2011qe,Kluson:2011rt,Golovnev:2011aa,Comelli:2012vz,Kluson:2012gz,Deffayet:2012nr,Deffayet:2012zc,Kluson:2012cqq,Comelli:2013txa}, among others, together with the existence of a secondary constraint \cite{Hassan:2011ea,Kluson:2012wf,deRham:2015cha}, (see also \cite{deRham:2014zqa} for a review). The form of the constraint was derived on arbitrary backgrounds, including on spherically symmetric ones as will be relevant to the study presented here \cite{Volkov:2013roa,Volkov:2012wp,Volkov:2014ida,Volkov:2014ooa,Volkov:2017gox}.
While the formulation of this primary constraint and the secondary one that follows has been well-established by now, how to implement it efficiently and in a well-posed way for the study of numerical evolution has proven more challenging.  Part of the purpose of this paper is to present the constraint in the full non-linear massive gravity in a formalism that can be directly implemented in a numerical evolution. Before carrying on with the full non-linear analysis in section~\ref{sec:constraints}, we first briefly discuss the physics behind the Vainshtein mechanism which signals a breakdown of linear perturbations and show how non-linear interactions play an essential role when considering the small mass limit of massive gravity.

\subsection{Decoupled modes and Vainshtein}
\label{sec:Vainshtein}
To better capture the essence of the Vainshtein mechanism,  it is convenient to  first identify the five propagating degrees of freedom more explicitly by introducing the \stu fields and writing the metric fluctuation as
\be
\label{eq:perth}
h_{\mu\nu} = a_{\mu\nu} + \frac{1}{m^2} \partial_{(\mu} A_{\nu)} + \frac{1}{2} \pi \eta_{\mu\nu} + \frac{1}{m^2} \partial_\mu \partial_\nu \pi\,.
\ee
The helicity-2, helicity-1 and helicity-0 degrees of freedom are then encapsulated in $a_{\mu\nu}$, $A_\mu$ and $\pi$ respectively. This formulation enjoys two local invariances,
\be
\begin{array}{ccc}
a_{\mu\nu} &\to& a_{\mu\nu} + \partial_{(\mu} v_{\nu)}  \\
A_\mu &\to& A_\mu  - m^2 v_\mu \\
\pi & \to & \pi
\end{array} \quad, \qquad
\begin{array}{ccc}
a_{\mu\nu} &\to& a_{\mu\nu} + \frac{\lambda}{2} \eta_{\mu\nu} \\
A_\mu &\to& A_\mu  + \partial_\mu \lambda \\
\pi & \to & \pi - \lambda
\end{array}
\ee
for $v_\mu$ and $\lambda$ a Lorentz covector and scalar respectively.

Now that these gauge invariances are made explicit, we can see that the gauge invariant rank-2 tensor $a_{\mu\nu}$ carries $10-4\times 2=2$ degrees of freedom (same as a massless graviton),
 $A_\mu$ carries $4-1\times 2=2$ degrees of freedom (same as any other gauge invariant vector field), and the last degree of freedom is carried by $\pi$, yielding a total of five degrees of freedom.

Endowed with these two sets of gauge invariances, we can freely pick the equivalent of the respective harmonic gauges (i.e. a slightly modified de Donder gauge for $a\mn$ and modified Lorenz gauge for $A_\mu$). Through appropriate gauge transformation, we can always set\footnote{
Any gauge transformation with parameters $v_\mu$ and $\lambda$ satisfying $\left( \nabla^2 - m^2 \right) v_\mu = 0$ and $\left( \nabla^2 - m^2 \right) \lambda = 0$ preserve that gauge choice, so there are some residual gauge freedom one could further set.}
\be
\partial^\rho a_{\rho\mu} - \frac{1}{2} \partial_\mu a = - \frac{1}{2} A_\mu \; , \quad   \partial \cdot A  = - m^2 \left( a + 3 \pi \right)\,.
\ee
In this gauge, the degrees of freedom entirely decouple and lead to the following set of three (dynamical) wave equations,
\be
- \frac{1}{2} \left( \nabla^2 - m^2 \right) a_{\mu\nu} = T_{\mu\nu} - \frac{1}{2} \eta_{\mu\nu} T \; , \quad
- \frac{1}{2} \left( \nabla^2 - m^2 \right) A_\mu = 0\; , \quad
- \frac{1}{2} \left( \nabla^2 - m^2 \right) \pi = \frac{1}{3} T \,.
\ee
Expressed in this form, it is clear that the helicity-0 mode $\pi$ remains coupled to matter (at the linear level), even when we take the massless limit, $m \to 0$ holding $T_{\mu\nu}$ fixed.  This explains why equation~\eqref{eq:lineq} does not lead to the same linearized equations as in GR, $- \frac{1}{2} \nabla^2 h_{\mu\nu} + \partial_{(\mu}  \chi_{\nu)} =  T_{\mu\nu} - \frac{1}{2} \eta_{\mu\nu} T$, due to the couplings to the trace of the stress tensor being different. The coupling of the helicity-0 mode to matter is responsible for an additional contribution of $T/6$ even in the small mass limit.
However it is also clear from the constraint \eqref{eq:linconstraint} on $h$ that the linearized theory breaks down in the massless limit. This is the essence of the Vainshtein mechanism pointed out in \cite{Vainshtein:1972sx}. Accounting for non-linear contributions under special conditions,  GR was recovered non-perturbatively in the massless limit of the DGP model \cite{Dvali:2000hr} in \cite{Deffayet:2001uk} and the same mechanism was proven to occur in dRGT massive gravity \cite{deRham:2010ik,deRham:2010tw}.\\

More precisely when taking the limit $m^2 \to 0$ limit, while keeping the stress energy source fixed, we may approximately solve the Einstein equations with a GR solution if we can find a coordinate system so that the vector equation $V_\mu = 0$ is satisfied. We may regard this vector equation as a `gauge condition' for the GR solution, and at least locally we have the correct number of coordinate degrees of freedom to solve it.
From the linear analysis we also know that generally the linear response differs to that of GR. This implies that while the solution of \eqref{eq:EinsteinEq}, $\mathcal{E}_{\mu\nu}=0$, resembles that of GR in the massless limit it is in a gauge where $| E^\mu_{~\nu} - \delta^\mu_\nu  | \gg 1$ so it is a non-linear deformation of Minkowski at the level of the metric (even though curvatures may be small). In this regime, the theory becomes classically `strongly coupled', in the sense that standard perturbation theory breaks down, without indicating a failure of predictability \cite{deRham:2014wfa}.
In particular,  subleading terms in the effective theory~\eqref{eq:mGREFT} involving higher derivatives of $\mathcal{K}^\mu_{~\nu}$ potentially become important (particularly if no Vainshtein resummation occurs), even though higher curvature terms may remain small.
Far outside the non-linear Vainshtein radius, the metric tends to Minkowski and the linear theory then applies, and hence leads to features that differ from GR. The matching region around the Vainshtein radius is then both a non-linear deformation of Minkowski, as well as having non-GR behaviour, and is subtle to track down precisely. Beyond its behaviour in the decoupling limit of the theory, it has remained challenging to follow this transition precisely other than for static and spherical symmetric situations as well as in other theories of massive gravity~\cite{Deffayet:2008zz}.\\

The Vainshtein region where non-linearities are important can be estimated by determining when the graviton mass $m$ is negligible compared to curvature invariants, $m^2\lesssim R$. For a compact matter source of mass $M$,  the corresponding Vainshtein radius   $R_V \sim \left( R_G /m^2 \right)^{1/3}$ where  $R_G \sim G M$ is the Schwarzschild radius associated  with that $M$. We note that the same non-linear terms that give rise to the Vainshtein mechanism and a smooth massless limit towards GR also lead to a breaking of perturbative unitarity at the scale $\Lambda_3 \sim \left(\mpl m^2 \right)^{1/3}$, with $\mpl$ the Planck mass. While this scale is naively very low, it is redressed within the Vainshtein radius, so that radiative corrections to the theory are irrelevant on scales where gravity may be probed \cite{deRham:2014wfa}. \\

Since all the non-linear pure helicity-0 interactions vanish for the minimal model, is not expected to  exhibit a standard Vainshtein mechanism (at the very least not without existing the helicity-1 modes non-trivially \cite{Renaux-Petel:2014pja}). There is therefore little known about the non-linear behaviour of this minimal theory in response to matter, even in the small mass limit.

\subsection{Vector equation and scalar constraint}
\label{sec:constraints}

We now return to the full non-linear theory and review the scalar and vector constraints.
We emphasize that the existence of these constraints has been discussed extensively in the literature and here we simply review these using the symmetric vierbein formulation. In the next section we will see that these constraints can be formulated algebraically in a $(3+1)$-decomposition non-perturbatively by appropriately identifying momentum variables.\\

In order to reveal the spin-1 and spin-0 constraints it is natural to consider (diffeomorphism) variations of the metric taking the form,
\be
\delta g_{\mu\nu} = \nabla_{(\mu} v_{\nu)}\,.
\ee
Now the action varies to give,
\be
\delta_v S = - \frac{1}{2}\int \d^4x \sqrt{-g} \, v_\alpha V^\alpha\,,
\ee
where $V_\mu$ is the vector defined in \eqref{eq:vectordef}. The equation of motion from varying $v_\mu$ is then the same as the vector constraint $V_\mu = 0$. In terms of the vector $\xi^\mu$ defined as
\be
\xi_\mu = E_{\mu\alpha}\eta^{\alpha\beta}V_\beta\,,
\ee
the constraint then takes the remarkably simple form,
\be
\label{eq:vectorconstraint}
\xi^\mu = V^{\mu\alpha\beta\sigma} \partial_{[\alpha} E_{\beta]\sigma}  = 0\,,
\ee
with
\be
V^{\mu\alpha\beta\sigma} = - 2 m_1^2  g^{\mu\alpha} (E^{-1})^{\beta\sigma}  + m_2^2 \left(E^{\mu\alpha}(E^{-1})^{\beta\sigma} + g^{\mu\alpha}g^{\beta\sigma} - E^{\rho}_{~\rho}g^{\mu\alpha}(E^{-1})^{\beta\sigma}\right)\,,
\ee
so we may view this as a linear constraint on the components of $\partial_{[\alpha} E_{\beta]\sigma}$.

Now we consider varying the action with respect to a scalar mode $\pi$ taking the form of a diffeomorphism combined with a conformal transformation and shift involving the reference metric,
\be
\delta g_{\mu\nu} = \frac{\pi}{2} \left( m_1^2 g_{\mu\nu} + m_2^2 E_{\mu\nu} \right) + \nabla_{(\mu} \tilde{v}_{\nu)} \; , \quad \tilde{v}_\mu = E_{\mu\alpha}  \eta^{\alpha\beta} \partial_\beta \pi\,.
\ee
This leads to the following variation of the action,
\be
\delta_\pi S =  \frac{1}{2}\int d^4x \sqrt{-g} \, \pi \Pi=0\,,
\ee
where
\be
\label{eq:pieq}
\Pi = \frac{1}{2} \left( m_1^2 g^{\mu\nu} + m_2^2 E^{\mu\nu} \right) \mathcal{E}_{\mu\nu} + \nabla \cdot \xi\,.
\ee
Some comments on this variation are in order. Linearizing about flat space, so $g_{\mu\nu} = \eta_{\mu\nu} + h_{\mu\nu}$ and $E_{\mu\nu} = \eta_{\mu\nu} + \frac{1}{2} h_{\mu\nu}$, we simply recover the spin-0 part of~\eqref{eq:perth}, namely $\delta g_{\mu\nu} = \frac{\pi}{2} m^2 \eta_{\mu\nu} + \partial_\mu \partial_\nu \pi$, but have now identified the fully non-linear equivalent excitation about an arbitrary background.
We note that in the `vielbein-like' language, this perturbation takes the simple form,
\be
\delta E_{\mu\nu} = \frac{\pi}{4} \left( m_1^2 E_{\mu\nu} + m_2^2 \eta_{\mu\nu} \right) + \frac{1}{2}\tilde{\nabla}_\mu\tilde{\nabla} _\nu \pi\,,
\ee
where the covariant derivative $\tilde{\nabla}$ is taken with the connection $\tilde{\Gamma}^\alpha_{~\mu\nu}$ expressed in terms of the standard dynamical metric connection by the relation,
\be
\Gamma^\alpha_{~\mu\nu} = E^{\alpha}_{~\beta}(E^{-1})^{\gamma}_{~(\mu|}\tilde{\Gamma}^{\beta}_{~|\nu)\gamma} + (E^{-1})^{\alpha\beta}\partial_{(\mu|}E_{|\nu)\beta}\,.
\ee
  In terms of $E_{\mu\nu}$, the scalar equation can be explicitly written as,
\be
\label{eq:Pi}
\Pi = \frac{1}{2} \left( m_1^2 \Pi^{(1)} + m_2^2 \Pi^{(2)} \right)=0\,,
\ee
where we have defined,
\ba
\Pi^{(1)} &=& A_{(1)}^{\alpha\beta\gamma\mu\nu\rho} \partial_{[\alpha} E_{\beta] \gamma}  \partial_{[\mu} E_{\nu]\rho} + m_1^2( 3 [E] - 12 )+m_2^2 \left( \frac{1}{2}[E]^2 - \frac{1}{2}[E^2] -6 \right)- T  \, ,\\
\Pi^{(2)} &=& A_{(2)}^{\alpha\beta\gamma\mu\nu\rho}  \partial_{[\alpha} E_{\beta] \gamma} \partial_{[\mu} E_{\nu]\rho} +
m_1^2\left(-[E^2]+ [E]^2 - 3[E]\right)  \\ &+& m_2^2 \left(\frac{1}{2}[E^3] - \frac{3}{4} [E] [E^2] + \frac{1}{4}[E]^3  - \frac{3}{2}[E]\right) - E^{\mu\nu} T_{\mu\nu}\,. \nonumber
\ea
Let us assume the matter is such that the stress tensor does not involve derivatives of the metric, as for example is the case for (minimally coupled) scalar or vector fields, Yang-Mills theories or perfect fluids.
Then the scalar equation $\Pi=0$ never involves terms with more than one derivative acting on the metric (or equivalently on $E_{\mu\nu}$), and thus is a constraint equation. Furthermore, the one derivative terms are determined by the tensors,
\ba
\label{eq:A}
A_{(1)}^{\alpha\beta\gamma\mu\nu\rho}
&=& \eta^{\gamma\rho} g^{\alpha[ \mu} g^{\nu ]\beta} - 2 (E^{-1})^{\rho[ \alpha} g^{\beta] [\mu} (E^{-1})^{\nu]\gamma}  + 4  (E^{-1})^{\gamma[ \alpha} g^{\beta] [\mu} (E^{-1})^{\nu]\rho} \, ,
\\
A_{(2)}^{\alpha\beta\gamma\mu\nu\rho} &=& [E]\left(\frac{1}{2}\eta^{\gamma\rho}g^{\alpha[\mu}g^{\nu]\beta}-(E^{-1})^{\rho[\alpha}g^{\beta][\mu}(E^{-1})^{\nu]\gamma} + 2(E^{-1})^{\gamma[\alpha}g^{\beta][\mu}(E^{-1})^{\nu]\rho}\right)  \\ &-& 2  \eta^{\gamma\rho}g^{\alpha[\mu}E^{\nu]\beta} + (E^{-1})^{\gamma\rho}g^{\nu[\alpha}g^{\beta]\mu} - 4(E^{-1})^{\rho[\mu}g^{\nu][\alpha}g^{\beta]\gamma} \notag \\ &+& 2 (E^{-1})^{\rho[\alpha}E^{\beta][\mu}(E^{-1})^{\nu]\gamma} - 4(E^{-1})^{\gamma[\alpha}E^{\beta][\mu}(E^{-1})^{\nu]\rho}\,.
\notag
\ea
An important point that will be relevant shortly is that due to the derivatives of $E_{\mu\nu}$ entering only via the combination $\partial_{[\alpha} E_{\beta]\gamma}$,  the scalar constraint contains no time derivatives of $E_{tt}$ at all. However it does contain spatial derivatives of $E_{tt}$.

\section{3+1 dynamical formulation}
\label{sec:formulation}

We now employ a 3+1 decomposition and since the map $E_{\mu\nu} \to g_{\mu\nu}$ is explicit, it will prove convenient to work with $E_{\mu\nu}$ as our dynamical variable. 
Given a novel choice for momentum variables, this 3+1 decomposition will allow us to solve the vector and scalar constraints explicitly.
Our starting point is the action, which written in terms of $E_{\mu\nu}$ takes the rather elegant form,
\be
\label{eq:action2}
S = \int \d^4 x |\text{det} E | \left( - \frac{1}{2} A_{(1)}^{\alpha\beta\gamma\mu\nu\sigma} \partial_{[\alpha} E_{\beta] \gamma}  \partial_{[\mu} E_{\nu]\sigma} - m^2\mathcal{L}_{\rm mass} + \mathcal{L}_{\rm matter} \right)\,.
\ee
Note the derivative term is identical to the one entering in $\Pi^{(1)}$. This is because, in the absence of matter, the terms containing derivatives of the $E$ matrix in $\Pi^{(1)}$ are simply equal to $-R + 2\nabla \cdot \xi^{(1)}$
with this last term being a total divergence. Hence we see the Einstein-Hilbert term in the action is just given by the derivative terms in $\sim | \det E | \Pi^{(1)}$ without matter. \\

Consider now the canonical conjugate momenta to $E_{\mu\nu}$. Firstly note that the action contains no momentum conjugate to $E_{tt}$ since $\dot{E}_{tt}$ does not appear in the Lagrangian.
Then the canonical momenta conjugate to $E_{ti}$ and $E_{ij}$ are given by,
\be
\pi^i = | E | A_{(1)}^{i t t \mu\nu\sigma} \partial_{[\mu} E_{\nu]\sigma} \; , \quad \pi^{ij} = -| E | A_{(1)}^{t i j \mu\nu\sigma} \partial_{[\mu} E_{\nu]\sigma}\,,
\ee
where we use the notation $|E| = |\det E|$. In what follows
we will work with the simpler momentum variables,
\be
\label{eq:momenta}
P_i =  \partial_{[t} E_{i]t} \; , \quad P_{ij} =  \partial_{[t} E_{i]j}\,,
\ee
which are linearly related to $\pi^i$ and $\pi^{ij}$ with coefficients depending only on $E_{\mu\nu}$.
An important point we return to later is that when the action is written in these variables, there are then no derivatives of $E_{tt}$ at all -- the only derivatives that enter above are spatial ones, and these can only occur in the combination $P_i =  \partial_{[t} E_{i]t}$.\\

Now using the (spatial part of the) reference metric we may decompose the spatial components $E_{ij}$ and our momenta $P_{ij}$ into their traceless parts, $\tilde{E}_{ij}$ and $\tilde{P}_{ij}$, and trace parts $\tilde{E}$ and $\tilde{P}$, as,
\be
E_{ij} = \tilde{E}_{ij} + \tilde{E} \delta_{ij} \; , \quad P_{ij} = \tilde{P}_{ij} + \tilde{P}\delta_{ij}\quad , \qquad \delta^{ij} \tilde{E}_{ij} = \delta^{ij} \tilde{P}_{ij} = 0\,.
\ee
We now regard the upper triangular components of the symmetric spatial traceless $\tilde{E}_{ij}$ (so $j \ge i$) as the dynamical variables of our massive gravity theory, in the sense that they have second order time evolution equations. As we will shortly discuss, the remaining components $\tilde{E}$ and $E_{it}$ have first order evolution equations from the vector equation, and the last component $E_{tt}$ is algebraically determined (at least for conventional matter) in terms of the other variables by the scalar constraint.
Thus we may write coordinates on the phase space as,
\be
\( {E}_{t i},  \tilde{E}, \tilde{E}_{ij}, P_i , \tilde{P}, \tilde{P}_{ij} \) \, ,
\label{eq:variables}
\ee
and then $E_{tt}$ is a function of these phase space variables and their first derivatives which we can regard as an auxilliary variable.
We now explicitly show how this works.

\subsection{Vector equation}

Focusing on the vector equation, $\xi_\mu = 0$, given in equation~\eqref{eq:vectorconstraint} then performing the 3+1 decomposition in phase space variables \eqref{eq:variables} and expanding  about flat space, so $E_{\mu\nu} \simeq \eta_{\mu\nu}$, we can write,
\ba
\label{eq:VectorSystem}
V^{t\alpha\beta\sigma}  \partial_{[\alpha} E_{\beta]\sigma}  &=&
2V^{t[ti]t}P_i + 2V^{t[ti]}_{~~~~~i}\tilde{P}+ 2V^{t[ti]j}\tilde{P}_{ij}  + V^{tij\sigma}\partial_{[i}E_{j]\sigma} \simeq  6 m^2 \tilde{P}  \nonumber \\
V^{i\alpha\beta\sigma}  \partial_{[\alpha} E_{\beta]\sigma}  &=&
2V^{i[tj]t}P_j + 2V^{i[tj]}_{~~~~~j}\tilde{P} + 2V^{i[tj]k}\tilde{P}_{jk} +V^{ijk\sigma}\partial_{[j}E_{k]\sigma} \simeq   -2  m^2 P_i\,,
\ea
where the approximation $\simeq$ is understood to mean up to terms that only involve spatial derivatives acting on $E_{\mu\nu}$.
Hence we may regard these 4 equations as linear constraints for the 4 momentum variables $\tilde{P}$ and $P_i$, and at least near flat space we may invert this linear system to solve for these momenta. These 4 momenta then depend on all the metric components $E_{\mu\nu}$, including $E_{tt}$, through the components of $V^{\mu\alpha\beta\sigma}$. They also depend linearly on spatial derivatives of metric components through $\partial_{[i}E_{j]\sigma}$, but crucially they do not depend on derivatives of $E_{tt}$.

\subsection{Scalar equation}

Now we turn to the scalar constraint $\Pi = 0$ given in \eqref{eq:Pi}, again assuming our matter is of a conventional type so that while the stress tensor depends on the metric, it does not explicitly involve metric connection terms.
As already observed above, this equation only depends on first derivatives of $E_{\mu\nu}$.
These enter through the combination $\partial_{[\mu} E_{\nu]\sigma}$, so as noted above, there are no $\dot{E}_{tt}$ terms. Furthermore spatial gradients of $E_{tt}$ come in the structure $\partial_{[t} E_{i]t}$, and hence are replaced  with the momenta $P_i$.
Therefore, we may write the scalar constraint in terms of our phase space variables~\eqref{eq:variables}, and their derivatives, together with $E_{tt}$ so that it contains no derivatives of $E_{tt}$ at all.

The dependence on the momenta is quadratic and will be given more explicitly below. First note that $(E^{-1})^{\mu\nu}$ can be written as,
\be
(E^{-1})^{\mu\nu} = \frac{1}{| E |} Q^{\mu\nu}\,,
\ee
where each component $Q^{\mu\nu}$ is a polynomial in those of $E_{\mu\nu}$, and linear in each one. Hence given the form of $A_{(1)}^{\alpha\beta\gamma\mu\nu\sigma}$ above we might have naively expected a quartic expansion of its derivative terms of the form,
\be
\label{eq:naivecoefficients}
| E |^4 A_{(1)}^{\alpha\beta\gamma\mu\nu\sigma} \partial_{[\alpha} E_{\beta] \gamma}  \partial_{[\mu} E_{\nu]\sigma}  & = & C_4 E_{tt}^4 + C_3 E_{tt}^3 + C_2 E_{tt}^2 + C_1 E_{tt}  + C_0\,,
\ee
with the coefficients $C_A$ depending on the components of $E_{\mu\nu}$ other than $E_{tt}$, together with the spatial gradients $\partial_{[i} E_{j] k}$ and also all the momenta, $P_i$, $\tilde{P}$ and $\tilde{P}_{ij}$, but no derivatives of these. Likewise taking $| E |^5 A_{(2)}^{\alpha\beta\gamma\mu\nu\sigma}$ we might have expected a quintic expansion in the component $E_{tt}$.
However, as we explain in detail in Appendix~\ref{app:scalardetails}, due to the 
index antisymmetries
of these two tensors, $A_{(1,2)}^{\alpha\beta\gamma\mu\nu\sigma}$, in fact we find simpler quadratic and cubic expansions going as,
\be
\label{eq:scalar_simplification}
| E |^2 A_{(1)}^{\alpha\beta\gamma\mu\nu\sigma} \partial_{[\alpha} E_{\beta] \gamma}  \partial_{[\mu} E_{\nu]\sigma}  & = & C'_2 E_{tt}^2 + C'_1 E_{tt}  + C'_0 \nl
| E |^3 A_{(2)}^{\alpha\beta\gamma\mu\nu\sigma} \partial_{[\alpha} E_{\beta] \gamma}  \partial_{[\mu} E_{\nu]\sigma}  & = & C''_3 E_{tt}^3 +C''_2 E_{tt}^2 + C''_1 E_{tt}  + C''_0\,.
\ee
Since the mass terms have similar structures, then for certain types of matter this constraint may determine $E_{tt}$ as the root of a polynomial.
As an example, consider matter that is a canonical scalar field $\phi$ with potential $V(\phi)$, so
\be
\label{eq:Tscalar}
T_{\mu\nu} = \partial_\mu \phi \partial_\nu \phi - \frac{1}{2} g_{\mu\nu} \left( (\partial \phi)^2 + V(\phi) \right)\,.
\ee
Now restricting ourselves to the case of a minimal mass term (so $m_2 = 0$), we can always scale the  scalar constraint by the determinant of $E$, and consider the constraint $| E |^2 \Pi=0$. The quadratic gradient term takes the form above. For the remaining terms, the explicit $E_{tt}$ dependence of the stress tensor term that enters takes an identical form,
\be
| E |^2 \left( m_1^2 \left( 3 [E] - 12 \right) - g^{\mu\nu} T_{\mu\nu} \right) = D'_2 E_{tt}^2 + D'_1 E_{tt}  + D'_0\,,
\ee
so in this minimal case with a canonical scalar field the scalar constraint is in fact a simple algebraic quadratic polynomial in $E_{tt}$.
We will give its explicit form in the case of spherical symmetry in our later numerical example, but emphasize that this reduction to a quadratic condition doesn't require any symmetry.
Including also the non-minimal mass term, so $m_1, m_2 \ne 0$, then considering $| E |^3 \Pi=0$ yields an algebraic cubic equation in $E_{tt}$ for such scalar field matter. \\

Since the scalar constraint is algebraic in $E_{tt}$ we may wonder whether we can solve it for real $E_{tt}$.
Near flat spacetime, $E_{\mu\nu} \simeq \eta_{\mu\nu}$, the scalar constraint reduces to the form in linear theory as written earlier in~\eqref{eq:linconstraint}, so in our variables,
\be
\label{eq:linscalarconst}
m^2 \left( - E_{tt} + 3 \tilde{E} \right) =  - \frac{2}{3} T\,,
\ee
and thus (given a non-zero mass) near flat spacetime we may always solve this for $E_{tt}$.
However when the geometry deforms away from flat spacetime non-linearly in $E_{\mu\nu}$ it is then an interesting question whether these algebraic relations can be solved for $E_{tt}$ (such that it is real). 
Since the EFT breaks down when the the algebraic relation can no longer be solved for real $E_{tt}$, this simply indicates a sensitivity on UV physics at that point. 
We will return to this issue in our explicit numerical example later.

\subsection{Physical degrees of freedom}

Starting with our phase space coordinates,
$
( \tilde{E}, E_{ti}, \tilde{E}_{ij}, \tilde{P}, P_i , \tilde{P}_{ij} )
$
and auxiliary $E_{tt}$,
the evolution of $\tilde{E}$, $E_{ti}$ and $E_{ij}$ is determined by,
\be
\label{eq:Kevolution}
\dot{\tilde{E}} = 2\tilde{P}+\frac{1}{3}\partial^i E_{ti} \; , \quad \dot{E}_{ti} = 2P_i + \partial_i E_{tt} \; , \quad \dot{\tilde{E}}_{ij} = 2 \tilde{P}_{ij} + \partial_i E_{tj} - \frac{1}{3}\delta_{ij} \partial^k E_{tk} \, .
\ee
However, we have now seen that $E_{tt}$ is determined algebraically by the phase space variables through the scalar constraint. This statement is fully non-linear and valid about any background.
 Further we have seen that the vector equation $\xi_\mu = 0$ gives linear constraints on $\tilde{P}$ and $P_i$, the coefficients in these linear equations again depending algebraically on the auxiliary variable $E_{tt}$. Thus the evolution of $\tilde{E}$, $E_{ti}$ is first order in time, and determined by this vector equation.
Hence we may reduce to the physical phase space of the theory, the component which enjoys a second order dynamics,
\be
( \tilde{E}_{ij}, \tilde{P}_{ij} )\,,
\ee
together with a first order dynamics for $( \tilde{E}, E_{ti} )$, given by the first two equations above in~\eqref{eq:Kevolution}, and the non-dynamical auxiliary $E_{tt}$ by solving the scalar and vector constraints simultaneously for $E_{tt}$, $P_i$ and $\tilde{P}$.\\

Writing $P_A = ( P_i, \tilde{P} )$, and taking $X$ to be the set of variables, $X = \{ \tilde{E}, E_{ti}, \tilde{E}_{ij} \}$, then schematically this system takes the form,
\be
\label{eq:ConstraintSystem}
M^{AB}(E_{tt}, X) P_A P_B + M^{A}(E_{tt}, X,  \partial_i X, \tilde{P}_{ij}) P_A + M(E_{tt}, X,  \partial_i X, \tilde{P}_{ij}) & = & 0 \nl
Q^{AB}(E_{tt}, X) P_B + Q^A(E_{tt}, X, \partial_i X, \tilde{P}_{ij}) & = & 0\,,
\ee
the former equation being the scalar constraint, the latter the vector one, and $\partial_i X$ denotes the set of spatial derivatives of $X$. We emphasize that this system is local to a spatial location $x^i$ since $P_A$ and $E_{tt}$ only enter algebraically, not through their derivatives.
Thus given the data $X$ on a time slice, hence we also have $\partial_i X$, then at any point $x^i$ on the slice we can consider this relatively simple system and solve for $P_A$ and $E_{tt}$. Explicitly one may solve the momenta as $P_B = - (Q^{-1})_{BA} Q^A$.
From the earlier equation~\eqref{eq:VectorSystem} we see we may always solve this linear system near flat spacetime.
 Then having solved for the $P_A$ we substitute them into the scalar equation to obtain a somewhat more complicated algebraic relation for $E_{tt}$ of the form,
\be
\label{eq:Kttconstraint}
\Pi(E_{tt}, X, \partial_i X, \tilde{P}_{ab}) = 0\,.
\ee
Thus given data on a time slice, $\tilde{E}$, $E_{ti}$, $\tilde{E}_{ij}$ and $\tilde{P}_{ab}$, we can then compute $\tilde{P}$, $P_i$ and $E_{tt}$ from this.
Having solved for these we have the evolution of the first order variables $E_{ti}$ and $\tilde{E}$, and what remains to close the dynamics is to obtain equations for the time derivatives $\dot{\tilde{P}}_{ij}$ which come from components of the Einstein equations. Then the second order dynamics $( \tilde{E}_{ij}, \tilde{P}_{ij} )$ gives the physical degrees of freedom of the theory, and given that $\tilde{E}_{ij}$ is the traceless part of the spatial components $E_{ij}$, correctly accounts for the 5 expected degrees of freedom.

\subsection{Evolution equations}
\label{sec:evolution}

To obtain the second order evolution equations we consider the components of the Einstein equation with one index raised, which we decompose as:
\be
\label{eq:decomp}
H_{\mu} = \mathcal{E}^t_{~\mu} \; , \quad  \mathcal{E}^i_{~j} = \tilde{\mathcal{E}}^i_{~j} + \tilde{\mathcal{E}} \delta^i_j\,,
\ee
so that $\tilde{\mathcal{E}}^i_{~i} = 0$. We note that since $g_{\mu\alpha} \mathcal{E}^\alpha_{~\nu} = g_{\nu\alpha} \mathcal{E}^\alpha_{~\mu}$ then $\mathcal{E}^i_{~t}$ is determined by a linear combination of the above components.

Now we consider the equation $\tilde{\mathcal{E}}^i_{~j}$ when it is written using our phase space variables $( {E}_{t i},  \tilde{E}, \tilde{E}_{ij}, P_i , \tilde{P}, \tilde{P}_{ij} )$ and $E_{tt}$.
The terms with time derivatives in the equations $\tilde{\mathcal{E}}^i_{~j}$ then take the form,
\be
\label{eq:linsys1}
 \tilde{\mathcal{E}}^i_{~j} =  \mathcal{F}^{i}_{~j} \dot{\tilde{P}} + \mathcal{F}^{ai}_{~~j} \dot{{P}_a} + \mathcal{F}^{abi}_{~~~j} \dot{\tilde{P}}_{ab} +  \mathcal{F}^{'i}_{~j} \dot{E}_{tt} + \ldots
 \ee
 where the coefficients depend on these phase space variables and $E_{tt}$, and the ellipses $\ldots$ are terms with no time derivatives (when written in these variables).
Near flat spacetime one simply finds,
$\tilde{\mathcal{E}}^i_{~j} \simeq \delta^{ia}\dot{\tilde{P}}_{aj} + \ldots$.
We note that one might have naively expected terms going as $(\dot{E}_{tt})^2$ as this would be a two derivative term, but we see from the form of the action in~\eqref{eq:action2} that it contains no time derivatives of $E_{tt}$, and thus in the equations of motion one can find at most one time derivative of $E_{tt}$. \\

Now since we impose vanishing constraints, $\xi_\mu$ and $\Pi$, which contain no time derivatives in our variables, we may also differentiate these with respect to time. Rewriting the time derivatives of $\tilde{E}$, $E_{ti}$ and $\tilde{E}_{ij}$ that are generated in terms of our momenta $\tilde{P}$, $P_i$ and $\tilde{P}_{ij}$,  they then have a similar form to the above, so,
\be
\label{eq:linsys2}
 0 = \dot{\xi}_\mu  &=&  \mathcal{G}_{\mu} \dot{\tilde{P}} + \mathcal{G}^{a}_{~\mu} \dot{{P}_a} + \mathcal{G}^{ab}_{~~\mu} \dot{\tilde{P}}_{ab} +  \mathcal{G}^{'}_{\mu} \dot{E}_{tt} + \ldots \nl
0 = \dot{\Pi}  &=&  \mathcal{H} \dot{\tilde{P}} + \mathcal{H}^{a} \dot{{P}_a} + \mathcal{H}^{ab} \dot{\tilde{P}}_{ab} +  \mathcal{H}^{'} \dot{E}_{tt} + \ldots\,.
 \ee
At every point on the  Cauchy surface, the equations~\eqref{eq:linsys1} and~\eqref{eq:linsys2} form a linear system which we may solve for $\dot{\tilde{P}}$, $ \dot{{P}_a}$, $ \dot{\tilde{P}}_{ab}$ and $\dot{E}_{tt}$.
Now having solved for $\dot{\tilde{P}}_{ab}$ we may use the definition of this momentum, in equation~\eqref{eq:Kevolution}, to find $\dot{\tilde{E}}_{ab}$, closing our dynamical system.
We note that while the solution of the above system gives $\dot{E}_{tt}$, 
$\dot{{P}_a}$ and $\dot{\tilde{P}}$ , since we are solving for these algebraically at every point, this information is redundant.

\subsection{Hamiltonian and momentum initial constraints}

While we have the expected 5 components of $\tilde{E}_{ij}$ which enjoy second order dynamics, we also have a first order dynamics in $\tilde{E}$ and $E_{ti}$ and one may wonder how to make sense of the data on the initial surface associated to these variables. However, as we now argue, just as in GR we have to satisfy the analogue of the Hamiltonian and 
momentum constraints on initial data, and we can take these to determine these 4 components $\tilde{E}$ and $E_{ti}$ in the initial data. Thus whilst these variables obey first order equations, since their initial data is constrained, there is no physical dynamics associated to them. \\

Recall in usual GR the Hamiltonian and momentum constraints are the Einstein equations $(\mathcal{E}_{(GR)})^t_{~\mu}$. Due to the contracted Bianchi identity and matter stress-energy conservation, we have,
\be
\label{eq:GRBianchi}
- \partial_t (\mathcal{E}_{(GR)})^t_{~\mu} = \partial_i (\mathcal{E}_{(GR)})^i_{~\mu} + \Gamma^\beta_{~\beta \alpha} (\mathcal{E}_{(GR)})^\alpha_{~\mu} - \Gamma^\alpha_{~\mu \beta} (\mathcal{E}_{(GR)})^\beta_{~\alpha}\,.
\ee
Provided the remaining Einstein equations $(\mathcal{E}_{(GR)})^i_{~j}$ hold, we see that $\partial_t (\mathcal{E}_{(GR)})^t_{~\mu} = c_\mu^{~i \nu} \partial_i (\mathcal{E}_{(GR)})^t_{~\nu} + c_\mu^{~\nu} (\mathcal{E}_{(GR)})^t_{~\nu}$ for some coefficients $c_\mu^{i \nu}$, $c_\mu^{~\nu}$. This, however, does not guarantee the constraints are satisfied, only that if they are initially true they will remain true. Thus we impose them as constraints on the initial data. \\

Returning to our massive gravity theory, having imposed the $\tilde{\mathcal{E}}^i_{~j}$, $\xi_\mu$ and $\Pi$ equations we may consider the analogous relations for the remaining Einstein equations $H_{\mu}$ and $\tilde{\mathcal{E}}$ in our decomposition~\eqref{eq:decomp}. Firstly given $\tilde{\mathcal{E}}^i_{~j} = 0$ we may write,
\be
g^{ij}g_{jk} \mathcal{E}^k_{~t}=g^{k\mu}g_{tt}H_{\mu}+\tilde{\mathcal{E}} g^{ij}g_{jt}\,,
\ee
and invert this as,
\be
\mathcal{E}^i_{~t} = \omega^{i\mu}H_{\mu} + \omega^i \tilde{\mathcal{E}}\,,
\ee
which implicitly defines the coefficients $\omega^i$ and $ \omega^{i\mu}$. At least near flat space this inversion is possible, and $\omega^{ij} \simeq - \delta^{ij}$, with $\omega^{it} = \omega^i \simeq 0$.
Using this expression we may write the vector and scalar constraint equations we are imposing as
\ba
0 &=& \nabla_\nu \mathcal{E}^\nu_{~t} = \dot{H}_t + \omega^{i\nu}\partial_i H_\nu +\omega^i \partial_i \tilde{\mathcal{E}}  + a_t^{~\nu} H_\nu + a_t \tilde{\mathcal{E}} \\
0 &=& \nabla_\nu \mathcal{E}^\nu_{~j} = \dot{H}_j + \partial_{j}\tilde{\mathcal{E}} + a_j^{~\nu} H_\nu + a_j \tilde{\mathcal{E}} \\
0 &=& \Pi = b^\mu H_\mu + b\, \tilde{\mathcal{E}}\,,
\ea
recalling that the vector equation comes from the divergence of the Einstein equations, and if it is satisfied, so $\xi_\mu = 0$, then from equation~\eqref{eq:pieq} $\Pi$ takes the above form when the equations $\tilde{\mathcal{E}}^i_{~j} = 0$ are satisfied. Here
 the various coefficients $a_\mu^{~\nu}$, $a_\mu$, $b^\mu$ and $b$ are functions of the metric variables and the connection, explicitly given as
\be
\begin{array}{rcl}
a_t^{~\nu} &=& \Gamma^{\rho}_{t\rho}\delta^{\nu}_t - \Gamma^\nu_{tt} + \omega^{i\nu}\Gamma^k_{ik} + \partial_i \omega^{i\nu} \\
a_t &=& \Gamma^k_{ik}\omega^i -\Gamma^k_{tk} + \partial_i \omega^i
\end{array}
\,, \quad
\begin{array}{rcl}
a_j^{~\nu} &=& \Gamma^{\rho}_{t\rho}\delta^{\nu}_j-\Gamma^\nu_{jt}-\Gamma^t_{ji}\omega^{i\nu} \nl
a_j &=& \Gamma^t_{jt} - \Gamma^t_{ji}\omega^i
\end{array}
\,, \quad
\begin{array}{rcl}
b^\mu &=& \frac{1}{2}m_1^2 \delta^\mu_t + \frac{1}{2}m_2^2\left(E^\mu_{~t}+E^t_{~j}\omega^{j\mu}\right) \\
 b &=& \frac{3}{2}m_1^2 + \frac{1}{6}m_2^2\left(\tilde{E} + 3E^t_{~j}\omega^j\right)\,.
\end{array}
\ee
Noting that we may eliminate $\tilde{\mathcal{E}}$ in terms of $H_\mu$ using the scalar constraint equation, which at least near flat space where $b \simeq \frac{3}{2} m_1^2 + \frac{1}{6} m_2^2$ will be possible, leaves the vector equation of the form,
\be
0 = \dot{H}_\mu + c_\mu^{i\nu}\partial_i H_\nu + c_\mu^{~\nu} H_\nu\,.
\ee
We see that this is precisely the same structure as in conventional GR in equation~\eqref{eq:GRBianchi} above -- to ensure that the full Einstein equation system $\mathcal{E}^\mu_{~\nu}$ is solved when evolving just the traceless spatial part $\tilde{\mathcal{E}}^i_{~j}$ and imposing the constraints $\xi_\mu$ and $\Pi$, it is sufficient and necessary to impose the Hamiltonian $H_t$ and momentum $H_i$ equations on initial data. Once imposed on the initial surface, they  remain true 
for all time without leading to any further constraints or needing to be imposed at each time. \\

These Hamiltonian and momentum constraints depend on the variables $E_{ti}$ and $\tilde{E}$, and thus we can think of using the freedom in their initial values to solve these initial constraints. Then, as claimed above, locally the free initial data is just that for the 5 second order degrees of freedom $( \tilde{E}_{ij}, \tilde{P}_{ij} )$.

\subsection{Summary}

Let us now summarize our evolution scheme. Starting with data on a Cauchy slice comprising the physical phase space variables $( \tilde{E}_{ij}, \tilde{P}_{ij})$ and first order variables $( \tilde{E}, E_i )$:
\begin{itemize}
\item
At each point on the Cauchy slice the constraint system~\eqref{eq:ConstraintSystem} can be solved for  $E_{tt}$, $P_i$ and $\tilde{P}$.
\item
These then yield the time derivatives of the first order variables $( \tilde{E}, E_i )$ using equation~\eqref{eq:Kevolution}.
\item
Now at each point the combination $\tilde{\mathcal{E}}^i_{~j}$, $\dot{\xi}_\mu$ and $\dot{\Pi}$ forms a linear system which may be solved for $\dot{\tilde{P}}_{ab}$ (and also
$\dot{\tilde{P}}$, $ \dot{{P}_a}$ and $\dot{E}_{tt}$, but these are not required). Then from~\eqref{eq:Kevolution} this yields the time derivatives of our dynamical variables $\tilde{E}_{ij}$.
\end{itemize}
If the data on the initial slice is chosen so that the Hamiltonian and 
momentum constraints $H = \mathcal{E}^t_{~t}$ and $H_i = \mathcal{E}^t_{~i}$ vanish, in addition to the vector and scalar constraints, then $H_\mu$ will remain zero under the evolution.

\section{Harmonic formulation for the minimal theory}
\label{sec:Harmonic}

Before turning to the issue of well-posedness, and demonstrating the above 3+1 decomposition scheme can be used to perform numerical gravitational collapse evolutions, we pause to briefly mention a different, elegant formulation of the dRGT theory where the vector constraint,
 if satisfied on the initial Cauchy surface,  is automatically satisfied for all times, and leads to no secondary constraint. 
We do this here only for the case of the minimal mass term. It can be done also for the quadratic mass term, but we do not detail this here as this alternate formulation is not our focus, and the resulting form is considerably more complicated than for the minimal mass term case. \\

For the minimal mass term, so setting $m_2 = 0$, there is a natural dynamical formulation that appears similar to the harmonic formulation for GR. For GR we may reformulate the (trace reversed) Einstein equation as (see for example~\cite{Wiseman:2011by} for a discussion of this),
\be
R^H_{\mu\nu} \equiv R_{\mu\nu} - \nabla_{(\mu} v_{\nu)} - \left( T_{\mu\nu} - \frac{1}{2} g_{\mu\nu} T \right) = 0\,,
\ee
which we can term the `harmonic' Einstein equation, where we have introduced terms involving a vector field constructed as,
\be
v^\mu = g^{\alpha\beta} \left( \Gamma^\mu_{~~\alpha\beta} - \bar{\Gamma}^\mu_{~~\alpha\beta} \right)\,,
\ee
where $\bar{\Gamma}^\mu_{~~\alpha\beta}$ is the connection of a smooth reference metric $\bar{g}_{\mu\nu}$. We note that since we have a difference of connections $v^\mu$ transforms correctly as a vector field globally. Now the principal part of the equation is,
\be
(R^H_{\mu\nu})_{\rm PP} = - \frac{1}{2} g^{\alpha\beta} \partial_\alpha \partial_\beta g_{\mu\nu}\,,
\ee
so that all components of the metric propagate according to the wave operator of the geometry itself. Thus initial data for the problem is $g_{\mu\nu}$ and $\dot{g}_{\mu\nu}$. However clearly generic initial data does not evolve to satisfy our original Einstein equation.
The key point is that the contracted Bianchi identity and matter stress-energy conservation implies,
\be
\nabla^2 v^\mu - R^\mu_{~\nu} v^\nu = 0 \, ,
\ee
so that provided $v^\mu = 0$ and $\dot{v}^\mu = 0$ on an initial Cauchy surface, then 
$v^\mu$
 will remain zero. If we choose $g_{\mu\nu}$ and $\bar{g}_{\mu\nu}$ and their first time derivatives to agree on the initial surface then $v^\mu = 0$ there. We may further choose our initial data $g_{\mu\nu}$ and $\dot{g}_{\mu\nu}$ so that the 4 conditions $\dot{v}^\mu = 0$ hold. These are precisely the usual Hamiltonian and momentum constraints. Then evolving for some choice of reference metric will yield a solution in the generalized harmonic gauge $v^\mu = 0$. \\

In massive gravity we naturally have a reference metric, and therefore one may wonder whether there is also such a harmonic formulation. For the mass terms considered here this is indeed the case.
We modify our Einstein equation~\eqref{eq:EinsteinEq} (with only the minimal mass term) using the quantity $\xi^\mu$ in equation~\eqref{eq:vectorconstraint}, whose vanishing gives the vector constraint, as follows;
\be
\label{eq:HarmEin}
\mathcal{E}^H_{\mu\nu} \equiv G_{\mu\nu} - \frac{2}{m_1^2} \left( \nabla_{(\mu} \xi_{\nu)} - \frac{1}{2} g_{\mu\nu}  \nabla \cdot \xi \right) + m_1^2 M^{(1)}_{\mu\nu} - T_{\mu\nu} = 0\,.
\ee
The principal part of this is,
\be
(\mathcal{E}^H_{\mu\nu})_{\rm PP} = \left(-\delta^{\beta}_{(\mu}\delta^{\gamma}_{\nu)}(E^{-1})^{\alpha\sigma}+\delta^{\alpha}_{\mu}\delta^{\beta}_{\nu}(E^{-1})^{\gamma \sigma}+E_{\rho(\mu}\delta^{\beta}_{\nu)}\eta^{\gamma\rho}g^{\alpha\sigma} - E_{\rho(\mu}\delta^{\gamma}_{\nu)}\eta^{\sigma\rho}g^{\alpha\beta}\right) \partial_\alpha \partial_\beta E_{\gamma\sigma}
\ee
and we see that this vanishes for the trace, $g^{\mu\nu} \mathcal{E}^H_{\mu\nu}$. Indeed the trace of this harmonic Einstein equation is precisely the scalar constraint $\Pi = 0$ (for $m_2 = 0$) and hence it contains no two derivative terms. For perturbations of flat space we see,
\be
(\mathcal{E}^H_{\mu\nu})_{\rm PP} &\simeq & - \partial^2 E_{\mu\nu} + \partial_\mu \partial_\nu E \; .
\ee
so the components of the traceless part of $E_{\mu\nu}$ obey wave equations.
Thus near flat space all but one linear combination of the $E_{\mu\nu}$ propagate by hyperbolic wave operators, with the remaining part being determined by the scalar constraint equation which is only first order in derivatives.
Rather than solving the vector equation $\xi_\mu = 0$ as a linear constraint on momenta as in our previous 3+1 decomposition, instead in this formulation
 the Bianchi identity implies,
\be
\nabla^2 \xi^\mu - R^\mu_{~\nu} \xi^\nu = 0\,,
\ee
so that if $\xi_\mu = 0$ and $\dot{\xi}_\mu = 0$ initially then the vector constraint $\xi_\mu = 0$  remains true under time evolution. The initial condition that $\xi_\mu = 0$ is  just the condition that the vector constraint is imposed on the initial data. Then the condition that $\dot{\xi}_\mu = 0$, together with the scalar constraint holding,  is  the condition that the Hamiltonian and momentum constraints hold initially.
Thus in this harmonic formulation,
we have a wavelike evolution for all but one linear combination of $E_{\mu\nu}$, and that is determined on every time slice by the scalar constraint. The vector constraint is only imposed initially, 
being automatically satisfied at all times, and no new constraint arises.

\section{A well-posed short distance completion}
\label{sec:wellposed}

Famously GR admits a well-posed hyperbolic formulation, as do some modified gravity theories, such as the Horndeski~\cite{Papallo:2017qvl,Kovacs:2020ywu} and Einstein-Aether~\cite{Sarbach:2019yso} theories.
When dealing with massive gravity, it is understood that even classically, it ought to be treated as a low-energy  effective field theory with operators entering at some cutoff scale $\Lambda_{\rm cutoff}$ as indicated in \eqref{eq:mGREFT} and hence only providing a meaningful description of the  long wavelength dynamics\footnote{Low energy gravity in string theory is another example where the leading low energy supergravity receives classical corrections; the $\alpha'$ higher derivative terms. However in this case the truncation to the leading supergravity (or at least its bosonic part) will be well-posed, but not at higher order. }.
 Since the issue of well-posedness is embedded in the short distance behaviour of the theory, whether the classical truncation of dRGT massive gravity without the higher order EFT operators is well-posed or not, is not a relevant physical question. However, when numerically simulating a theory, it is important to have  p.d.e.s
that are well-posed in the continuum limit.
Without a well-posed formulation, it is unclear what a numerical discretization of the system represents -- while a finite discretization will give a unique time evolution from initial data, one would not expect to be able to refine the discretization and obtain numerical solutions in a continuum limit. Hence the interpretation of any numerical solution is not well-defined (see earlier footnote~\ref{foot:cutoff}).
Even though we are only interested in the long wavelength dynamics of the system which should remain insensitive to the higher order EFT operators included in \eqref{eq:mGREFT}, these operators are of paramount importance to the well-posedness of the theory, and hence to its numerical simulation, if the truncation to the leading low energy theory is not well-posed itself. \\

Thus we should ask whether our 3+1 formulation is expected to give a well-posed initial value problem. We believe this is generally unlikely.
Certainly linearized perturbations of flat spacetime obey a well-posed hyperbolic system. 
However the scalar constraint changes away from flat space -- for linear perturbations about flat space $E_{tt}$ is simply determined by the stress tensor and $\tilde{E}$, as in~\eqref{eq:linscalarconst}, but non-linearly it involves terms quadratic in first derivatives of the second order dynamical variables $\tilde{E}^i_{~j}$ (as seen explicitly in~\eqref{eq:Pi}). Thus solving for $E_{tt}$ and then substituting its form into the evolution equations for $\tilde{E}^i_{~j}$ will change the derivative structure, and likely will lead to ill-posedness for dynamics away from flat spacetime.\\

While the truncated theory (very likely) lacks a well-posed initial value formulation,
 the higher order EFT operators naturally provide a  short distance completion to achieve well-posedness, see for instance \cite{Allwright:2018rut,Bernard:2019fjb,Cayuso:2020lca,Lara:2021piy,
 Figueras:2021abd,Gerhardinger:2022bcw,
Franchini:2022ukz,Barausse:2022rvg,Franchini:2022ukz}.
The precise effects of higher order EFT operators on the long-wavelength modes is irrelevant. However on short-distance modes these operators affect the behaviour in a way that can naturally lead to well-posedness of the theory as a whole. The aim of this work is not to prove that {\it every} completion leads to well-posedness (an unlikely outcome particularly when focusing on a specific formulation and gauge choice), but rather to show that the low-energy EFT we consider {\it can in principle} be embedded within a well-posed formulation and that low-energy observables are immune to the details of this high-energy-inspired formulation.\\

Instead of going back to the covariant formulation of higher order operators at the level of the action, a more pragmatic approach we will follow here is to include dissipative contributions directly at the level of our 3+1 formulation. These are understood to mimic the effect of higher order (covariant) EFT operators on short distance modes entering at the cutoff scale  $\Lambda_{\rm cutoff}$. In doing so, we will need to ascertain that adjusting the precise value of that scale bears little effects on low-energy physics. This will then ensure that the completed theory is diffusive, rather than hyperbolic, and well-posed. More details on how diffusive or higher order gradient terms arise from the UV completion of related types of theories are found in \cite{Gerhardinger:2022bcw}.  We now discuss the concrete inclusion of these terms and their effects on the posedness of the system. \\

Our dynamical system comprises the fields $\tilde{E}_{ij}$, ${E}_{j}$, $\tilde{E}$ and momenta $\tilde{P}_{ij}$ once we have algebraically eliminated $E_{tt}$, $P_i$ and $\tilde{P}$, and we may write this system as,
\be
\partial_t \tilde{P}_{ij} = \mathcal{S}_{ij} \; , \quad
\partial_t \tilde{E}_{ij} = \mathcal{U}_{ij}\; , \quad
\partial_t {E}_{i} = \mathcal{V}_{i}\; , \quad
\partial_t \tilde{E} = \mathcal{W} \, .
\ee
The latter three relations simply follow directly from the definition of our momenta in~\eqref{eq:momenta}. The first derives from solving~\eqref{eq:linsys1} and~\eqref{eq:linsys2} for $\partial_t \tilde{P}_{ij}$.\\

Focussing on the highest spatial derivative terms in these evolution equations, the equation for the time evolution of the momenta $\tilde{P}_{ij}$ contains second spatial derivatives of these fields,
\be
\mathcal{S}_{ij} = \mathcal{J}_{ij}^{~~klmn} \partial_k \partial_l \tilde{E}_{mn} + \mathcal{J}_{ij}^{~~klm} \partial_k \partial_l \tilde{E}_{m} + \mathcal{J}_{ij}^{~~kl} \partial_k \partial_l \tilde{E} +\ldots
\ee
where the ellipses include  terms with only first spatial derivatives acting on $\tilde{E}_{ij}$, ${E}_{j}$, $\tilde{E}$ and $\tilde{P}_{ij}$. The coefficient functions, the $\mathcal{J}$'s above, depend on the fields and also their first spatial derivatives (as they generally depend on the $E_{tt}$, $P_i$ and $\tilde{P}$, which when eliminated introduce first derivatives of the other fields).
The remaining evolution equations for the fields $\tilde{E}_{ij}$, ${E}_{j}$, $\tilde{E}$ only contain first order spatial derivatives.\\

While the Einstein equations are second order in spatial derivatives, the structure above is quite non-trivial in the sense that one might imagine $\mathcal{S}_{ij}$ should contain spatial derivatives of higher order for two reasons. Firstly, as discussed in section~\ref{sec:evolution} it derives not just from the Einstein equations $\tilde{\mathcal{E}}^i_{~j}$, as in~\eqref{eq:linsys1}, but also from time derivatives of the scalar and vector constraints, as in~\eqref{eq:linsys2}. Secondly we might imagine $\mathcal{S}_{ij}$ should contain spatial derivatives of higher order than two once $E_{tt}$, $P_i$ and $\tilde{P}$ are eliminated, since we know that these depend quadratically on first derivatives of the fields $\tilde{E}_{ij}$, ${E}_{j}$, $\tilde{E}$. \\

To address these points, we recall the fact noted earlier, that the action~\eqref{eq:action2} when written in our momentum variables $P_i$ and $P_{ij}$ is only algebraic in $E_{tt}$, having no terms with derivatives (time or space) acting on it. Further it is clearly algebraic in the $P_i$ and $\tilde{P}$. Hence being first order in derivatives, when this action is varied to obtain the Einstein equations, and in particular the components $\tilde{\mathcal{E}}^i_{~j}$, and these are written in our momentum variables, these contain at most first derivative terms in $E_{tt}$, $P_i$ and $\tilde{P}$.
In addition, since the scalar and vector constraints contain no derivatives in $E_{tt}$, $P_i$ and $\tilde{P}$, and only first derivatives in the other fields, then solving~\eqref{eq:linsys1} and ~\eqref{eq:linsys2} for $\partial_t \tilde{P}_{ij}$ still gives an expression that contains at most second spatial derivatives in $\tilde{E}_{ij}$, ${E}_{j}$, $\tilde{E}$, and first spatial derivatives in the $E_{tt}$, $P_i$ and $\tilde{P}$.
Now finally given the structure~\eqref{eq:ConstraintSystem}, so that $E_{tt}$, $P_i$ and $\tilde{P}$ depend on first  derivatives (albeit quadratically) in  the other dynamical fields, when they are eliminated to yield $\mathcal{S}_{ij}$ they will generate at most second derivative terms. \\

As motivated by the previous discussion and by the inclusion of higher order operators in our EFT \eqref{eq:mGREFT}, we now simply include additional  spatial diffusion terms given by the flat reference metric into each evolution equation as,
\ba
\partial_t \tilde{P}_{ij} = \mathcal{S}_{ij} + \ell^2 \delta^{mn} \partial_m \partial_n \tilde{P}_{ij} \; , \quad &&
\partial_t \tilde{E}_{ij} = \mathcal{U}_{ij} + \ell^2 \delta^{mn} \partial_m \partial_n \tilde{E}_{ij}\; ,\label{eq:diff1} \\
\partial_t {E}_{i} = \mathcal{V}_{i} + \ell^2 \delta^{mn} \partial_m \partial_n E_{i}\; , \quad &&
\partial_t \tilde{E} = \mathcal{W} + \ell^2 \delta^{mn} \partial_m \partial_n \tilde{E}\,,\label{eq:diff2}
\ea
where the scale $\ell$ is the length scale associated to the short distance completion.
Then for time scales $T$ and length scales $L$ such that,
\be
T \ll 1/\ell^2 \; , \quad L \gg \ell\,,
\ee
these diffusion terms will be irrelevant. Now to understand the character of the system we should linearize about a general background, and consider the highest derivative terms for a perturbation about this. We then see the highest derivative terms, which are those of second order, take the form,
\be
\partial_t \left(
\begin{array}{c}
\delta \tilde{P}_{ij} \\
\delta \tilde{E}_{ij} \\
\delta {E}_{i} \\
\delta \tilde{E}
\end{array}
\right) =
\left(
\begin{array}{cccc}
\ell^2 \delta_{i}^{m} \delta_{j}^{n}  \delta^{kl} & \mathcal{J}_{ij}^{~~klmn}  & \mathcal{J}_{ij}^{~~klm}  & \mathcal{J}_{ij}^{~~kl}  \\
 0 & \ell^2 \delta_{i}^{m} \delta_{j}^{n}  \delta^{kl} & 0 & 0 \\
 0 & 0 & \ell^2 \delta_{i}^{m}  \delta^{kl} & 0  \\
 0 & 0 & 0 & \ell^2  \delta^{kl}
\end{array}
\right)
\partial_k \partial_l \left(
\begin{array}{c}
\delta \tilde{P}_{mn} \\
\delta \tilde{E}_{mn} \\
\delta {E}_{m} \\
\delta \tilde{E}
\end{array}
\right) + \ldots\,,
\ee
where the ellipses represent terms that are lower order in derivative terms. In the previous expression, the coefficients $\mathcal{J}$ are understood to be evaluated on the background. Then on short scales the two derivative terms dominate, and we may think of the coefficient functions in the matrix controlling this term as approximately constant. To elicit the local behaviour, we write the perturbation in Fourier space as,
\be
\delta \tilde{P}_{ij} = a_{ij} e^{- \omega t} e^{i k_m x^m} \; , \quad \delta \tilde{E}_{ij} = b_{ij} e^{- \omega t} e^{i k_m x^m} \; , \quad \delta {E}_{i} = c_{i} e^{- \omega t} e^{i k_m x^m} \; , \quad \delta \tilde{E} = c e^{- \omega t} e^{i k_m x^m}
\ee
so that on small scales, locally we have,
\be
 \omega \left(
\begin{array}{c}
a_{ij} \\
b_{ij} \\
c_{i} \\
c
\end{array}
\right) \simeq
 \left(
\begin{array}{cccc}
\ell^2 \delta_{i}^{m} \delta_{j}^{n}  k^2 & \mathcal{J}_{ij}^{~~klmn} k_k k_l  & \mathcal{J}_{ij}^{~~klm} k_k k_l  & \mathcal{J}_{ij}^{~~kl} k_k k_l \\
 0 & \ell^2 \delta_{i}^{m} \delta_{j}^{n}  k^2 & 0 & 0 \\
 0 & 0 & \ell^2 \delta_{i}^{m} k^2 & 0  \\
 0 & 0 & 0 & \ell^2 k^2
\end{array}
\right)
 \left(
\begin{array}{c}
a_{mn} \\
b_{mn} \\
c_{m} \\
c
\end{array}
\right) + \ldots\,,
\ee
with $k^2 = \delta^{ij} k_i k_j$. Clearly $\omega$ is given by the eigenvalues of the matrix on the righthand side. However its upper triangular form implies that its eigenvalues are simply given by its diagonal entries. Hence we have,
\be
\label{eq:omega}
\omega = \ell^2 k^2\,,
\ee
for all the eigenvectors of this system, and thus all the field and momentum perturbations diffuse on small scales, governed by the diffusion length scale $\ell$.
Thus this diffusive short distance completion has a well-posed initial value formulation, for any positive diffusion constant $\ell$. \\

The formulation (\ref{eq:diff1}, \ref{eq:diff2}), motivated by the existence of a meaningful  completion allows the theory to enjoy a well-posed continuum that can then be discretized and numerically solved. We are taking here a pragmatic (unashamedly artificial) approach which should not be regarded as {\it the} actual physical completion, such as, for example one arising from integrating out additional massive degrees of freedom\footnote{A specific covariant example of how integrating out additional massive degrees of freedom leads to EFT operators that change the nature of the dispersion relation and ultimately lead to a trivial eigenvalue for the system was presented in \cite{deRham:2018red,deRham:2019wjj}.}. Rather it is a pragmatic proxy for what one would expect to arise.  We emphasize that classical dRGT must be completed by \emph{something}, but the precise details of what this completion is, is irrelevant to the  description of long wavelength phenomena. Despite being more artificial (and not formulated covariantly),  our formulation is very attractive from a numerical perspective, being simple, and also is very natural when using a (3+1)-phase space formulation -- the decomposition in time naturally defining the frame for diffusion.
We also emphasize that with a short distance completion, the theory is not guaranteed to be free from instabilities and pathologies. This is an independent question from that of well-posedness. Instabilities or pathologies may still arise in the long wavelength dRGT dynamics, but they will not be associated with arbitrarily short scales, and instead will be associated to the dynamical length scales in the problem, set by the length scale of the graviton mass, as well as the scales included in the initial data. Such instabilities or pathologies, should they exist, would be independent of the irrelevant diffusion terms of the completion, and then interpreted as physical phenomena of the long wavelength description, signalling its breakdown. If such phenomena arise, then a physical short distance completion would be required to continue dynamical evolution, rather than the artificial one we have introduced. We now turn to explicit simulation of the dRGT theory to illustrate  the above formulation with its diffusive short distance contributions.

\section{Spherical collapse in the minimal theory}
\label{sec:sphericalcollapse}

While the minimal theory is not thought to exhibit the non-linearity required to switch on  an active  Vainshtein mechanism, it is nonetheless interesting to explore what happens under gravitational collapse, even though we do not expect it to exhibit four-dimensional GR-like behaviour in the small mass limit.
Even when the mass is not small it is still a theory of gravity, by which we mean a theory of a dynamic spacetime, and hence it is interesting to understand its behaviour. For example do (exotic) black holes form when matter collapses? Can naked singularities form?
In what follows we therefore explore some aspects of its dynamics under the assumption of  spherical symmetry and, as expected, we shall indeed see very different dynamics to that of usual GR. The point of the following work is not to disfavor the minimal theory against known gravitational dynamics but rather to show a proof of concept of how the dynamical evolution can be followed through numerically in that simple (minimal) example. The application to the non-minimal model and to other physically relevant situations will be explored elsewhere. \\
To proceed, we commit to a specific matter content and consider gravitational collapse of a massless scalar field, $\Phi$, in spherical symmetry. The scalar equation of motion is $\nabla^2 \Phi = 0$ and
gives the matter stress tensor,
\be
T_{\mu\nu} = \partial_\mu \Phi \partial_\nu \Phi - \frac{1}{2} g_{\mu\nu} (\partial \Phi)^2 \; .
\ee
We use the coordinate invariance of the theory to choose spatial polar coordinates for the metric and Minkowski reference metric, $x^\mu = (t, r, \theta, \phi)$ writing,
\be
E_{\mu\nu} = \left(
\begin{array}{cccc}
c & r\, h & 0 & 0 \\
& b + 2 r^2 a & 0 & 0 \\
& &  r^2 \left( b - r^2 a \right) & 0 \\
& & & r^2 \left( b - r^2 a \right) \sin{\theta}^2
\end{array}
\right)
\,, \qquad
f_{\mu\nu} = \left(
\begin{array}{cccc}
-1 & 0 & 0 & 0 \\
& 1 & 0 & 0 \\
& & r^2 & 0 \\
& & & r^2 \sin{\theta}^2
\end{array}
\right)\,,
\ee
so that regularity of the metric at the origin implies that $a, b, c, h$ should be smooth functions of $r^2$ there.
Earlier, our Minkowski reference metric was expressed in Minkowski coordinates. The only change to the previous discussion from the use of spherical spatial coordinates is at the level of the scalar and vector constraints, and in the definition of the momenta in sections~\ref{sec:constraints} and~\ref{sec:formulation}, where
the spatial partial derivatives, $\partial_i$,
 are  now replaced by spatial covariant derivatives with respect to these spherical coordinates in Minkowski. We have,
\be
\bar{\nabla}_{[t} E_{i]\mu} = \partial_{[t} E_{i]\mu} - \tilde{\Gamma}^\nu_{\mu[t} E_{i]\nu}\,,
\ee
where $\bar{\nabla}$ is the covariant derivative with respect to the reference metric $f_{\mu\nu}$, so the Minkowski metric in the spherical spatial coordinates. Since  $\tilde{\Gamma}^\nu_{\alpha\beta}$ vanishes if any of its indices equal time, then the connection terms vanish in the momenta $P_i = \bar{\nabla}_{[t} E_{i]t}$.
However they do contribute to, 
${P}_{ij} =\bar{\nabla}_{[t} E_{i]j}$ and its trace part $\tilde{P} = \frac{1}{3} f^{ij} \bar{\nabla}_{[t} E_{i]j}$.
We then find first order dynamical variables and momenta,
\be
E_{tr} &=& r \, h \; , \quad P_r = \frac{1}{2}\left(\partial_t E_{tr} - \partial_r E_{t t}\right)  = \frac{1}{2} \left( r \dot{h}  - \partial_r c \right) \nl
 \tilde{E} &=& b \; , \quad \tilde{P} =  \frac{1}{6}\left(3 \partial_t \tilde{E} - \partial_r E_{t r} + f^{ij}\tilde{\Gamma}^\nu_{ij}E_{t\nu}\right) = \frac{1}{2} \left(  \dot{b}  - \frac{1}{3}\partial_r h -  h \right)\,,
\ee
with the second order geometric variable being,
\be
\tilde{E}_{rr} = 2 r^2 a \; , \quad \tilde{P}_{rr} = \frac{1}{2} \left(\partial_t E_{rr}  -\partial_r E_{tr} +\tilde{\Gamma}^\nu_{rr}E_{t\nu} \right) - \tilde{P}f_{rr} = r^2 \dot{a} - \frac{1}{3} r \partial_r h\,,
\ee
together with the matter scalar $\Phi$ and its momentum $p_\Phi = \dot{\Phi}$. The remaining component,
\be
E_{tt} = c \, ,
\ee
will be algebraically determined by the scalar constraint. \\

For convenience we define the following quantities from the momenta,
\be
 p_h = \frac{2}{r} P_r \; , \quad
    p_b =  2 \tilde{P}  \; , \quad
    p_a =  \frac{1}{r^2}\tilde{P}_{rr}\,,
\ee
so that $\dot{h} = p_{h}  + \ldots$, $\dot{b} = p_{b}  + \ldots$ and $\dot{a} = p_{a} + \ldots$, where terms with no time derivatives are included in the ellipses $\ldots$. This will allow us to formulate the evolution equations for the metric functions $a$, $b$ and $h$. We see that the metric functions $h$ and $b$ are associated to $P_i$ and $\tilde{P}$ and thus have a first order dynamics, since these momenta are linearly determined by the vector constraint and we may algebraically eliminate them. Thus it is the function $a$ with its momentum $p_a$ that embodies the second order dynamics in the metric, and $c$ is determined algebraically.

In this formulation, the vector equation has non-trivial components,
\be
V_t &=& \frac{3 p_b \left(a r^2+b\right)+2 h r \left(5 a r+r^2 \partial _ra-\partial _rb\right)-6 a r^4 p_a}{\left(b-a
   r^2\right) \left(r^2 \left(2 a c-h^2\right)+b c\right)} \nl
V_r &=& \frac{10 a c r-2 h r^3 p_a-a r^3 p_h+2 h r p_b+b r p_h+2 c r^2 \partial _ra-2 c \partial _rb}{\left(b-a r^2\right)
   \left(r^2 \left(2 a c-h^2\right)+b c\right)}\,,
\ee
which we note are linear in $p_h$ and $p_b$,
and the scalar constraint takes the quadratic form, $a_2 c^2 + a_1 c + a_0 = 0$, with the precise formulae for the coefficients $a_i$ given in Appendix \ref{app:scalardetails}.

These vector and scalar constraints then form an algebraic system for determining $c$, $p_h$ and $p_b$ in terms of the other variables.
While the scalar constraint takes a form above that is quadratic in $c$, it is worth emphasizing that since it depends on $p_b$ and $p_h$, once we solve for these from the vector constraint (which also contains $c$), then the resulting algebraical equation for $c$ takes a complicated form. Thus in what follows we solve this system for $c$, $p_b$ and $p_h$ numerically at each point, rather than analytically eliminating these variables. \\

The Einstein equation $\tilde{\mathcal{E}}^r_{~r}$ together with the scalar field equation then determines the second order dynamics giving $\dot{\tilde{P}}_{rr}$ and $p_\Phi$.

\subsection{Strong coupling}

Since the scalar constraint is a quadratic in $c$, we may formally write its solution as,
\be
\label{eq:quadratic}
c = - \frac{1}{2} \left( \frac{a_1}{a_2} \pm \sqrt{\Delta} \right) \; , \quad \Delta = \frac{a_1^2}{a_2^2} - 4 \frac{ a_0}{a_2} \; .
\ee
We will call $\Delta$ the `discriminant' even though it differs in normalization from the usual definition.
We call the `positive branch' the solution with the `$+$' sign and the `negative branch' that with the `$-$' sign.
Linearizing about flat spacetime,
\be
c = -1 - \epsilon \delta c \; , \quad b = 1 + \epsilon \delta b \; , \quad h= \epsilon \delta h \; , \quad a = \epsilon \delta a \; , \quad \Phi &=& \sqrt{\epsilon} \delta \Phi \, ,
\ee
then we find,
\be
a_2 & \simeq & - 3 m^2 + \epsilon \left(  (\partial_r \delta \Phi)^2 - 21 m^2 \delta b - 6 m^2 r^2 \delta a \right) \nl
a_1 & \simeq & - 3 m^2 - 6 \epsilon m^2 \left( 2 \delta b + r^2 \delta a \right) \nl
a_0 & \simeq & - \epsilon (\partial_t \delta \Phi)^2 \nl
\Delta & \simeq & 1 - \frac{2 \epsilon}{3 m^2} \left(  9 m^2 \delta b  + 2 (\partial_t \delta \Phi)^2 - (\partial_r \delta \Phi)^2 \right)\,.
\ee
We recall that we are considering just the minimal mass $m_2=0$ and we will rescale units such that $m=m_1=1$.  Thus $\Delta \simeq 1$ for small perturbations about flat spacetime, and furthermore, $c$ is given by the \emph{positive} solution in~\eqref{eq:quadratic} above. \\

When non-linear effects associated to strong coupling conspire so that the vierbein components strongly deviate from their flat spacetime values then two interesting pathologies may occur with the scalar constraint.
Firstly $\Delta$ may appear to become negative, indicating that no (real) solution to the scalar constraint can exist for $c$. More precisely as $\Delta \rightarrow 0^+$ the solution will become infinitely strongly coupled and the EFT breaks down before this point. This is easy to see if we naively perturb around the solution with $\Delta =0$ then we would obtain say $\delta c \sim \mp\frac{1}{2} \sqrt{\delta \Delta}$ which is inconsistent in perturbation theory if $\delta \Delta$ has a first order perturbation. Requiring that $\delta \Delta$ starts at second order imposes a restriction on the variables that is indicative of a degree of freedom being lost, i.e. its kinetic term vanishing. This is the tell-tale sign of infinite strong coupling, \\

A second feature that may occur is that the quadratic form linearizes with $a_2 \to 0$ (with $a_1$ remaining finite). Whilst we would not normally regard this as a pathology for a quadratic equation, depending on the sign of $a_1$ this then picks a particular branch. The `correct' branch for $a_1 > 0$ is the positive branch, and for $a_1 < 0$ it is the negative branch, and then the usual linear solution is reproduced $c \simeq -a_0/a_1$ in this limit $a_2 \to 0$. However being on the opposite `wrong' branch implies $c \simeq - a_1/a_2$ and diverges as $a_2 \to 0$.
We note that for flat spacetime we have $a_1 < 0$ and are on the positive branch. If a situation arose where $a_2 \to 0$ with $a_1$ staying the same sign this would correspond to being on the `wrong' branch, and the solution for $c$ would diverge which in itself indicates infinite strong coupling and the breakdown of the EFT. \\

We may identify these two pathologies by either $\Delta$ tending to zero (strictly small values) or alternatively $\Delta$ diverging positively as $a_2 \to 0$. In this latter case $c$ would diverge positively. Later we will see that for certain choices of initial data indications of both these behaviours in the non-linear collapse dynamics.

\subsection{Initial data}

We begin with initial data that is an approximately in-going pulse of the scalar field, starting initially away from the origin. We must then solve the Hamiltonian and momentum constraints as well as the scalar and vector constraints. We choose the width of the scalar pulse to be approximately the length scale associated to the graviton mass, $\sim 1/m_1$. In doing so we depart very much from the phenomenologically interesting regime of massive gravity since this would presuppose a spherical symmetric source of the size of the Hubble radius. For our purposes this is merely a proof of principle that the dynamical formulation we have developed is well defined. It is beyond the scope of this paper to consider the type of hierarchies and boundary conditions needed for phenomenological applications. \\

Thus from now on we choose units so that $m_1 = 1$. For the results we present here we take an initial approximately Gaussian profile for the scalar, localized at a radius of $r \simeq 2$,
with a momentum profile that in flat spacetime would give a purely in-going pulse,
\begin{align}
    \Phi(t = 0, r) &=  \frac{A}{A_0} r^4 \exp{\left(-\frac{(r^2 - 2)^2}{10} \right)} \\
    p_{\Phi}(t = 0,r) & = \partial_r \Phi(t = 0, r) + \frac{1}{r} \Phi(t=0,r) \; .
\end{align}
Here $A_0$ is chosen so that $A = \max\left( \Phi(t=0,r) \right)$ is a constant giving the maximum amplitude of the scalar profile.
The metric function $a$ is the one that has second order dynamics, and we initially choose it to have its flat space value (which is zero), and vanishing momentum, so that,
\be
a(t = 0,r) = p_a(t = 0,r) = 0 \; .
\ee
Recalling that the momenta $p_h$ and $p_b$ are eliminated using the vector constraint, then it remains to give the metric functions $c$, $h$ and $b$ to determine the initial data. Now $c$ is determined from the scalar constraint, but we must also solve the Hamiltonian and momentum constraints, giving the two conditions for $h$ and $b$. These
involve second spatial derivatives of $b$, and first derivatives of $h$. \\

We may solve the non-linear system of scalar, Hamiltonian and momentum constraints by using an iterative relaxation method or Newton's method -- we have implemented both. Since our scalar pulse is quite far from the origin, and we start it with relatively small amplitude, the solution is close to the solution to the linearized system which is easily determined by taking the metric functions close to their flat spacetime values,
\be
c(t=0,r) &=& -1 - \epsilon \delta c \nl
b(t=0,r) &=& 1 + \epsilon \delta b \nl
h(t=0,r) &=& \epsilon \delta h \nl
\Phi(t=0,r) &=& \sqrt{\epsilon} \delta \Phi\,,
\ee
where one then finds $\delta b$ is determined from the o.d.e.,
\be
\label{eq:linb}
   2  \delta b'' +\frac{4 \delta b'}{r} -3 m^2 \delta b = -\frac{\delta \Phi^2}{2 r^2} -\frac{\delta \Phi \delta \Phi'}{r}-\delta \Phi'^2
\ee
which may be solved by quadrature with the boundary condition that $\delta b \to 0$ as $r \to \infty$, and is regular at the origin.
The remaining $\delta h$ and $\delta c$ are given algebraically in terms of $\delta \Phi$ and the solution to $\delta b$ as,
\be
 -3 m^2 r^2 \delta c = 9 m^2 r^2 \delta b+2 r \delta \Phi  \delta \Phi' +\delta \Phi^2 \; , \quad
   m^2 r \delta h = \delta \Phi' \left(\delta \Phi' +\frac{\delta \Phi}{r}\right) \; .
\ee
We may use this linearized approximate solution as an initial guess to solve the scalar, Hamiltonian and momentum constraints by an iterative relaxation or Newton's method. \\

For convenience we compactify the radial coordinate as $\tilde{r} = r/(1-r^2)$, so that the real axis is compactified to the interval $[0, 1]$ for $\tilde{r}$. We then employ 6th order spatial differencing. For the iterative relaxation scheme we use a method analogous to Gauss-Seidel for the Poisson equation, solving each of the three equations in turn at each lattice point, then moving to the next until the whole grid has been covered, and then we repeat until convergence is reached. This method, while crude, works well and is straightforward to implement, giving the same results as the Newton solver. \\

\begin{figure}
 \centerline{
  \includegraphics[width=8cm]{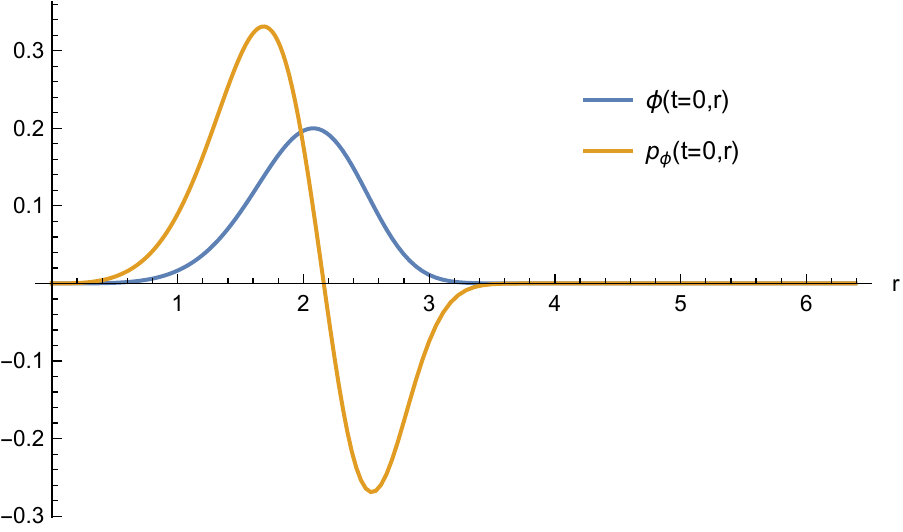}    \hspace{0.5cm}  \includegraphics[width=8cm]{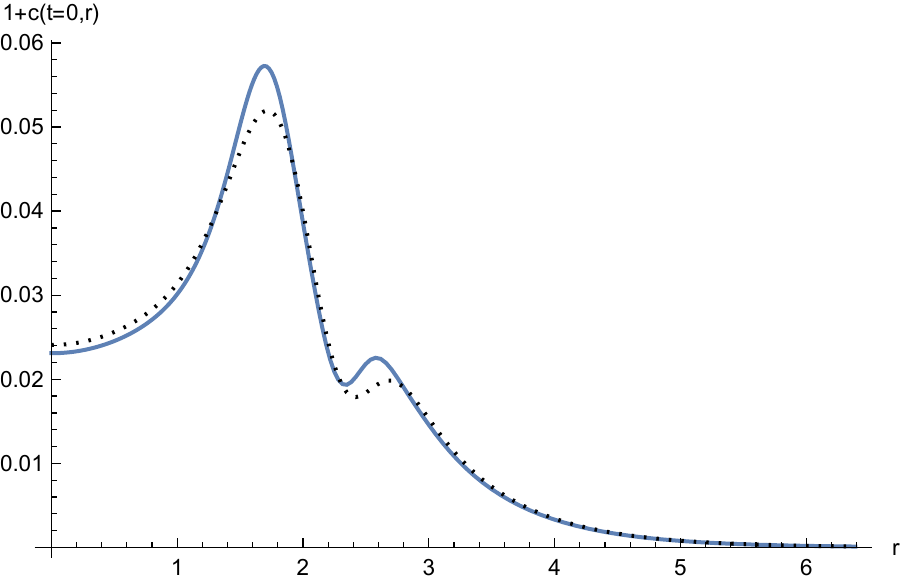}
  }
  \vspace{0.25cm}
  \centerline{
  \includegraphics[width=8cm]{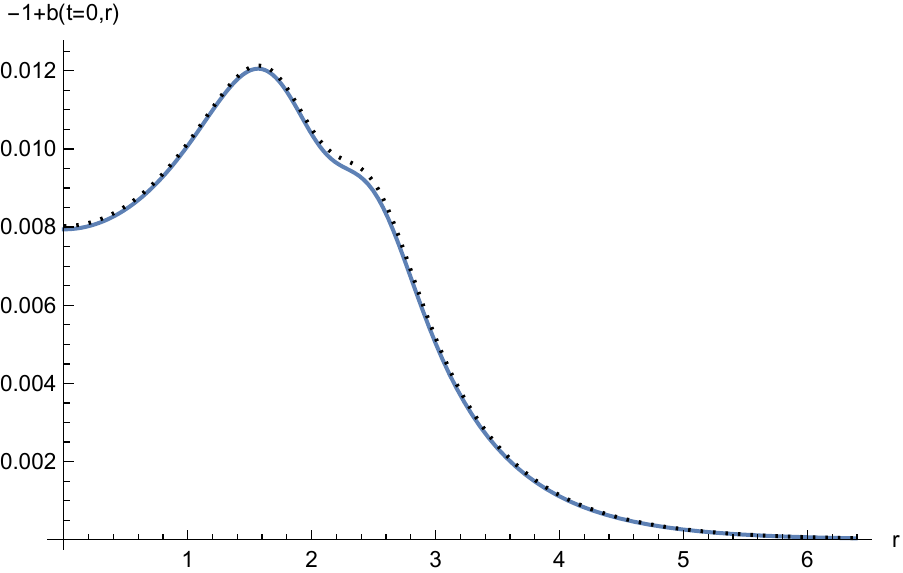}   \hspace{0.5cm}   \includegraphics[width=8cm]{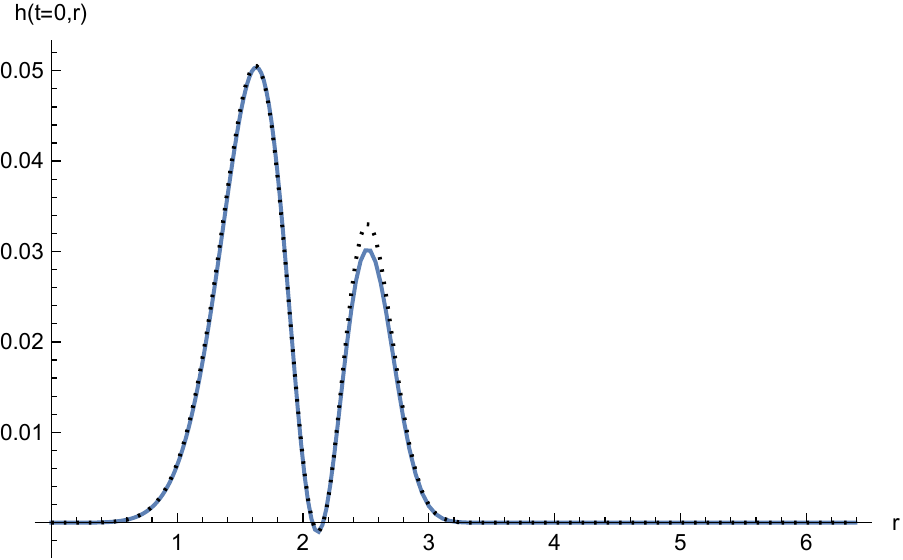}
  }
  \caption{ \label{fig:initialdata}
The top left plot shows the initial radial profile for the scalar $\Phi$ and its momentum $p_\Phi$ for amplitude $A = 0.20$ (in our units where the mass $m=1$). Solving the scalar, Hamiltonian and momentum constraints with this matter, we then obtain the initial metric functions $c$, $b$ and $h$, shown as solid curves in the top right, bottom left and bottom right frames respectively. Recall we are choosing the remaining metric function $a$ and its momentum to vanish. In order to solve this coupled non-linear system we use relaxation or Newton's method, with an initial guess given by the linear approximation -- these linear solutions are shown as dotted curves for each of these metric functions, and since the scalar pulse starts quite far from the axis with relatively small amplitude, it is close to the full non-linear solution.
}
\end{figure}

An example of this initial data if given in Fig.~\ref{fig:initialdata} for one of the largest amplitudes, $A = 0.2$ used later in the discussion.
In the figure we have plotted both the full non-linear solution as well as the linearized approximation, which can be seen to be close.

\subsection{Evolution}

We then evolve this initial data by imposing the scalar constraint, vector constraint and evolution equation $\tilde{\mathcal{E}}^r_{~r}$ together with the matter scalar equation.
As for finding the initial data, we use the compactified radial coordinate $\tilde{r} = r/(1-r^2)$, and 6th order finite differencing for spatial derivatives on the interval $[0, 1]$ in $\tilde{r}$. For time derivatives we use an implicit Crank-Nicolson differencing scheme. We solve this implicit system using iterative relaxation. As mentioned above, while we may in principle solve the vector and scalar constraints to eliminate $c$, $p_h$ and $p_b$ from the remaining equations, in practice since we implement an implicit scheme which must be solved at each time step anyway, we have found it convenient to simply solve the constraints as part of this implicit system.

Recall that we are using units for which the mass $m_1 =m=1$. We take initial data to be a pulse with approximately unit width. Thus all scales in the problem are comparable, and the parameter we now vary is the scalar pulse amplitude $A$. For small amplitude we expect the theory to be well described by linear dynamics, where the pulse will travel in to the origin, pass through it and then disperse to infinity. As we increase the amplitude $A$ we expect to see non-linear behaviour, and it is this that is our focus. The natural question is whether in this massive theory of gravity we see a horizon form, or some different non-linear phenomena. \\

In order to ensure that our continuum 3+1 system is well-posed under time evolution, we complete it at short distance by adding the diffusion terms as discussed in the previous section.
Focussing on the long wavelength physics and typical dynamical timescales we will see we are insensitive to this term, provided the diffusion constant, $D = \ell^2$, is sufficiently small.
Without the diffusion term, so setting $D = 0$, we find some short distance instability on the scale of the lattice spacing when employing resolutions greater than $N > 100$ points, and becoming more severe for evolutions that deviate further from flat space. The instability typically arises near the origin, and it is unclear at this stage whether this is a result of the ill-posedness of the continuum equations (without diffusion) or whether it is an artefact of the numerical discretization -- recall that even well-posed continuum equations may have lattice scale pathologies depending in detail on the numerical scheme chosen to discretize them\footnote{It is interesting to note that the most well behaved p.d.e., the diffusion equation itself, it unstable numerically when using explicit time differencing, and even using an implicit Crank-Nicholson scheme it requires the time step to be sufficiently small to avoid lattice scale instabilities.}.
 Taking $D = 0.001$ stabilizes the system for all resolutions considered here, up to the highest we have implemented $N = 1600$, and this value of $D$ is the one used to make the figures presented here unless otherwise stated. As discussed later, by varying the value of this diffusion constant, we can confirm that this value is sufficiently small that is has essentially no impact on the solutions we find, and the diffusion term is irrelevant on the length scales and time scales of interest, ie. those of order $\sim \mathcal{O}(1)$ in size.\\

We are able to simulate for a range of lattice spacings. The data we show here is for $N = 400$ lattice points, which gives very good accuracy with our 6th order spatial finite difference. A small time step is required for stability of the Crank-Nicholson scheme, and in the data shown we typically take $\Delta t = 0.00002$.
Refining $N$ indeed shows our code converges to a good continuum limit, and we give more details of this convergence in Appendix~\eqref{app:numdetails}. However, since we modify the short distance physics using the diffusion terms, and we regard these short scales as being beyond the validity of our EFT, in practice we find our resolution of $N = 400$ is sufficient to reproduce the long distance physics of interest for $D = 0.001$.
Taking higher resolutions shows our discretization properly approaches the continuum given by the well-posed low energy truncation together with short distance diffusive completion, but refining past $N = 400$ accesses the scales dominated by this diffusion and does not reveal the low energy physics we are interested in more accurately. \\

Finally the Hamiltonian and momentum constraints, once satisfied for the initial data, are preserved during the evolution if one takes only the low energy truncation. However, with the diffusion terms added then already at the level of the continuum p.d.e.s the constraints will no longer be exactly preserved under evolution.
In Appendix~\eqref{app:numdetails} we study the violation of these constraints under evolution as we change the resolution, $N$, and also the diffusion constant $D$. As expected we find that for a given small diffusion constant $D$, refining $N$ makes this constraint violation smaller until some value of $N$ past which the violation is caused by the diffusion terms at the level of the continuum p.d.e.s and is not due to numerical discretisation error. For smaller $D$, a larger $N$ is reached before diffusion dominates the violation, and the smaller the constraint violation becomes. Again for our typical choice of $D = 0.001$ we find that for $N = 400$ the constraint violation is small, and is dominated by the diffusion terms rather than numerical discretization error.

\subsection{Low amplitude initial data}

\begin{figure}
\centerline{
  \includegraphics[width=7cm]{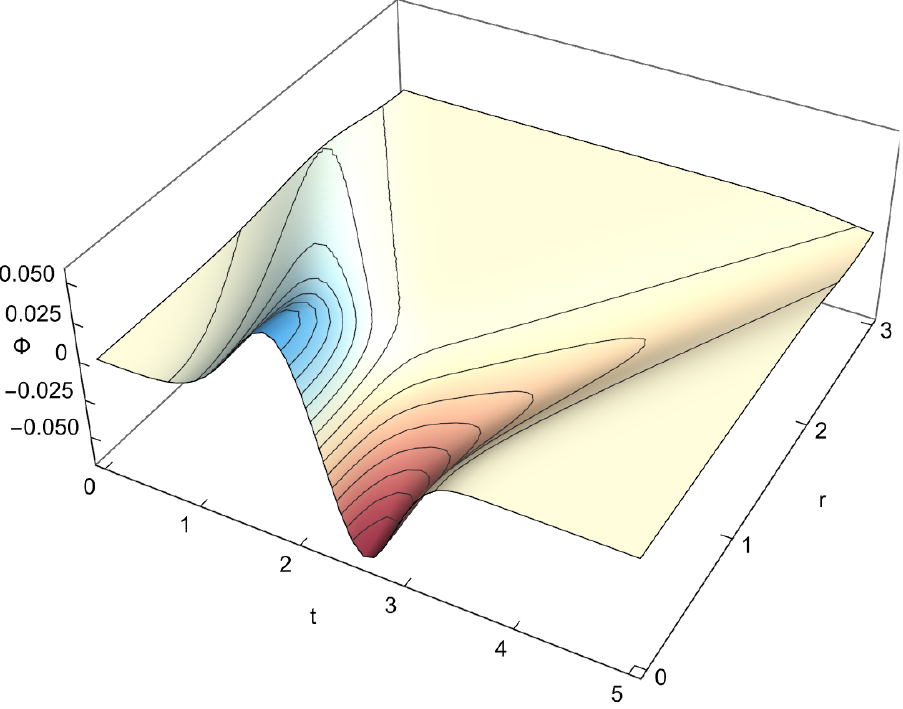}  \hspace{0.5cm}  \includegraphics[width=7cm]{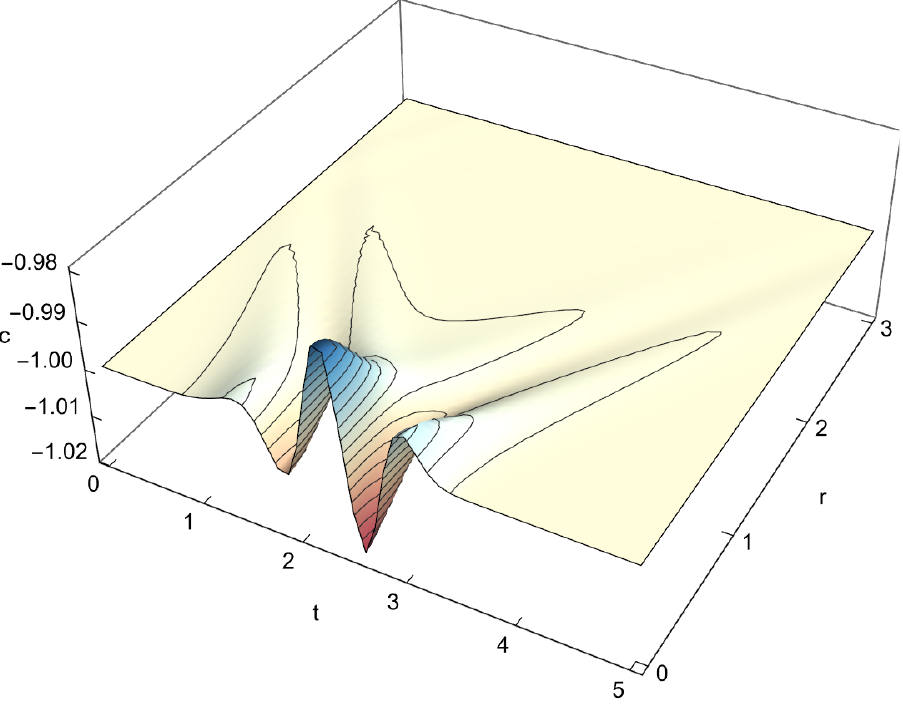}
  }
  \centerline{
\includegraphics[width=7cm]{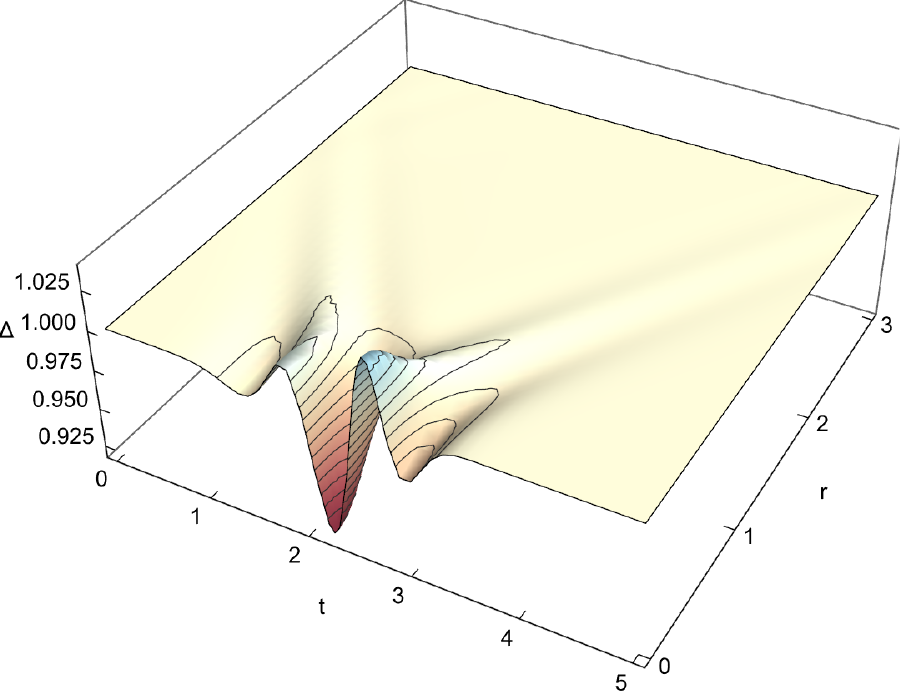}  \hspace{0.5cm}   \includegraphics[width=7cm]{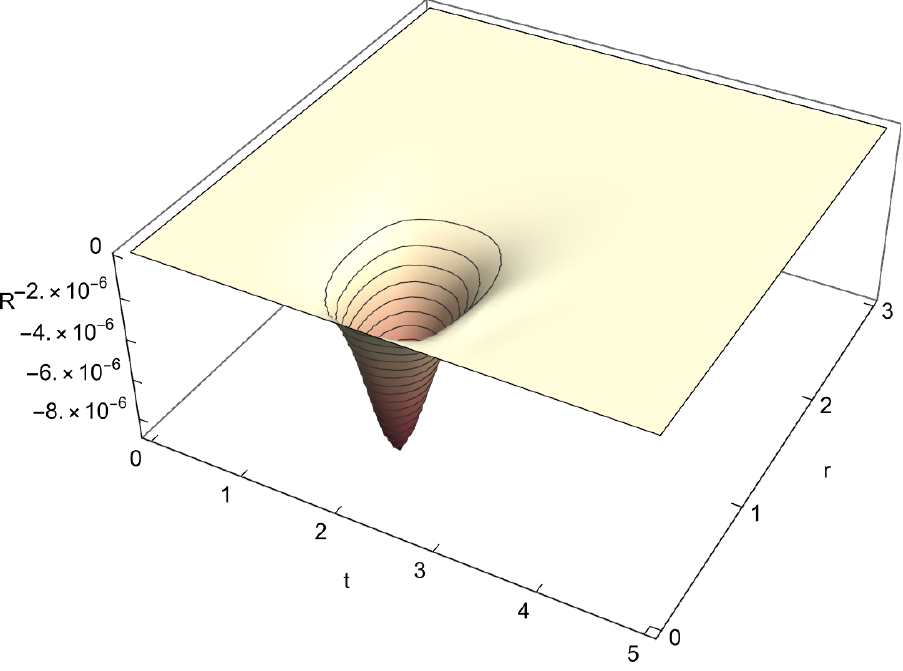}
  }
    \centerline{
    \includegraphics[width=7cm]{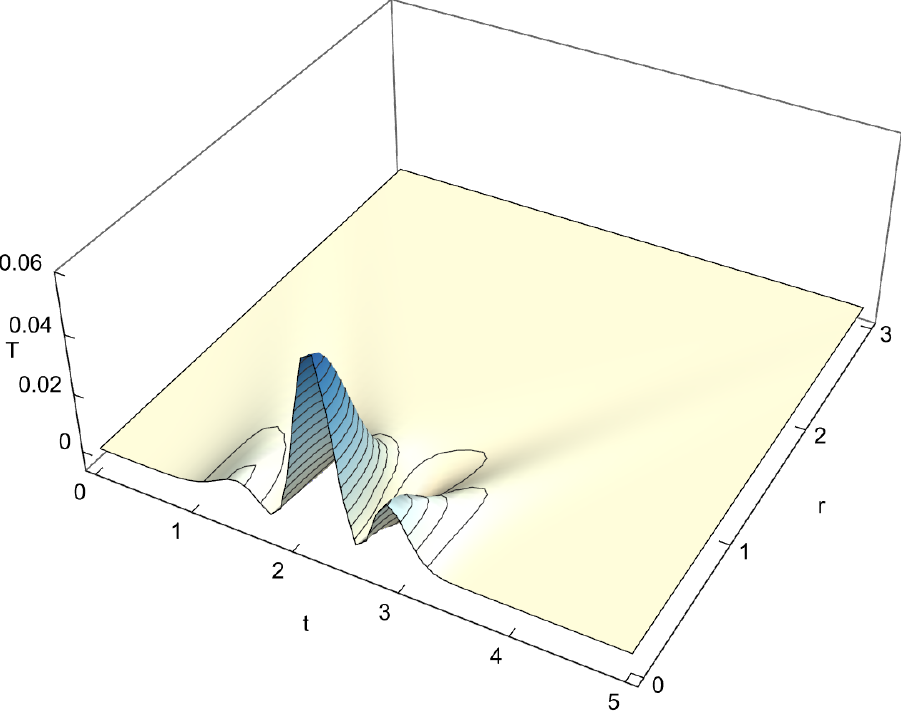}
  }
  \caption{\label{fig:weak}
  Figure showing the scalar field for weak initial data with $A = 0.01$ in the top left hand frame. During the evolution all the metric components remain close to their values for Minkowski spacetime, with the in-going scalar pulse passing through the origin and then dispersing. The vierbein component $c$ is shown in the top righthand frame, and the other metric components have a response of similar amplitude. The middle left figure shows the discriminant function, $\Delta$, for the quadratic scalar constraint.  The middle righthand figure shows the Ricci scalar $R$ and the bottom plot shows trace of the stress tensor $T$. For linear perturbations the Ricci scalar vanishes in massive gravity (unlike GR where $R = - T$).
    }
\end{figure}

Firstly let us consider the collapse of a small amplitude scalar pulse, taking small $A = 0.01$, so that the subsequent evolution is a weak perturbation of Minkowski spacetime, bearing in mind that this is 
far from a phenomenologically realistic situation.
 The initial in-going pulse of the scalar traverses towards the centre, increasing in amplitude as it becomes increasingly focussed. It then reaches the origin, with still a relatively small peak height of $|\phi| \sim 0.05$,  passes though, and subsequently disperses. In Fig.~\ref{fig:weak}, we show the scalar field as a function of time and the radial coordinate. The behaviour is qualitatively similar to that for GR, although obviously in detail differs due to the mass. Also in the figure the vierbein function $c(t,r)$ is shown, and we see that this metric component deviates from its flat space value of zero with a maximum amplitude of $\sim 0.02$.
The same is true for the other components. As a result the discriminant function, also shown, remains very close to its Minkowski vacuum value $\Delta \simeq 1$. \\

To examine the nature of the spacetime, we further show that the Ricci scalar deviates from its trivial zero value at order $\mathcal{O}\(10^{-5}\)$, and the trace of the stress tensor $T^\mu_{~\mu}$ has amplitude $\sim 0.05$ as the pulse traverses the origin. Recall that $m = 1$ in our units, and all scales in the initial data are $\sim \mathcal{O}(1)$ too. We see that the Ricci scalar is much smaller than this stress tensor trace, as we expect from linear theory where it vanishes for massive gravity. (Recall that in GR we would simply have $R = -T$ in our units.) \\

\begin{figure}
\centerline{
  \includegraphics[width=7cm]{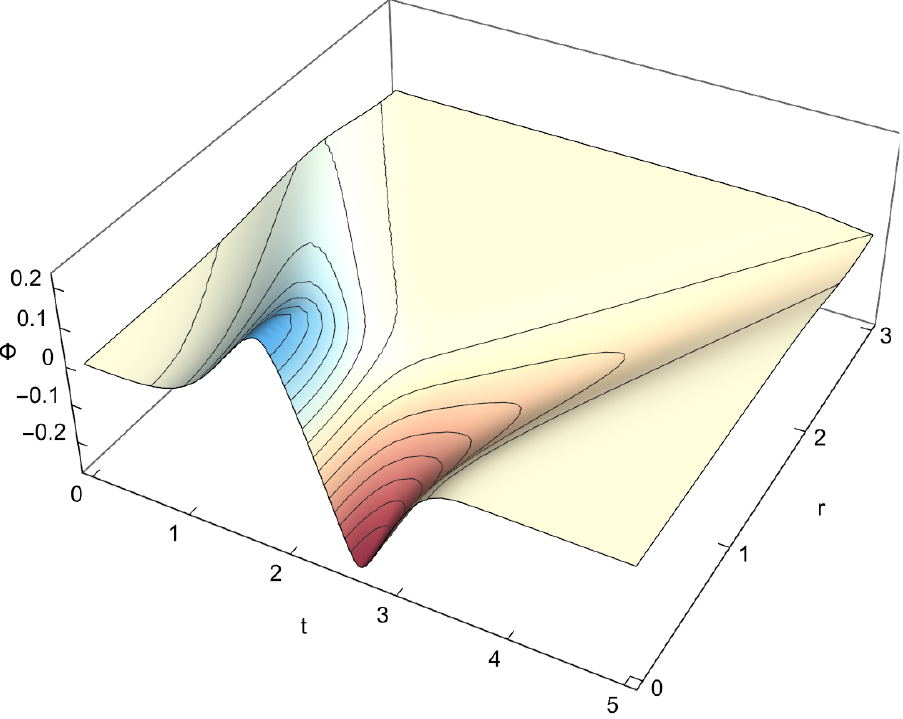} \hspace{0.5cm}    \includegraphics[width=7cm]{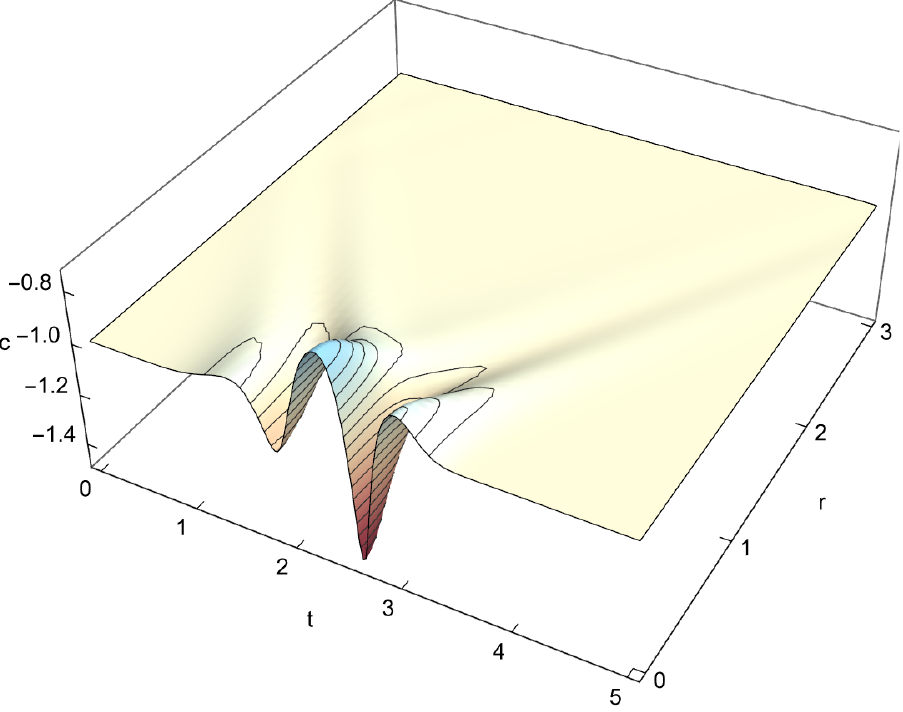}
  }
  \centerline{
\includegraphics[width=7cm]{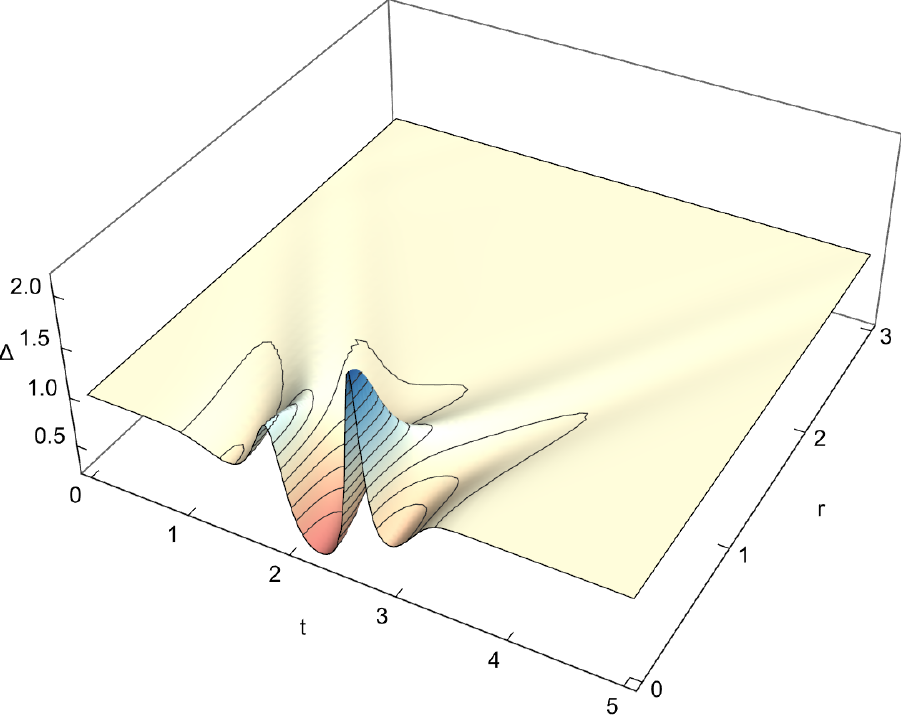}  \hspace{0.5cm}   \includegraphics[width=7cm]{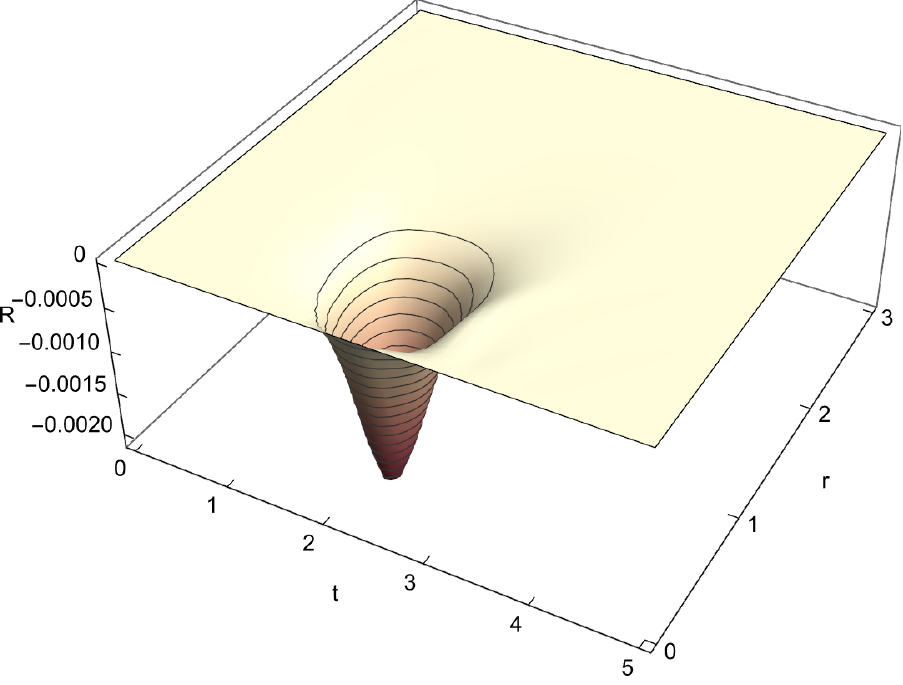}
  }
    \centerline{
    \includegraphics[width=7cm]{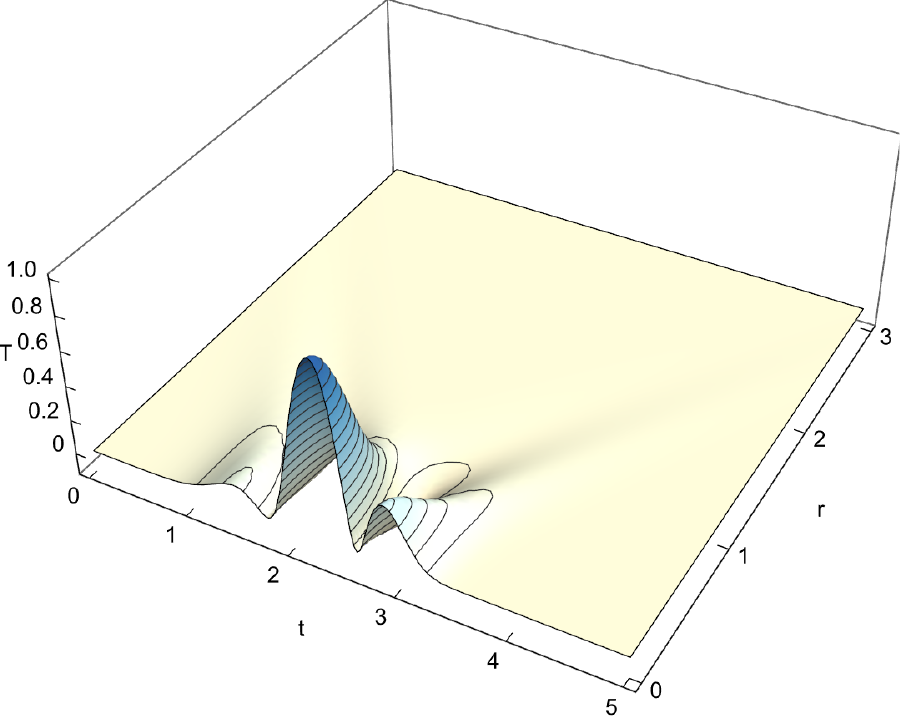}
  }
  \caption{\label{fig:moderate}
  The same quantities are shown as in the previous figure, now for an in-going scalar pulse with greater initial amplitude $A = 0.04$. We see a greater response in the metric function $c$, and correspondingly in $\Delta$, and likewise for the trace of the stress tensor. We see a much larger response in the Ricci scalar, of order $10^3$ larger than for $A = 0.01$,  reflecting that its dependence on amplitude is non-linear (recall it vanishes in linear theory).}
\end{figure}

For comparison purposes, we then show in Fig.~\ref{fig:moderate} the collapse of an in-going pulse with a greater initial amplitude of $A = 0.04$. Here we see a stronger non-linear response to the pulse, although a similar qualitative behaviour as the pulse reaches the origin with a height of $\sim 0.2$ and again then disperses. Now the metric function $c$ is clearly perturbed from its zero Minkowski value by an $O(1)$ amount as the matter traverses the origin. The same is true for the other vierbein components. While the Ricci scalar remains small compared to the characteristic scales of the problem, we see a much stronger response than for the pulse with $A = 0.01$, reflecting the fact that the Ricci scalar reacts non-linearly (recall it vanishes in linear theory).
We see that the discriminant $\Delta$ now reaches much smaller values, dipping to $\sim 0.1$ near the origin as the pulse transits through. As already discussed, if $\Delta$ becomes very small,  the system becomes strongly coupled and is classically no longer under control. For $\Delta\sim \mathcal{O}(1)\gg \ell^2$, one can in principle still expect to trust the theory, but as we carry on with stronger initial data, as we shall do shortly, we will  reach a point where
the EFT truncation can no longer be used to describe the system dynamics.

\subsection{Strong initial data and breakdown due to strong coupling}

\begin{figure}
\centerline{
  \includegraphics[width=7cm]{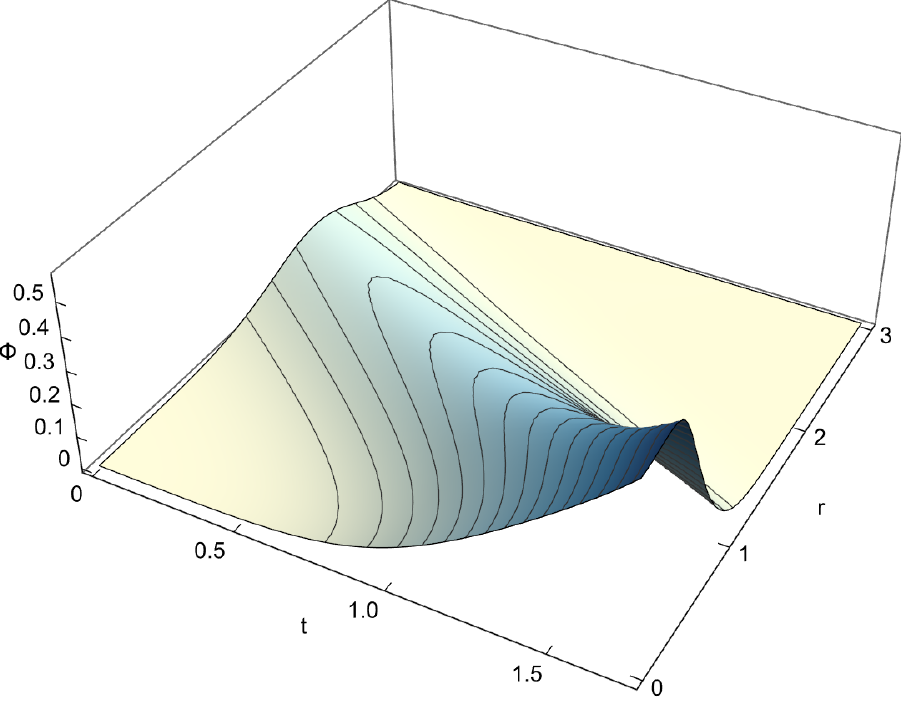}   \hspace{0.5cm}  \includegraphics[width=7cm]{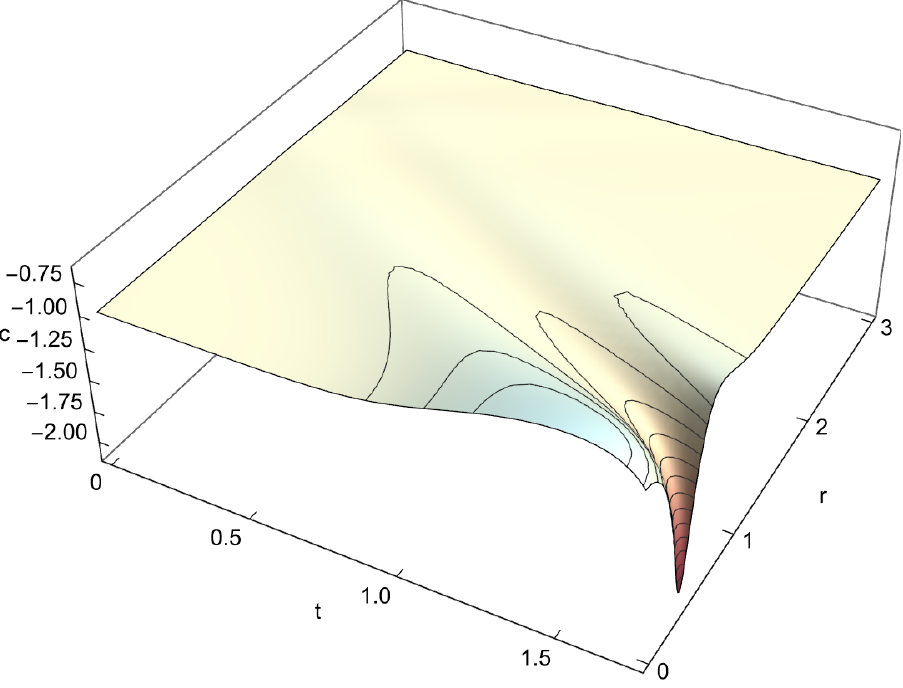}
  }
  \centerline{
\includegraphics[width=7cm]{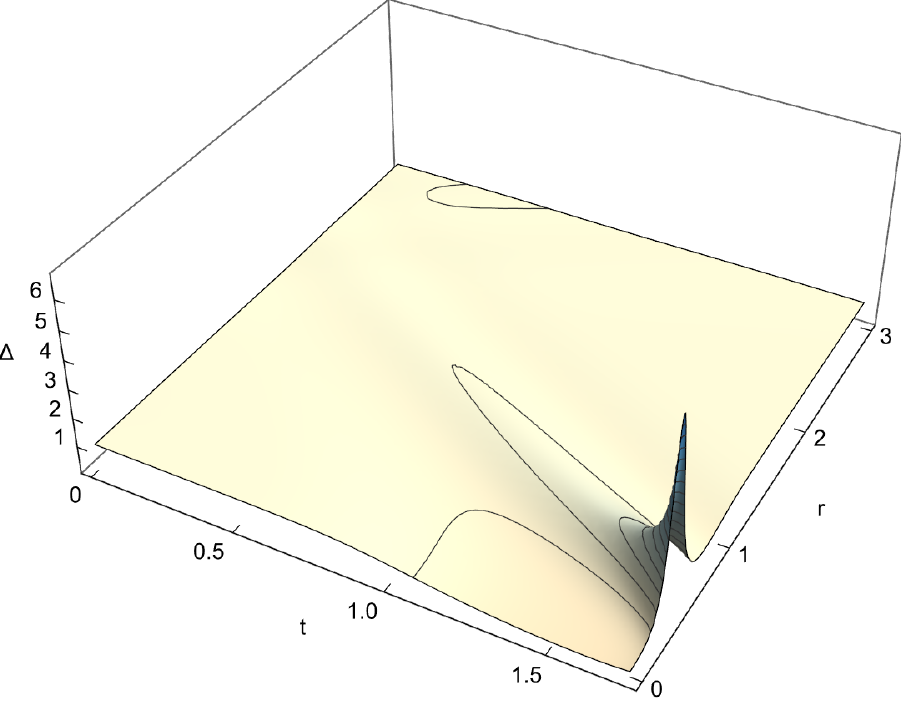} \hspace{0.5cm}    \includegraphics[width=7cm]{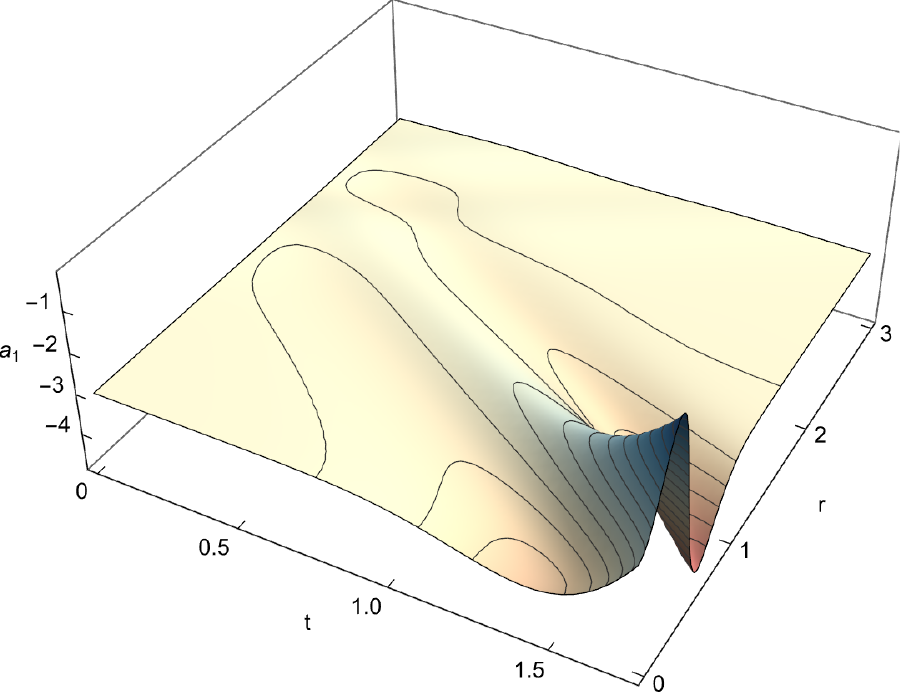}
  }
  \caption{\label{fig:high1}
  For an initial amplitude $A = 0.12$ the evolution breaks after a time $t \simeq 1.7$. The top plots show the matter field $\Phi$ (left) and $c$ (right), and the bottom ones show $\Delta$ (left) and the coefficient $a_1$ from the scalar constraint (right). These are plotted over the whole range where the evolution exists.
  We see that near the end of the evolution $\Delta$ appears to be vanishing near the origin, whereas it appears to diverge where the matter pulse is located. The function $a_1$ is negative everywhere, and hence the divergence in $\Delta$ corresponds to being on the `wrong' branch of the quadratic solution as the equation linearizes. Indeed we see $c$ appear to diverge 
negatively where $\Delta$ does so.
  Thus the time evolution ends as the scalar constraint becomes pathological, and cannot be solved for a finite real $c$.
    }
\end{figure}

\begin{figure}
\centerline{
  \includegraphics[width=7cm]{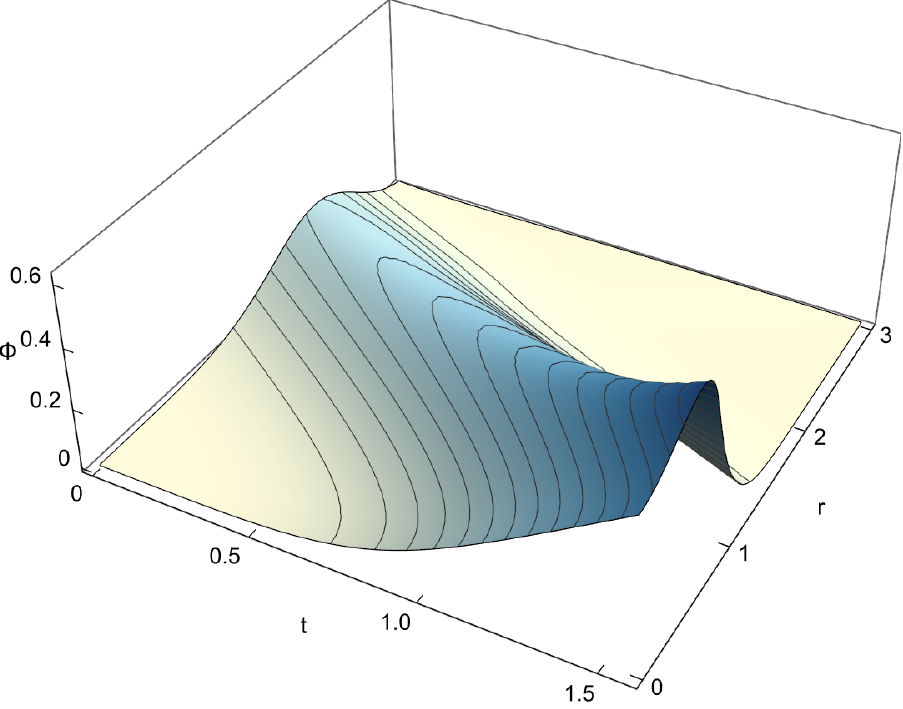}  \hspace{0.5cm}   \includegraphics[width=7cm]{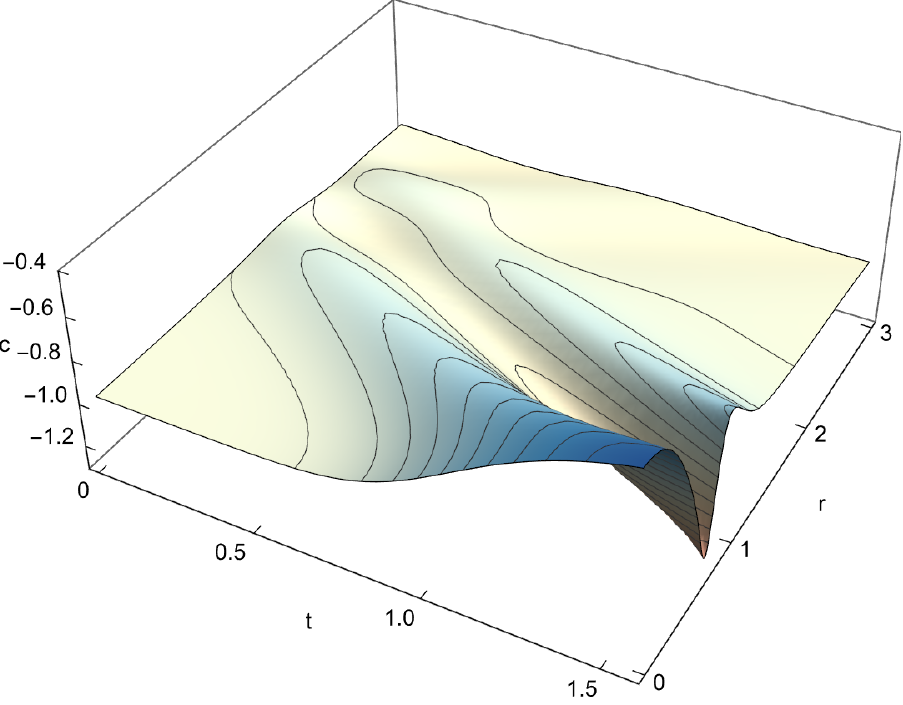}
  }
  \centerline{
\includegraphics[width=7cm]{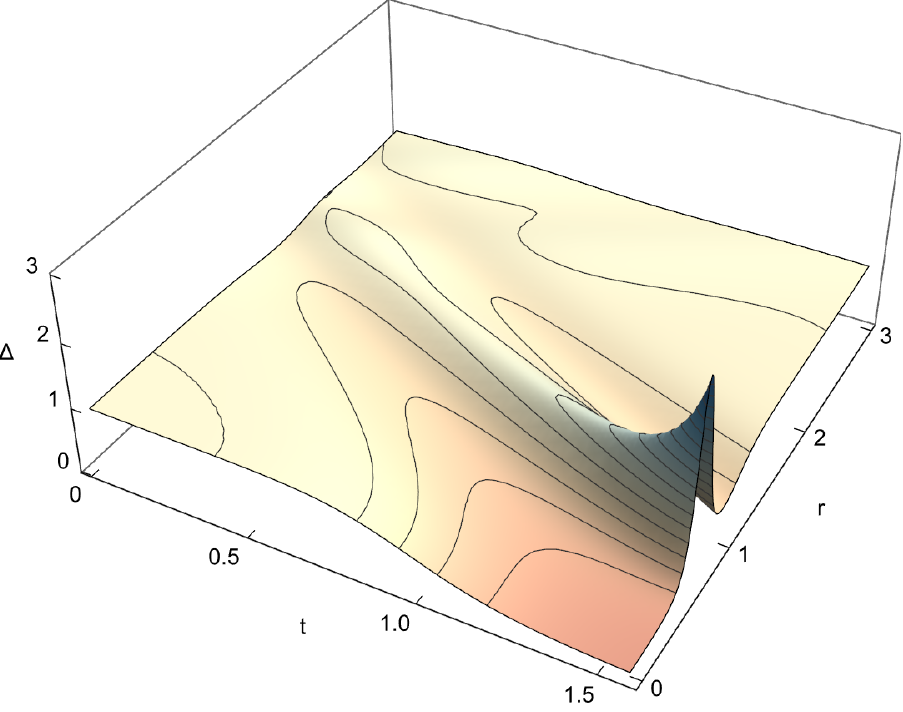}  \hspace{0.5cm}   \includegraphics[width=7cm]{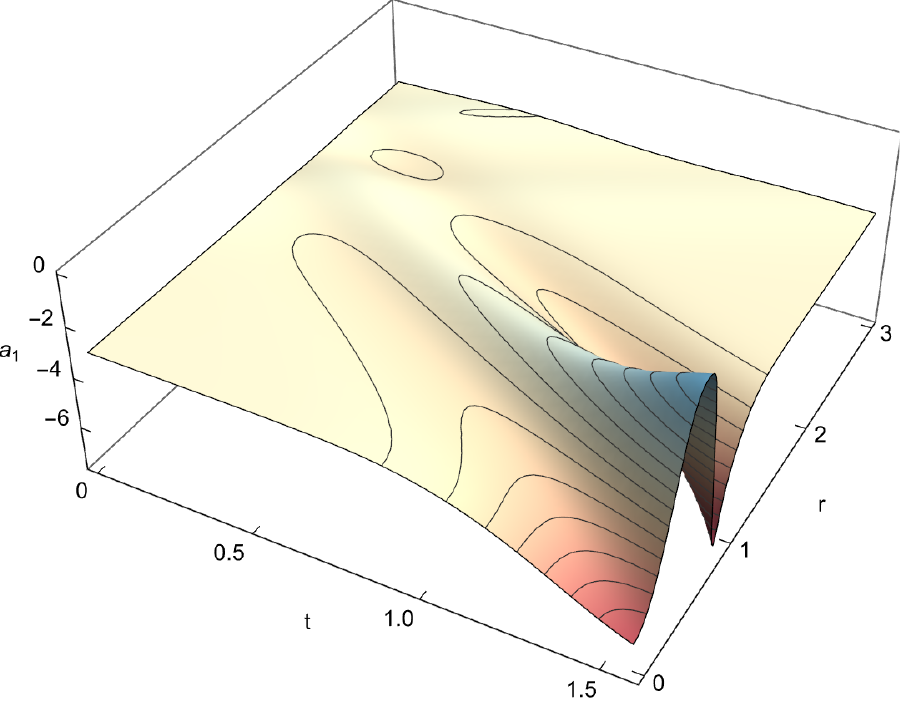}
  }
  \caption{\label{fig:high2}
  This figure shows the same plots as in the previous figure for a larger amplitude $A = 0.2$. Here we see the time evolution breaks even earlier at $t \simeq 1.5$. We see the same behaviour in the various functions plotted, and in particular again we see a divergence in $\Delta$ which we believe ultimately signals the scalar constraint becomes pathological and cannot be satisfied.
    }
\end{figure}

We now consider larger amplitudes of initial data. In figures~\ref{fig:high1} and~\ref{fig:high2} we plot $\Phi$, $c$, $\Delta$ and also the quadratic coefficient $a_1$ in the scalar constraint for the amplitudes $A = 0.12$ and $A = 0.20$ respectively. Both numerical evolutions break down \emph{before} the matter pulse reaches the origin, in the sense that the implicit system governing the timesteps cannot be solved.
The figures show the evolutions up until the timeslice when the simulation breaks.
Looking at the discriminant $\Delta$ and the quadratic coefficient $a_1$ we can see why the evolutions break down. In both cases as the matter pulse traverses towards the origin and grows due to the focussing, we see $\Delta$ become small near the axis, but also appear to diverge positively away from the axis approximately at the location of the peak of the scalar field amplitude. We can see that although the coefficient $a_1$ is decreasing in the region where $\Delta$ becomes large, it remains positive, and hence the diverging of $\Delta$ signifies being on the `wrong' branch -- as a result we see that the vierbein function $c$ appears to diverge in the region where $\Delta$ does. Since this phenomenon occurs with a change of branch, perturbations become infinitely strongly coupled on that solution and it can hence no longer be trusted before $c$ even diverges. \\

It is difficult to definitively diagnose whether the simulation is breaking due to $\Delta$ diverging where the matter pulse is located, or whether it is due to $\Delta$ going towards zero near the origin. We believe it is the former.
Either way the breakdown is associated with the solution becoming infinitely strong coupled and
hence becoming meaningless before reaching those singularities. Details of the microscopic physics need to be folded in order to potentially perform an evolution past those points.

\begin{figure}
\centerline{
  \includegraphics[width=7cm]{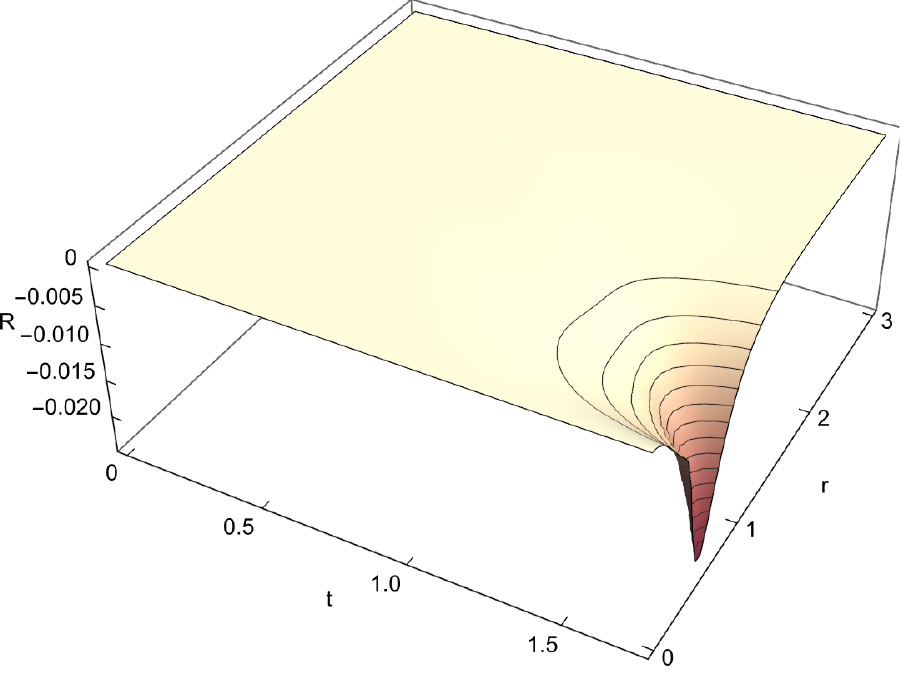}  \hspace{0.5cm}   \includegraphics[width=7cm]{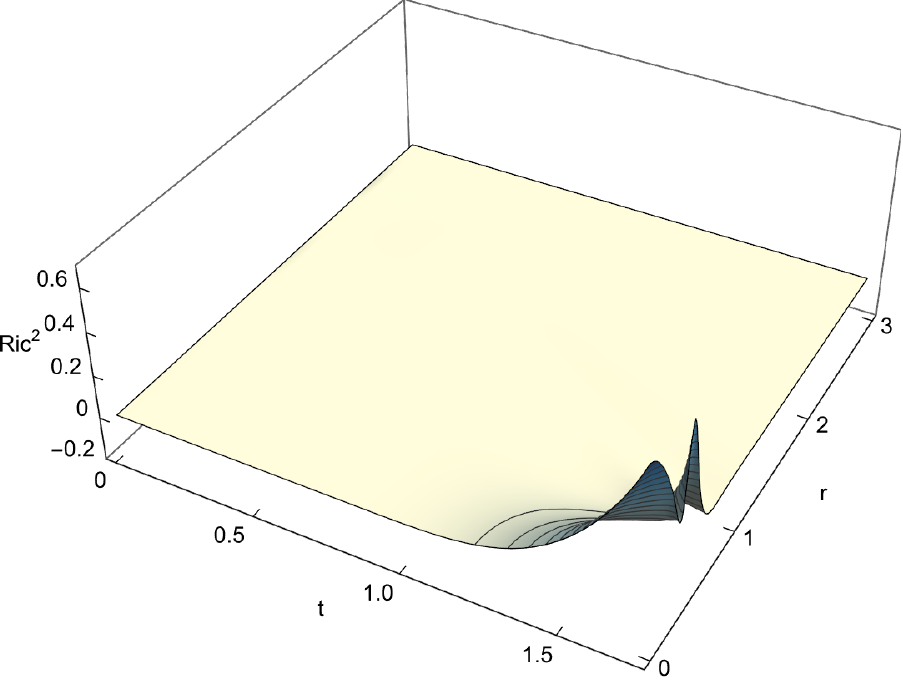}
  }
  \centerline{
\includegraphics[width=7cm]{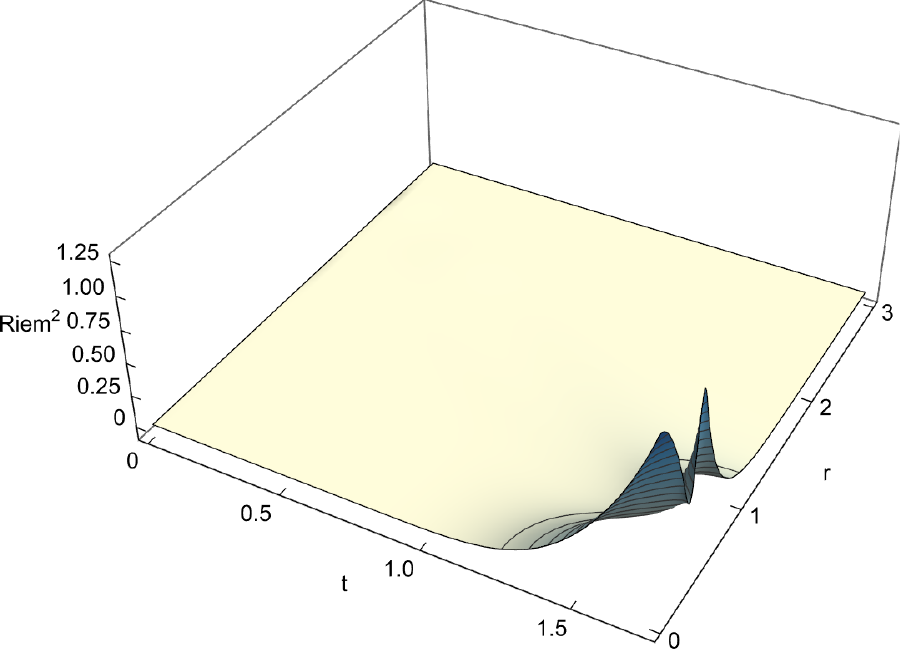} \hspace{0.5cm}    \includegraphics[width=7cm]{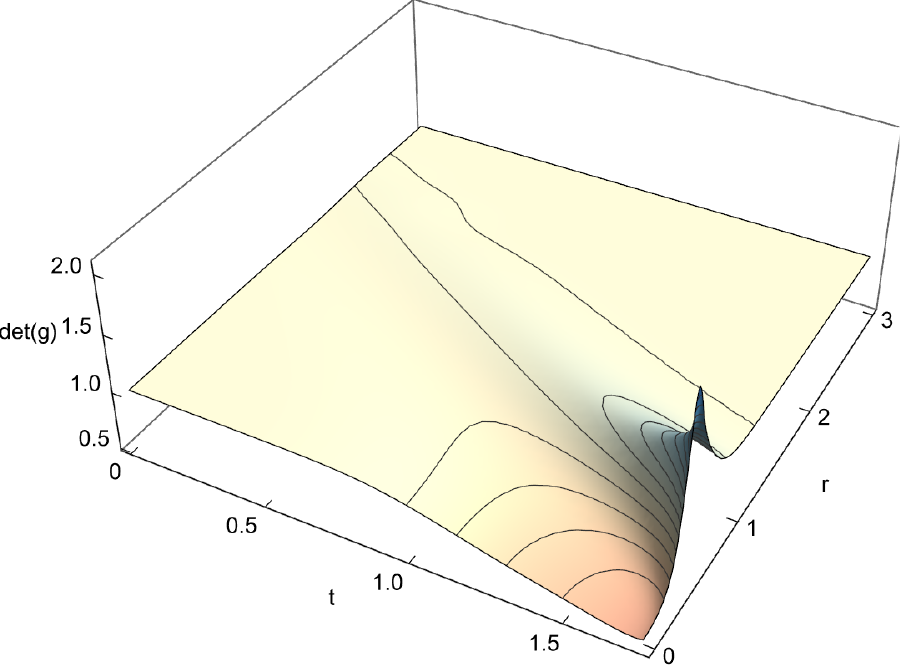}
  }
\caption{\label{fig:invariants1}
  This figure shows the following curvature invariants for the $A = 0.12$ evolution: Ricci scalar (top left), $Ric^2 = R_{\mu\nu} R^{\mu\nu}$ (top right) and $Riem^2 = R_{\mu\nu\alpha\beta} R^{\mu\nu\alpha\beta}$ (bottom left). While these increase in the region where the strong coupling develops, 
they remain quite small. It is therefore unclear whether the strong coupling is associsated to a curvature singularity developing.
In the bottom right frame we plot the ratio of determinants $\det(g_{\mu\nu})/\det(f_{\mu\nu})$, which is a coordinate invariant, and this appears to become singular, diverging positively in the region where $\Delta$ and $c$ look to be blowing up.
   }
\end{figure}

\begin{figure}
\centerline{
  \includegraphics[width=7cm]{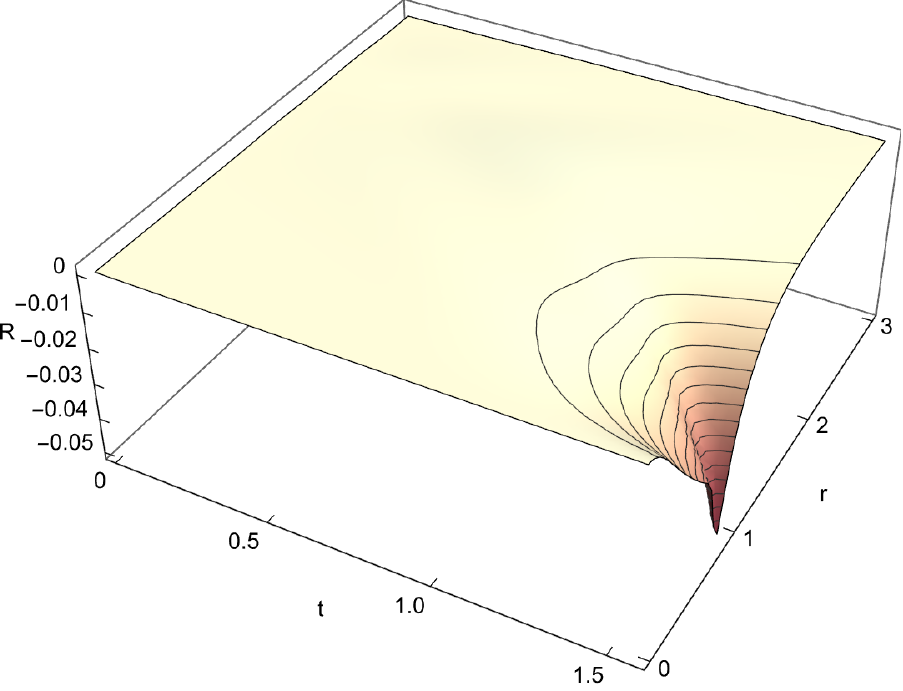}  \hspace{0.5cm}   \includegraphics[width=7cm]{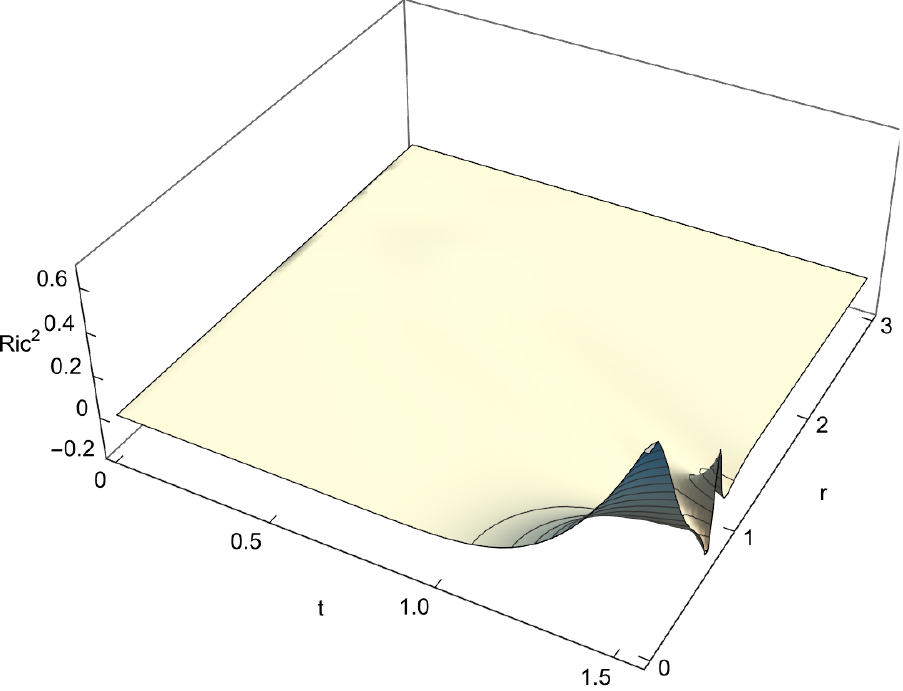}
  }
  \centerline{
\includegraphics[width=7cm]{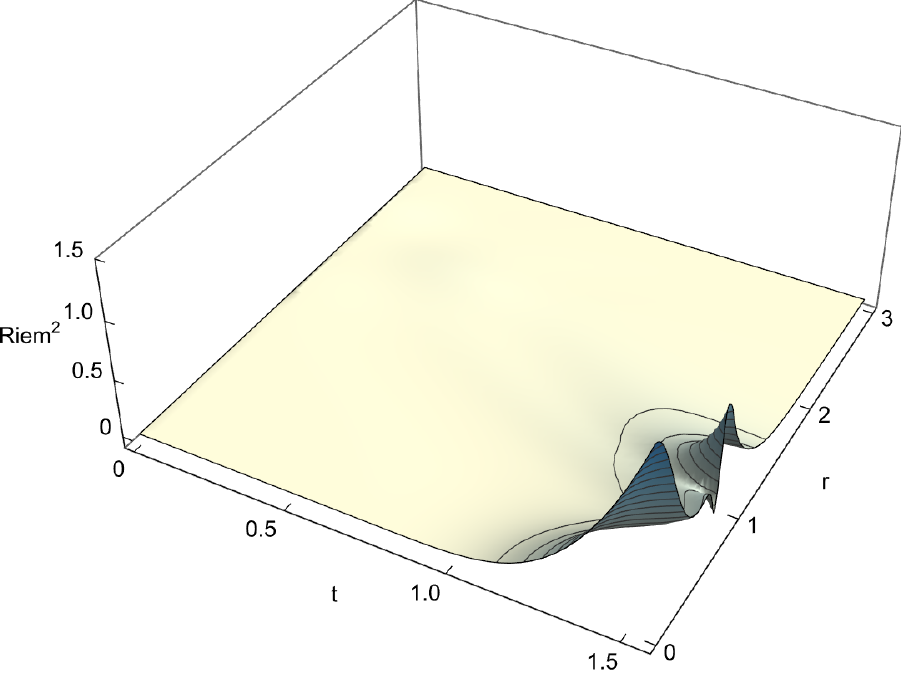} \hspace{0.5cm}    \includegraphics[width=7cm]{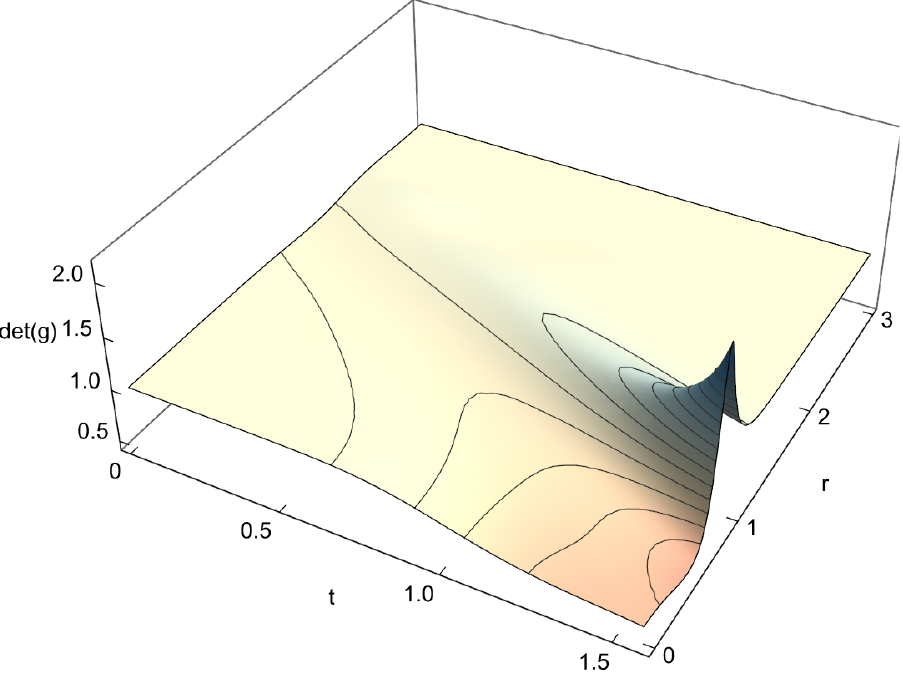}
  }
\caption{\label{fig:invariants2}
  This figure shows the same curvature and coordinate invariants as in the previous one, now for $A = 0.20$. We see that while $Riem^2$ remains relatively small, it does appear to grow quickly in the region where $\Delta$ and $c$ diverge. Thus in this case it also remains ambiguous whether curvatures remain bounded as the evolution breaks down.
   }
\end{figure}

 A natural question is whether this  infinitely strong coupling is also associated to the development of a curvature singularity. As usual, the Einstein equation determines the Ricci tensor of the metric, and from this we may compute the Ricci scalar $R$ and the invariant $Ric^2 = R_{\mu\nu} R^{\mu\nu}$. However, again as for GR, the Kretschman invariant $Riem^2 = R_{\mu\nu\alpha\beta} R^{\mu\nu\alpha\beta}$ is not determined by the Einstein equation, but may be computed directly from the metric. We plot these quantities in figure~\ref{fig:invariants1} for the evolution $A = 0.12$ and in figure~\ref{fig:invariants2} for $A = 0.20$. In both cases we see the curvature invariants increase near the axis, and the region where $\Delta$ becomes large. In the case $A = 0.12$ they do not appear to be strongly diverging. The case $A = 0.20$ is less clear, as while Ricci and $Ric^2$ look well behaved, $Riem^2$ appears as if it is quickly increasing as $\Delta$ looks to diverge, although its value is still relatively small when the evolution breaks down.
Thus the $A = 0.12$ example shown suggests that the pathology may be due to strong coupling that is not associated to curvature. Such a result would be interesting as it would indicate how strong coupling in the reference metric sector (or its \stu fields) may not necessarily propagate in the standard gravitational sector.
The $A = 0.20$ is less conclusive. Other amplitudes have similar behaviours to the two cases shown, and thus we refrain from making a strong statement about whether strong coupling here is associated to a curvature blow up. \\

The theory is invariant under a combined coordinate transformation of the metric and reference metric. Clearly if curvature invariants diverge, there is no way to remove a singularity by changing coordinates. If instead the strong coupling is not due to curvature diverging, one might naively wonder whether these pathologies are then simply a result of a bad unitary gauge coordinate choice. Also in the figures~\ref{fig:invariants1}  and~\ref{fig:invariants2}  we show the ratio of determinants of the metric and reference metric,  $\det(g_{\mu\nu})/\det(f_{\mu\nu})$ for the two cases $A = 0.12$ and $A = 0.20$. We see that this ratio appears to be diverging in the region where $\Delta$, and consequently $c$ becomes large. Crucially this ratio of  determinants is a coordinate invariant, and hence its divergence signifies this behaviour cannot be removed by a coordinate transformation. Hence even if it is the case that curvature remains bounded, the singular behaviour we are seeing associated to $\Delta$ and $c$ diverging is definitely not a coordinate artefact. \\

In fact we can argue that neither pathology of the scalar constraint, so the behaviour associated to $\Delta \to \infty$ or that where $\Delta$ becomes negative, can be removed by a diffeomorphism. Consider the case of $\Delta$ diverging, resulting in $c$ diverging as well (which we believe is driving the break down in the evolutions presented here). 
To render the metric components finite would then require a singular gauge transformation which would then yield a divergent reference metric. One can move the divergence between the metric and the reference metric, but there will be no choice of coordinates where both are smooth.
 Now consider the other case, that $\Delta$ shrinks to zero. Extending past this point presumably requires negative $\Delta$ and hence complex $c$ (although it may be bounded in magnitude), and consequently a complex metric. Again a complex coordinate transformation might restore the metric to a real form, but would then render the reference metric to be complex. Thus both pathologies would represent genuine physical breakdown of the EFT, and are not simply coordinate artefacts, even if they don't give rise to curvature blow up. \\

\begin{figure}
  \centerline{
    \includegraphics[width=11cm]{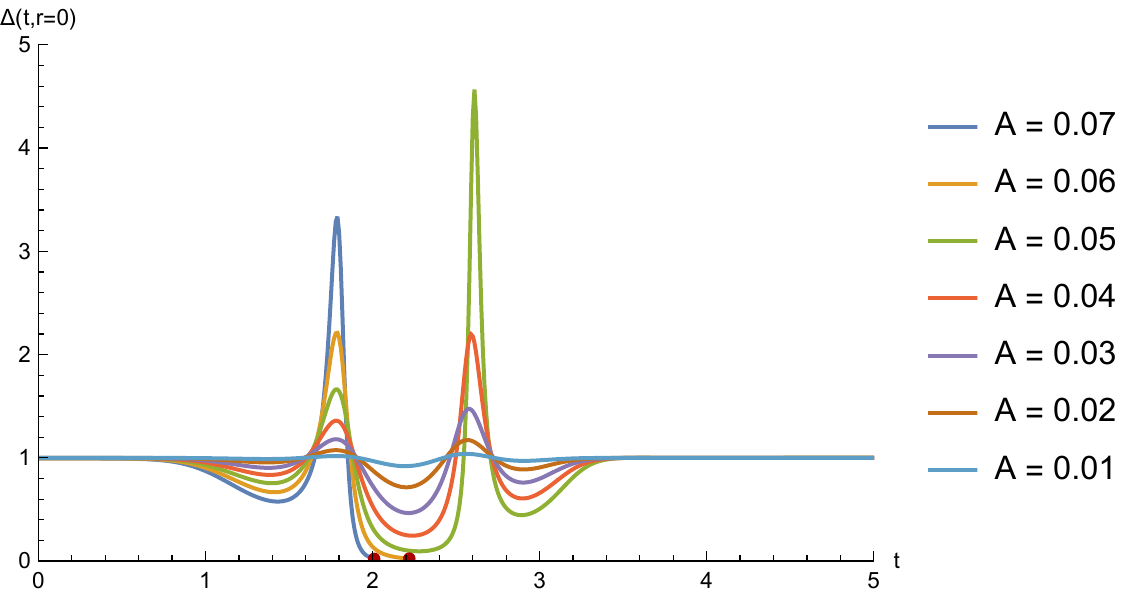}
    }
    \caption{\label{fig:disc_with_amplitude}
    Plot showing the value of the scalar constraint discriminant, $\Delta$, at the origin, as a function of time for evolutions with different initial amplitudes $A$. For evolutions with initial $A \le 0.05$ the scalar pulse passes through the origin at $t \sim 2$ and subsequently disperses, and $\Delta$ returns to its Minkowski value of $\Delta_{\rm Mink} = 1$. However for larger values, $A = 0.06$ and $0.07$ (or indeed greater values) the numerical evolution breaks down before $t \sim 2$ due to strong coupling, and the curves are plotted only up to the point the simulation fails, indicated by the red markers.
      }
\end{figure}

As the initial scalar amplitude is increased from $A = 0.2$ the evolution breaks earlier, and as it is decreased the evolution runs longer, with the apparent breakdown in the scalar constraint occurring when the pulse is nearer to the origin. 
An amplitude of approximately $A \simeq 0.06$ appears
 to divide the weak field dispersive behaviour from the large amplitude evolutions which break down. This is shown in figure~\ref{fig:disc_with_amplitude} where we plot the value of $\Delta$ at the origin as a function of time for increasing initial amplitude. For the curves with $A = 0.06$ and $0.07$ then $\Delta$ is only plotted for times up until the evolution breaks down. For these smaller amplitudes it is unclear whether the evolution breaks due to the scalar constraint becoming pathological precisely at the origin, or close to it. In the previous figures, say for $A = 0.12$, we clearly see a candidate pathology away from the axis of symmetry.
It would be interesting to explore whether there is a critical behaviour associated to this transition regime in between dispersion and breakdown due to strong coupling (ie. examine whether the two behaviours are continuously connected as in a `second order transition').

\subsection{Independence on the diffusion terms}
\label{sec:independence_diffusion}

As stated earlier, the previous figures were made for evolutions with diffusion constant $D = \ell^2 = 0.001$. A key concept is that this diffusion constant must be sufficiently small to ensure good behaviour on small scales, whilst being irrelevant for the low energy physics we are interested in. If we try to remove the diffusion altogether, setting $D = 0$, simulations where we see reasonable deviations from flat spacetime break down due to lattice scale instabilities. As with any numerical discretization it is difficult to say whether this is a result of the low energy truncation being ill posed, or whether it is our numerical scheme introducing lattice pathologies. However taking  $D = \ell^2 = 0.001$ controls lattice scales up to the highest resolutions we probed ($N = 1600$), and is sufficiently small that it is irrelevant for the low energy behaviour we are studying. \\

\begin{figure}
  \centerline{
    \includegraphics[width=9cm]{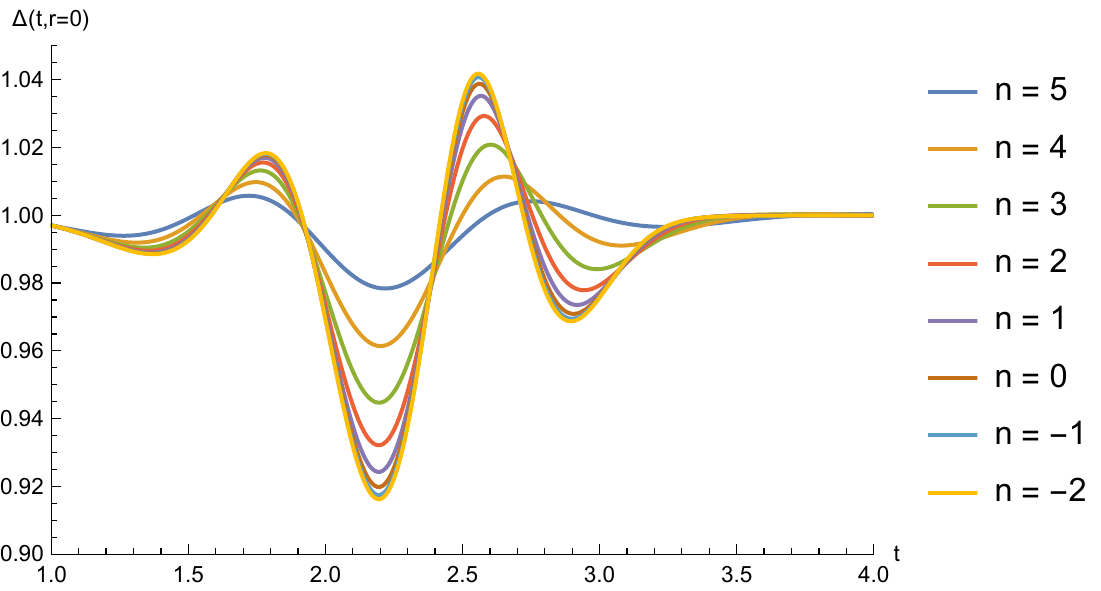}\hspace{0.5cm}  \includegraphics[width=9cm]{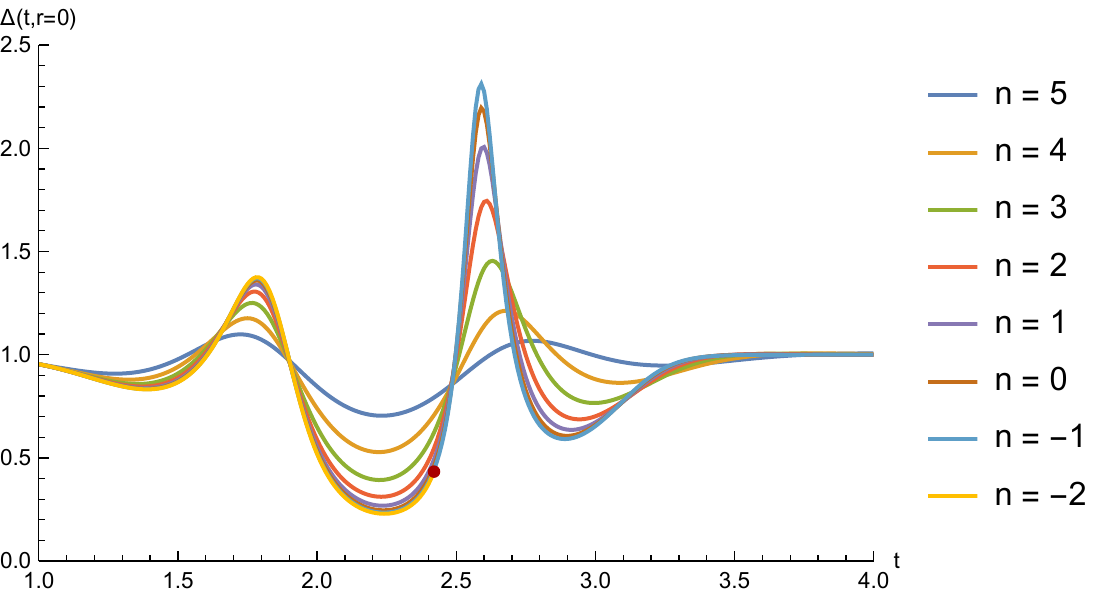}
    }
    \caption{\label{fig:diffusion}
   Figure showing the effect of varying the diffusion coefficient $D = \ell^2$ on the time dependence of $\Delta$ evaluated at the origin, for an initial amplitude $A=0.01$ (left), or $A=0.04$ (right), shown at times $1 \leq t \leq 4$. We take values $D = 2^n \times 10^{-3}$ and show curves for $-2 \le n \le 5$.  We see that as $D$ approaches zero, so $n$ gets smaller, the curves appear to converge on a (putative) continuum value.  However if $D$ is taken too small, such as for $n=-2$ in the righthand plot (marked by the red dot), then the simulation breaks due to lattice scale instabilities.
   We see that setting $n = 0$, so $D = 0.001$, as we have for most of the plots in this work, is sufficiently small to 
   allow long time simulations, and yet keep corrections from the diffusion terms being irrelevant.
      }
  \end{figure}

To demonstrate this we may run the same initial data with different diffusion constants to see a good limiting behaviour as $D$ becomes small. In figure~\ref{fig:diffusion} we plot the function $\Delta$ at the axis as a function of the time coordinate for both $A = 0.01$ and $A = 0.04$ for a sequence of diffusion constants, each half the previous one (all for a fixed resolution of $N = 400$ points). We plot the sequence ${D} = 2^n \times 10^{-3}$ for integer $-2 \le n \le 5$ so that $D$ spans the range from $\sim 0.0002$ up to $\sim 0.03$. We see that by eye the resulting curves are very similar for $n \le 0$, so $D \le 0.001$. Indeed the same is true for all the functions plotted above -- the effect of the diffusion term for $D = 0.001$ is nearly everywhere less than percent level. The exception to this, as we can see in the figure, is that for $A = 0.04$ near the second peak of $\Delta$, the effect is slightly larger, at a few percent. This is consistent with the expectation that where the theory starts to develop strong coupling we should become increasingly sensitive to these higher order operators.

\subsection{Lack of horizon formation}

\begin{figure}
  \includegraphics[width=9cm]{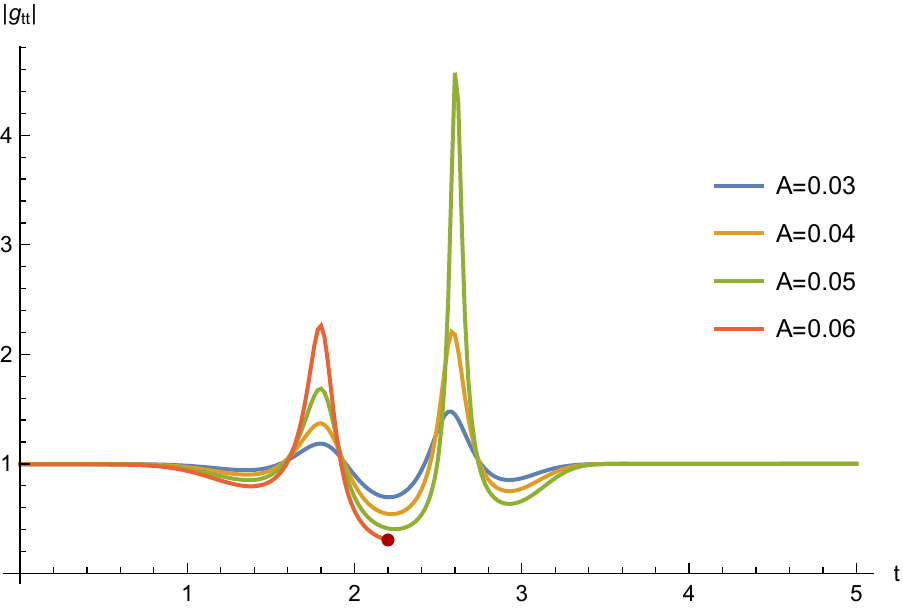}
  \caption{\label{fig:gtt}
    The behaviour of the $g_{tt}$ component of the physical metric at origin, where it is equal to $-c^2$. When $r = 0$ the off-diagonal terms in the metric vanish, so we can take this to be a diagnostic of the formation of an apparent horizon,
    which would happen for $g_{tt} = 0$. As we see $| g_{tt} |$ remains positive for the case $A = 0.06$ where the evolution breaksdown, as indicated by the red marker. It doesn't appear to tend to zero, but instead we see periods of time where there is a strong blueshifting with $| g_{tt} |$ becoming quite large for the stronger amplitudes $A = 0.05$ and $0.06$.
    }
\end{figure}

We have seen for sufficient amplitude initial data we apparently see a break down in evolution due to strong coupling. The dynamics in such a region of strong coupling will then sensitively depend on the short distance completion of the theory. A crucial question is then whether this region is visible to an asymptotic observer. An interesting quantity to compute is the value of $g_{tt}$, the time-time component of the metric at the origin. Since the curve $r =0$ at the origin is a timelike geodesic, then $g_{tt} = - c^2$ is a physical quantity that measures the relative redshift/blueshift of this geodesic as seen by an asymptotic observer. In figure~\ref{fig:gtt} we plot $| g_{tt} |$ at the origin against time for increasing amplitudes of initial data, up to $A = 0.06$ where the evolution just breaks down, apparently at or very near to the origin. \\

We see a very different pattern to that in GR, where collapsing matter leads to a decrease in $| g_{tt} |$, corresponding to a redshift of physics  at the origin as seen by an asymptotic observer, and for sufficient amplitudes $| g_{tt} |$ vanishes as a horizon forms. Instead in this minimal massive theory we see the collapsing matter shell leads to a significant blue shift just before the matter reaches the origin at $t \sim 2$, when there is then a redshifting, before another period of even larger blue shift while 
the matter disperses. Looking at this redshifted period, the minimum value of $| g_{tt} |$ attained for each evolution doesn't seem to be tending to zero as the amplitude is increased to the limit value of $A = 0.06$ when the evolution breaks. For $A = 0.06$ the minimum value is $| g_{tt} | \simeq 0.3$, occurring when the evolution breaks due to strong coupling at or near the origin, so still somewhat greater than zero. Thus there is no suggestion of a horizon developing. \\

\begin{figure}
\centerline{
  \includegraphics[width=7cm]{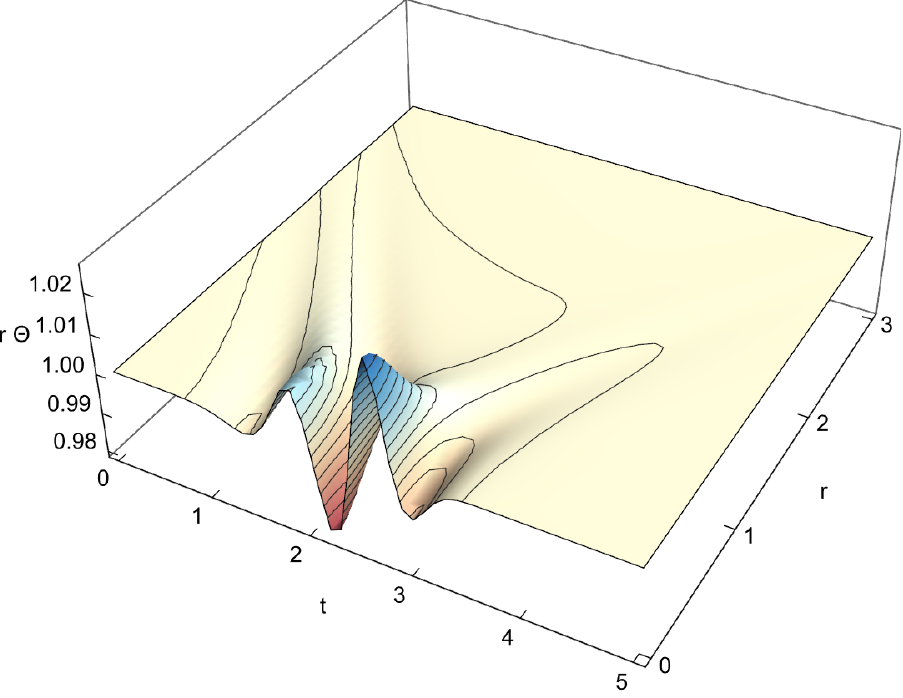}   \hspace{0.5cm}  \includegraphics[width=7cm]{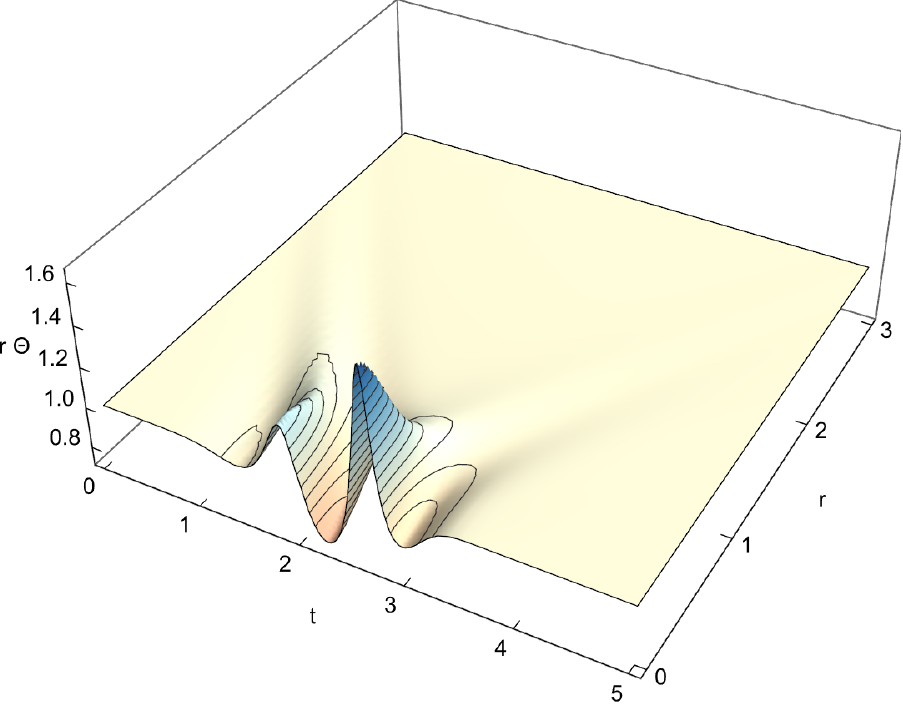}
  }
  \centerline{
\includegraphics[width=7cm]{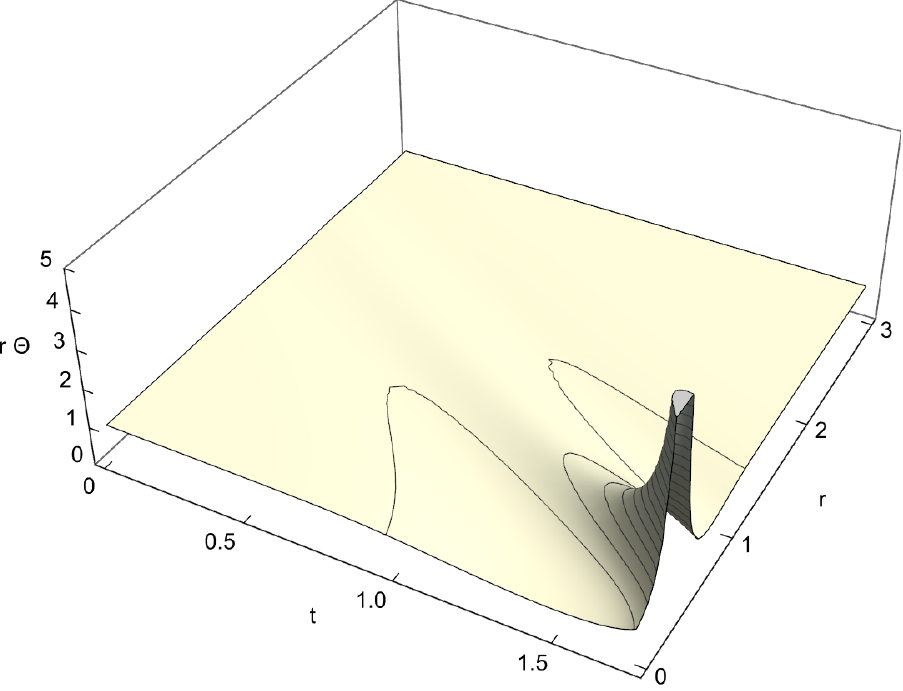}  \hspace{0.5cm}   \includegraphics[width=7cm]{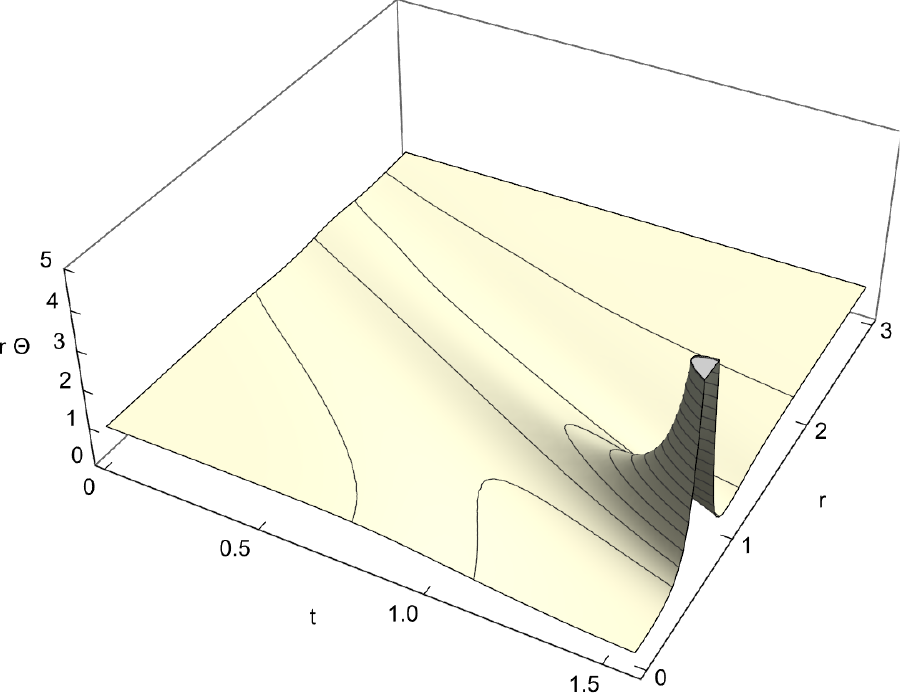}
  }
  \caption{\label{fig:expansion}
    Plots of the outgoing null expansion $\Theta$ plotted as $r \, \Theta$ to make the behaviour more evident. In the top left frame and right frames we plot this for $A = 0.01$ and $A = 0.04$ respectively, and in the bottom left and right frames are plotted the cases $A = 0.12$ and $A = 0.2$. Vanishing $\Theta$ indicates a trapped surface. Rather than $\Theta$ becoming small, instead we see $\Theta$ becoming very large in the region where strong coupling occurs, indicating no horizon is forming. The peak of $\Theta$, which is cut-off in the plot, is $\sim 100$ when evolution stops in the case $A = 0.12$ and even greater in the case $A = 0.2$.
    This suggests the region of strong coupling will be visible to asymptotic observers.
    }
\end{figure}

For larger amplitudes the evolution apparently breaks due to strong coupling away from the axis and we should consider whether a trapped surface forms at any radius, not just consider $g_{tt}$ at the origin.
In our spherically symmetric context a trapped surface occurs when the expansion of outward radially directed null rays vanishes. This expansion is computed as,
\be
\Theta = \frac{1}{2} k^\mu \partial_\mu \log\left( g_{\theta\theta} g_{\phi\phi}  \right) \, ,
\ee
where $k^\mu$ is the outward directed null vector with non-vanishing components $k^t = 1$ and $k^r$ solving the outward null condition. In figure~\ref{fig:expansion} we show this expansion $\Theta$ as a function of time and radial coordinate for the various amplitudes displayed as examples here, so $A = 0.01$, $0.04$, $0.12$ and $0.2$. We plot $r \Theta$ rather than $\Theta$ as near the origin $\Theta \sim 1/r$ which obscures the behaviour. We see that rather than the expansion becoming smaller and vanishing to give a trapped surface, instead it becomes very large in the region of strong coupling in the cases $A = 0.12$ and $A = 0.2$ where the evolution breaks down, presumably reflecting the diverging $c$ component, and suggesting a blueshifting associated to the strong coupling, rather than redshifting. \\

An important conceptual point is that for conventional GR and matter the existence of a trapped surface proves the region lies within an event horizon -- however for massive gravity it is unclear such a statement would hold. Thus finding a trapped surface or otherwise in this theory cannot prove or disprove the existence of an event horizon -- for that presumably one would need to construct the full spacetime and consider the past of $\mathcal{I}^+$. However the lack of a trapped surface is suggestive that no event horizon has formed.

\subsection{Varying mass}

\begin{figure}
  \includegraphics[width=10cm]{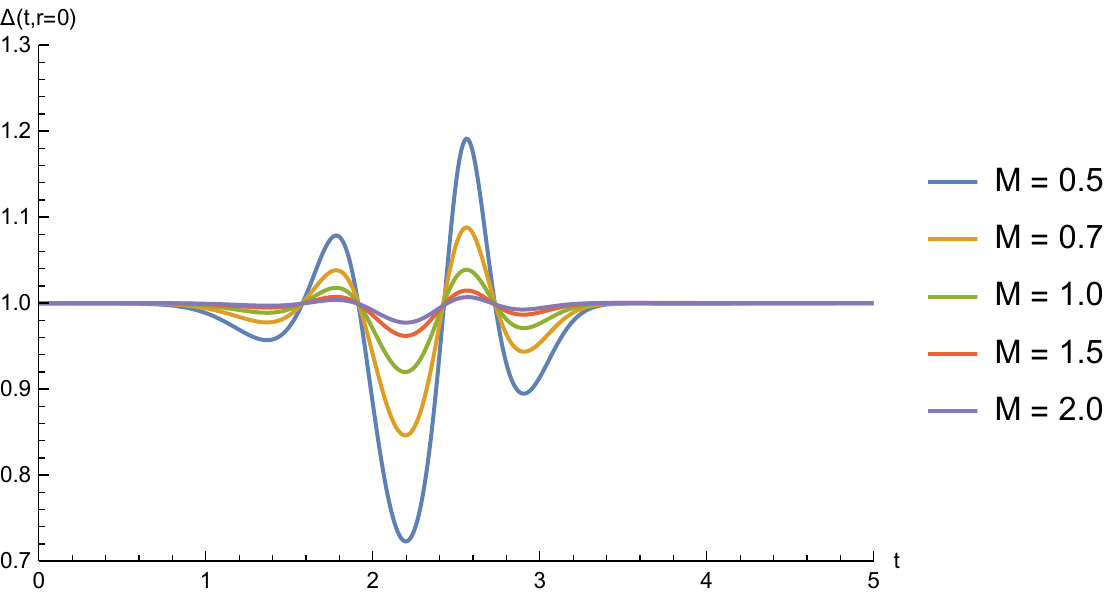}
  \caption{\label{fig:vary_mass}
The effect of varying the graviton mass $m$ whilst keeping the amplitude $A=0.01$ and initial scalar profile constant. We again concentrate on the behaviour of $\Delta$ at the origin, and note that it takes the value $\Delta = 1$ for flat spacetime for all values of mass. We see that decreasing the mass is analogous to increasing the amplitude, indicating non-linear effects become increasingly important.
  }
\end{figure}

While exploring the full phenomenology of collapse for the minimal theory is beyond the scope of this work, we briefly consider how the behaviour changes as we vary the graviton mass from $m = 1$, whilst keeping the profile of the initial ingoing scalar pulse the same. In figure~\ref{fig:vary_mass} we show the behaviour of $\Delta$ at the origin as a function of time for the same initial data $A = 0.01$ and masses ranging from $m^2 = 4$ to $m^2 = 0.25$. For smaller values of $m^2$ we see the same initial data elicits a stronger response. This presumably reflects the behaviour of the linear theory, which breaks down as we take $m \to 0$, again indicating non-linear effects become more important in the small mass limit for the same matter behaviour.

\section{Conclusion}
\label{sec:Conclusion}

For dRGT massive gravity with its minimal and quadratic mass terms we have provided a 3+1 dynamical formulation of the theory using a symmetric vierbein. With an appropriate choice of momentum variables the vector and scalar constraints, which are second class and thus must be applied on every time slice, can be solved algebraically for the time-time vierbein component together with certain momenta, yielding a system of evolution equations that can straightforwardly be implemented numerically. With no symmetry assumption this then leaves 5 second order geometric dynamical variables, which correctly counts the number of massive gravitational degrees of freedom. \\

As a low-energy EFT, the truncated theory is not expected to be well-posed on its own. While remaining agnostic 
as to the details of its high energy embedding, one can however include spatial diffusion terms to the low-energy EFT which are natural in such a 3+1 formulation. In the presence of these terms, one can successfully prove that the formulation gives a well-posed initial value problem of diffusive character.  We demonstrate this framework by numerically evolving dynamical collapse of massless scalar field matter in spherical symmetry, restricting for simplicity to the theory with minimal mass term. This minimal model is not expected to exhibit a smooth massless limit to GR due to the absence of a Vainshtein mechanism present in more general massive gravity models. Moreover in our numerics we consider sources of comparable size to the graviton Compton wavelength which is far from the regime of phenomenological interest. Nonetheless being an EFT of gravity in its own right it is interesting to ask how the minimal model responds to collapsing matter. \\

We focus on the situation when all scales in the problem -- ie. the mass and the characteristic length scales in the initial data -- are approximately equal, so there is no parametric scale separation.
Our simulations reveal the expected dispersive behaviour for low amplitude initial data. However as the amplitude is increased, and the response becomes non-linear, we see pathologies occur in the scalar constraint equation, which taking a quadratic form is not guaranteed to have solutions. It appears that the theory becomes strongly coupled, and the evolution breaks down, before the matter reaches the origin. This signals that the EFT is breaking down, and evolution past this point would become sensitive to the precise form of the short distance completion. Our results remain ambiguous over whether the breakdown is  associated to large curvatures. While we see curvatures are relatively small up until the break down, for some cases such as the case $A = 0.20$ shown above, the curvature does appear to be increasing in the regions where strong coupling appears to develop. Interestingly we see no formation of trapped surfaces, which is suggestive that the region where the strong coupling develops is not shielded from asymptotic observers by a horizon. \\

In future work we will examine collapse in the theory with quadratic mass term, which is expected to exhibit a Vainshtein mechanism that will yield GR behaviour for small graviton masses.
Another interesting direction is to refine understanding of the strong coupling break down in the minimal theory. In particular it would be interesting to determine whether it is the divergence of $\Delta$ (being on the `wrong' branch as the quadratic constraint linearises) or $\Delta$ being driven negative (associated to the solution for $c$ becoming complex) that finally breaks the evolution -- both seem to occur in different locations at very similar times. Another question to resolve is whether the curvatures seen by the matter degrees of freedom remain bounded or not. It would also be interesting to find coordinates which excise the region where the theory breaks down, and allow evolution outside the future lightcone of the pathology, rather than have to halt evolution on our timeslice as soon as one location becomes strongly coupled. Finally, ultimately we would like to know what happens for more phenomenologically relevant physical situations where there is typically a large hierarchy between the size of the source and the Compton wavelength of 
the graviton.
For example, in realistic applications of massive gravity, the graviton Compton wavelength 
should be taken to be comparable
to the Hubble scale, many orders of magnitude larger than the astrophysical scales of collapsing matter distributions.

\appendix

\section{Scalar constraint details}
\label{app:scalardetails}

Here we present some further details. Firstly we give an argument for the simplification that leads to the quadratic and cubic forms in equation~\eqref{eq:scalar_simplification}. We then give the explicit expressions for the coefficients of the quadratic scalar constraint in the minimal model for our spherically symmetric ansatz.

\subsection{Polynomial expansion in $E_{tt}$ of scalar constraint}

We may understand the simplification that we saw in equation~\eqref{eq:scalar_simplification} relative to the naive expectation in~\eqref{eq:naivecoefficients} by writing the 3+1 decomposition of the symmetric vierbein as,
\be
\label{eq:ADM}
E_{\mu\nu} = \left( 
\begin{array}{cc}
\phi + n_i n^i & n_i \\
& h_{ij} 
\end{array}
\right)
\ee
where now indices are raised/lowered wrt $h_{ij}$ and its inverse $h^{ij} = (h_{ij})^{-1}$. Then the inverse can be written as,
\be
(E^{-1})^{\mu\nu} = \frac{1}{\phi} n^\mu n^\nu + h^{\mu\nu}
\ee
where $h^{tt} = h^{ti} = 0$ and $n^\mu = (1 , - n^i)$. We note that $| E | = \phi \det{ h_{ij} }$ and the inverse metric is,
\be
g^{\mu\nu} = \frac{J}{\phi^2} n^\mu n^\nu + \frac{2}{\phi} n^{(\mu} v^{\nu)} + H^{\mu\nu} \; , \quad v_\mu = \eta_{\mu\nu} n^\nu \; , \quad J = n^\mu v_\mu \; , \quad H^{\mu\nu} = h^{\mu\alpha} \eta_{\alpha\beta} h^{\beta\nu} \; .
\ee
Consider $A_{(1)}^{\alpha\beta\gamma\mu\nu\rho}$ as given in equation~\eqref{eq:A}. It comprises a term $\sim \eta^{\gamma\rho} g^{\alpha[ \mu} g^{\nu ]\beta}$, which  one would naively expect to go as $\eta^{\gamma\rho} g^{\alpha[ \mu} g^{\nu ]\beta} \sim \frac{1}{\phi^4}$ in an expansion in inverse powers of $\phi$. However due to the antisymmetry in the $[\alpha\beta]$ and $[\mu\nu]$ index pairs, and the structure of the leading terms in $g^{\mu\nu}$ above, we see such a term vanishes. Likewise the subleading $1/\phi^3$ behaviour also vanishes, again due to this index antisymmetry. Thus in fact $\eta^{\gamma\rho} g^{\alpha[ \mu} g^{\nu ]\beta} \sim \frac{1}{\phi^2}$. Noting that $| E | \sim \phi$, then $| E |^2 \eta^{\gamma\rho} g^{\alpha[ \mu} g^{\nu ]\beta}$ is a quadratic polynomial in $\phi$, and hence is also a quadratic polynomial in $E_{tt}$ (since $E_{tt}$ is linear in $\phi$). A similar argument applies to the other two terms in $A_{(1)}^{\alpha\beta\gamma\mu\nu\rho}$, showing they also give quadratic contributions in $\phi$ when multiplied by $| E |^2$ due to the index antisymmetries. Likewise analogous arguments show that $| E |^3 A_{(2)}^{\alpha\beta\gamma\mu\nu\rho}$ is a cubic in $E_{tt}$.

\subsection{Scalar constraint coefficients}

The coefficients $a_2$, $a_1$ and $a_0$ of the scalar constraint in the spherically symmetric minimal model of Section \ref{sec:sphericalcollapse} are given by,
\be
a_2 &=& -48 a^4 m^2 r^8+6 a^3 (8 b-3) m^2 r^6+a^2 r^2 \left(36 b^2 m^2 r^2-9 b m^2 r^2+r^2 \partial _r\Phi {}^2-50\right) \\
  &-& 2 a r \left(r \left(12 b^3 m^2-9 b^2 m^2+b \partial _r\Phi {}^2+10 r \partial _ra\right)-10 \partial _rb\right)-12 b^4 m^2+9 b^3 m^2+b^2 \partial _r\Phi {}^2-2 \left(\partial _rb-r^2 \partial _ra\right){}^2 \nl
a_1 &=& -\left(b-a r^2\right) \Big(3 m^2 \left(-4 a^3 r^6+b^2 r^2 \left(3 a-8 h^2\right)+b h^2 r^2 \left(5-8 a r^2\right)+a h^2 r^4 \left(16 a r^2+7\right)+b^3\right) \\
  &+& 2 h r \partial _r\Phi  p_{\Phi } \left(b-a r^2\right)+4 r p_h (r (5 a+r \partial _ra)-\partial _rb)\Big)+4 h r p_b (\partial _rb-r (5 a+r \partial _ra)) \nl
  &+& 4 h r^3 p_a (r (5 a+r \partial _ra)-\partial _rb) \nl
a_0 &=& -4 a^4 r^8 p_{\Phi }^2+2 a^3 r^6 \left(2 b p_{\Phi }^2+3 h^2 m^2 r^2+2 h r \partial _r\Phi  p_{\Phi }\right)+a^2 r^4 \Big(3 \left(-8 r^2 p_a p_b+8 r^4 p_a^2+b^2 p_{\Phi }^2\right) \\
  &-& 2 h r \left(40 r p_a+3 b \partial _r\Phi  p_{\Phi }\right)+h^2 \left(-9 b m^2 r^2+r^2 p_{\Phi }^2+r^2 \left(-\partial _r\Phi {}^2\right)+50\right)+20 h p_b-12 h^4 m^2 r^4\Big) \nl
  &+& b^2 \left(-6 r^4 p_a^2+6 p_b^2+h^2 r^2 \left(-12 h^2 m^2 r^2+p_{\Phi }^2-\partial _r\Phi {}^2\right)\right)+2 a r^2 \Big(-b \left(6 r^2 p_a p_b+b^2 p_{\Phi }^2-6 p_b^2\right) \nl
  &+& 2 h \left(r^2 p_a \left(5 b-r^2 p_h-4 r^3 \partial _ra+4 r \partial _rb\right)+p_b \left(10 b+r^2 p_h+r^3 \partial _ra-r \partial _rb\right)\right)+3 (4 b-1) h^4 m^2 r^4 \nl
  &+& h^2 r \left(-b r p_{\Phi }^2+r \left(b \partial _r\Phi {}^2+10 r \partial _ra\right)-10 \partial _rb\right)\Big)+2 b h r \Big(2 r^2 p_a \left(r p_h+r^2 \partial _ra-\partial _rb\right) \nl
  &-& 2 p_b \left(r p_h-2 r^2 \partial _ra+2 \partial _rb\right)+3 h^3 m^2 r^3\Big)+2 h^2 r^2 \left(2 r^2 p_a p_b+r^4 \left(-p_a^2\right)-p_b^2+\left(\partial _rb-r^2 \partial _ra\right){}^2\right) \nl
  &-& b^4 p_{\Phi }^2+b^3 h r \left(3 h m^2 r+2 \partial _r\Phi  p_{\Phi }\right) \notag\,.
\ee

\section{Numerical details and convergence}
\label{app:numdetails}

As discussed in the main text, the dynamical evolution is performed with a compactified radial coordinate, $\tilde{r} = r/(1-r^2)$ on the interval $[0,1]$. Spatial differences are approximated using 6th order finite differencing, and time derivatives using a Crank-Nicholson scheme so that the system is implicit. The exception to this is that the diffusion terms are differenced using the same spatial scheme, but forward Euler differencing in time. We are not concerned in accurately simulating the diffusion terms which should be irrelevant on the scales we are interested in, and taking second order Crank-Nicholson differencing in time would  necessitate an analogue of the Courant condition for the time step associated to these terms. \\

\subsection{Iterative solution of the implicit time step}

We solve the implicit system as follows: suppose we have the data $\vec{X}=(\tilde{E}^i_{~j},E_i,\ldots)$ on some time slice $t$. The equations of motion then allow us to determine the time derivative $\vec{X}_t$ at $t$.
With that we make an initial guess of the data at $t+\Delta t$ as $\vec{Y}=\vec{X}+\Delta t \vec{X}_t$. The method then enters an iterative stage with the following steps:
\begin{itemize}
  \item Estimate the data at $t + \Delta t/2$ by $\vec{Z} = (\vec{X} + \vec{Y})/2$.
  \item Compute the equations of motions to find $\vec{Z}_t$.
  \item Update  $\vec{Y}_{new} = \vec{Z} + \Delta t \vec{Z}_t/2$.
\end{itemize}
This continues until the maximal difference between $\vec{Y}$ and $\vec{Y}_{new}$ becomes smaller than a set error tolerance, which for the results in this paper was taken to be $10^{-8}$.

\subsection{Convergence}

\begin{figure}
  \centerline{
   \includegraphics[width=9cm]{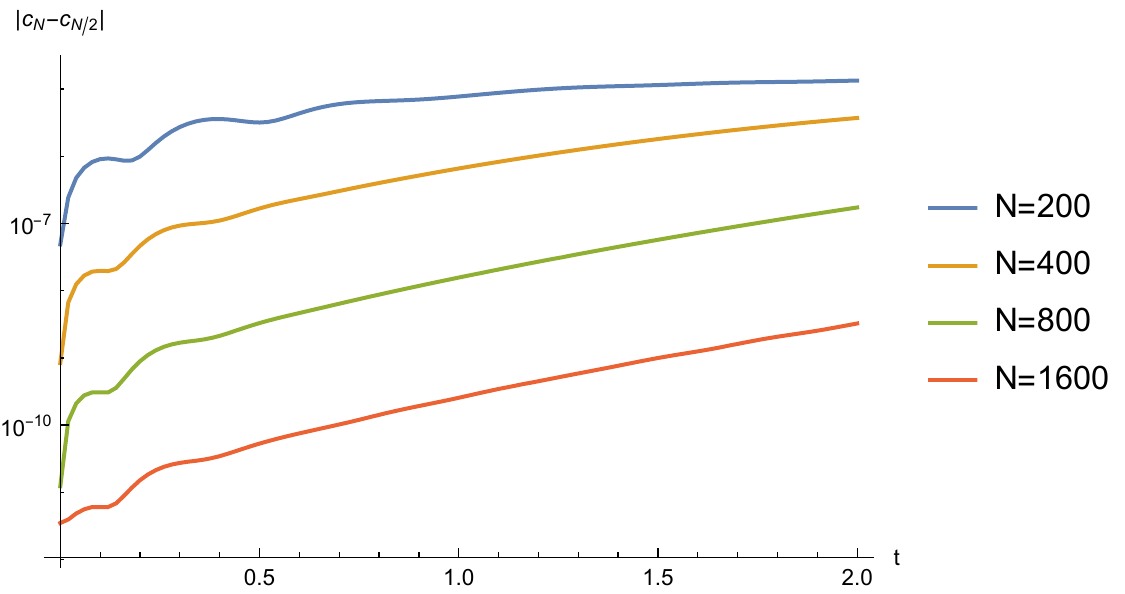} \hspace{0.5cm}   \includegraphics[width=9cm]{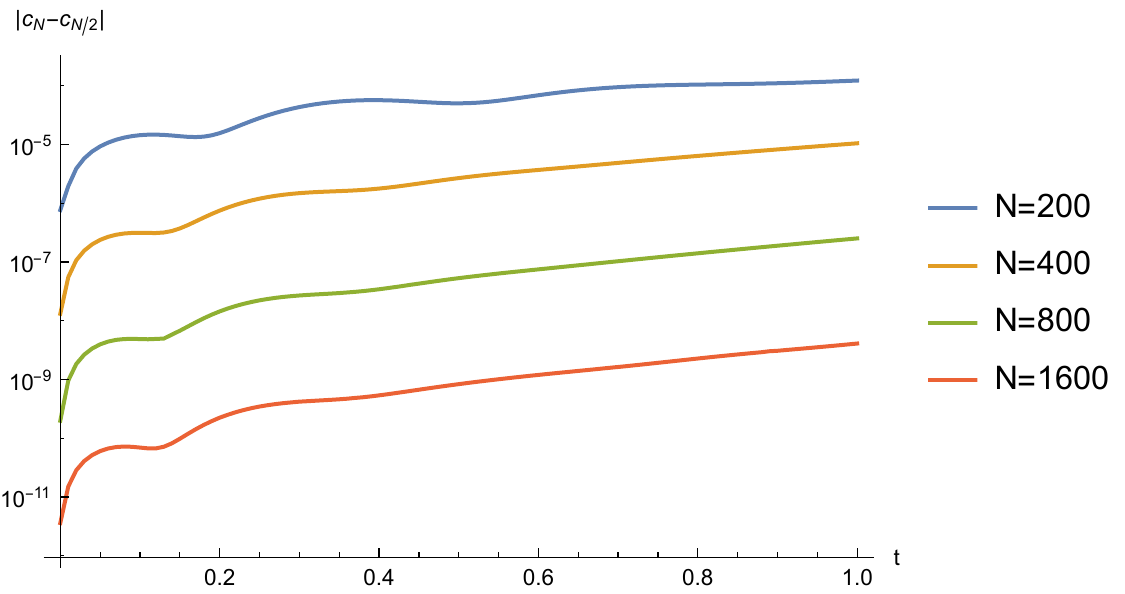}
      }
      \caption{\label{fig:convergence_bigN}
          Figure showing the effect of varying the spatial resolution $N$ while keeping all other parameters constant, in particular $D = 0.001$. The plots display the norm of the difference between the values for the
      $c$ vierbein component obtained for a resolution $N$ and half that resolution, so $| c_N - c_{N/2} |$, as a function of time. The left hand side has $A = 0.01$, and the right $A = 0.04$. In both cases these differences decrease with resolution, with the decrease being consistent with our sixth order finite differencing -- the difference decreases by a factor of $\sim 2^6$ for a doubling of the resolution $N$.
      }
\end{figure}

In this section, we examine the convergence of our numerical results as we vary the spatial resolution $N$ and the diffusion coefficient $D$.  While there is also numerical error associated to our discretization in time, in practice the time step size we use is sufficiently small that discretization error is dominated by the spatial resolution. To compare the results across the entirety of the spatial slices, we use the $L^2$ norm in spherical symmetry, given by
\be
\| f \|_{L^2} :=\left( \int_{0}^{\infty}r^2|f(r)|^2  \,\d r \right)^{1/2}
\ee
As a first check, figure~\ref{fig:convergence_bigN} examines the behaviour of the time-time component of the vierbein $E_{\mu\nu}$, so the function $c$, as $N$ is varied between $100$ and $1600$ by doubling it, keeping the diffusion constant fixed at $D = 0.001$. We use the notation that $c_N$ is the solution for $c$ at resolution $N$, and then plot the norm of the difference of $c$ between a resolution $N$ and $N/2$ as a function of coordinate time, so $| c_N - c_{N/2} |$ against time $t$. In the figure we show this for two initial amplitudes, $A = 0.01$ and $A = 0.04$.
We see that this difference decreases as the resolution increases, and is consistent with the sixth order spatial finite differencing we employ -- the differences decrease by a factor $\sim 2^6$ upon doubling the resolution.
Thus the solutions are consistent with convergence to a continuum solution. However, we emphasize that this continuum solution is that of massive gravity together with our diffusive term higher order terms, and not of the `bare' system of equations (\ref{eq:EinsteinEq}).

\begin{figure}
\centerline{
  \includegraphics[width=9cm]{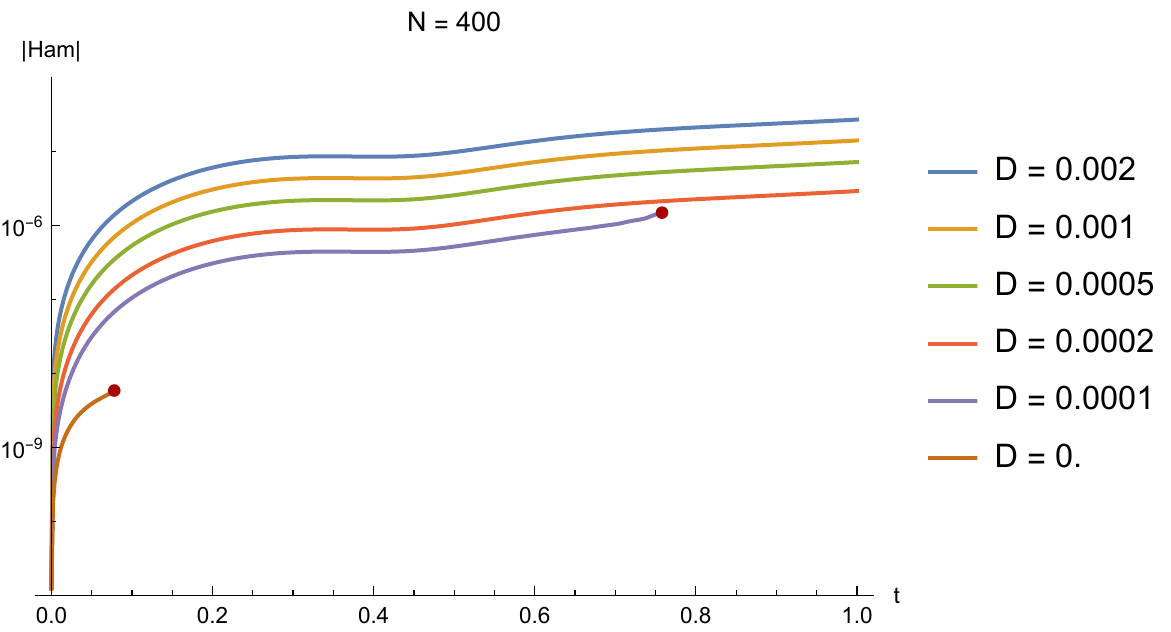}  \hspace{0.5cm}   \includegraphics[width=9cm]{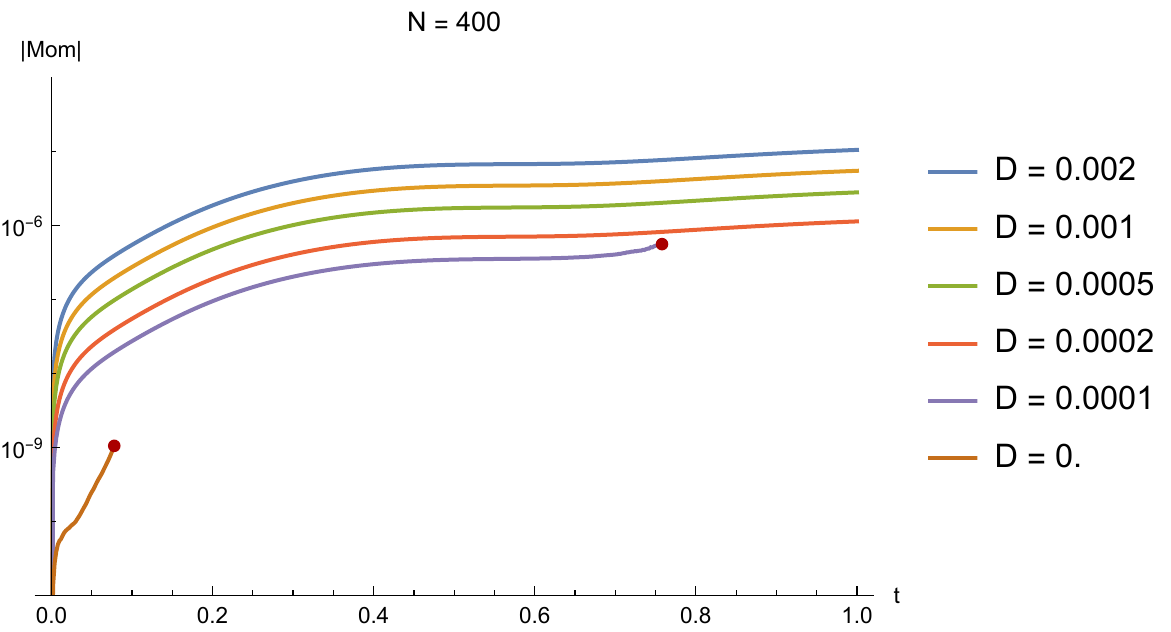}
  }
  \centerline{
  \includegraphics[width=9cm]{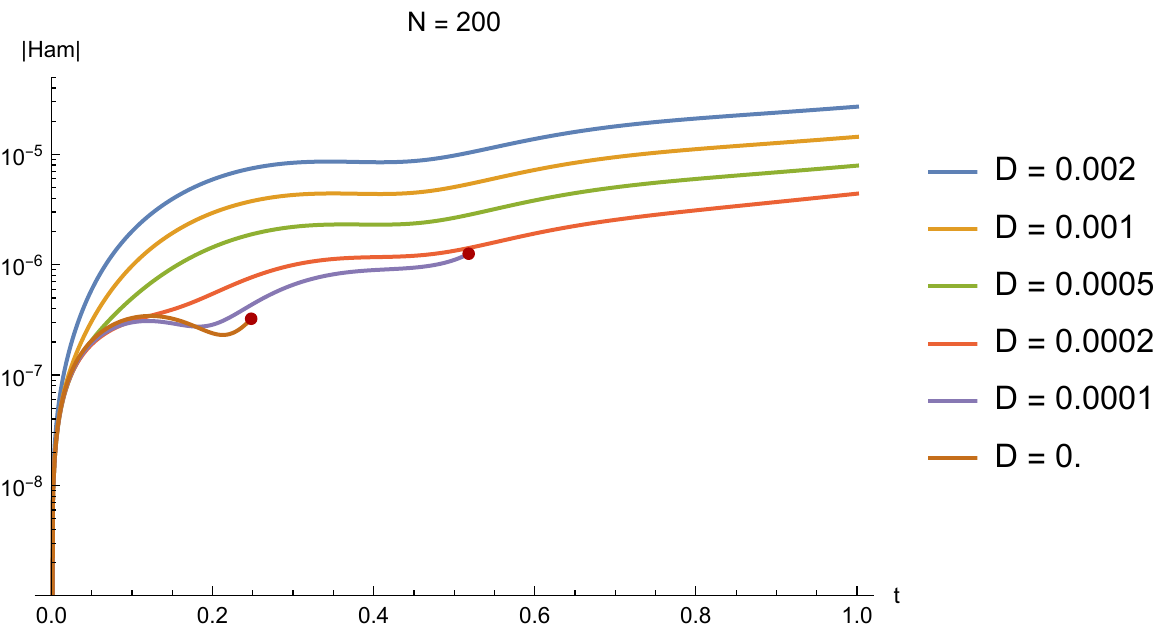}  \hspace{0.5cm}   \includegraphics[width=9cm]{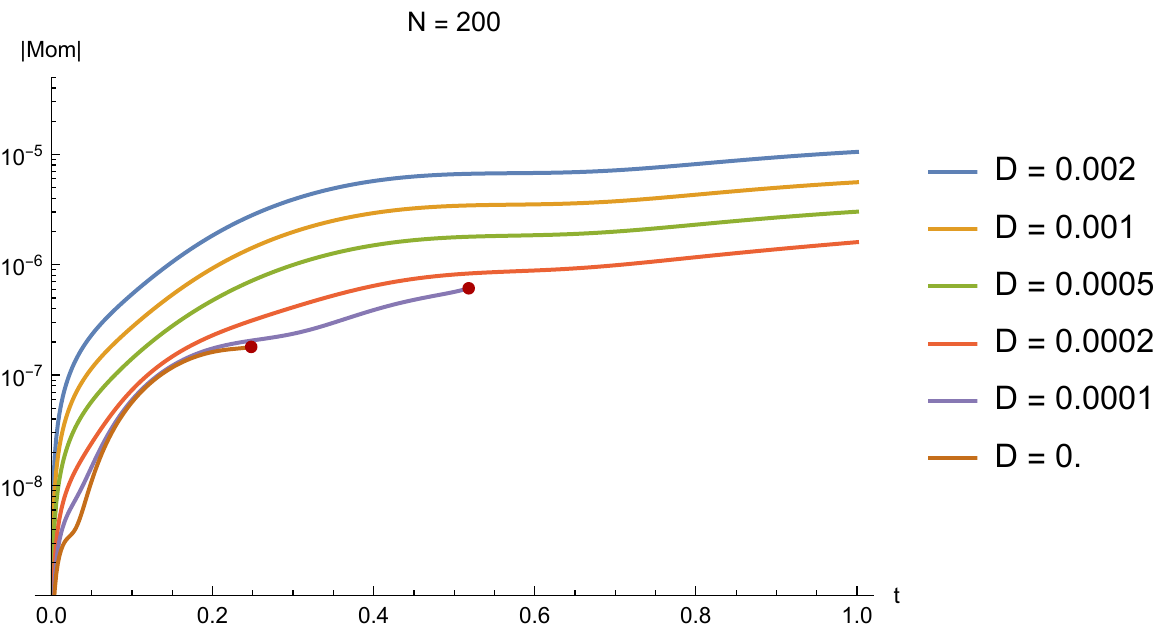}
  }
  \caption{\label{fig:convergence_withD}
    The behaviour of the $L^2$ norm of the Hamiltonian (left) and momentum (right) constraints against time for $A = 0.01$ as $D$ is varied between $0$ and $0.002$. The top figures have a resolution $N = 400$, whilst for the lower ones $N = 200$. We see the expected convergence of these constraints to zero as $D$ is initially decreased, but the simulations become increasingly susceptible to lattice scale instabilities in that limit, and so for $D$ that is too small the simulations break down,  as indicated by the red markers. For $N = 200$, eventually there is little benefit in further decreasing $D$, signalling that the numerical errors associated with the discretization dominate over the effects of diffusion.
    }
\end{figure}

We now turn to the (non-)conservation of the Hamiltonian and momentum constraints during evolution. The reason for this constraint violation is twofold: it includes the errors introduced by the discretization of the PDEs, as well as the effects of the diffusion term not present in the Einstein equations (\ref{eq:EinsteinEq}).
Above  in section~\ref{sec:independence_diffusion} we have argued that the precise value of the diffusion coefficient $D$ does not affect the long-range dynamics of our system. In figure~\ref{fig:convergence_withD} we further examine the effect of varying $D$, for a small amplitude $A = 0.01$ and resolutions $N = 200$ and $400$. In both cases we see that  decreasing the diffusion coefficient initially leads to an improvement in the constraint violation. However, for the smallest values considered ($D = 0.0001$ and $0$) the simulation breaks before $t = 1$ due to lattice scale instabilities -- some amount of diffusion is indispensable if we want to evolve our system for an appreciable amount of time. Also for $N = 200$ we see that eventually the curves for decreasing values of $D$ coincide (before the onset of instability), indicating that the discretization error dominates over the effects of diffusion in causing the constraint violation.\\

\begin{figure}
\centerline{
  \includegraphics[width=9cm]{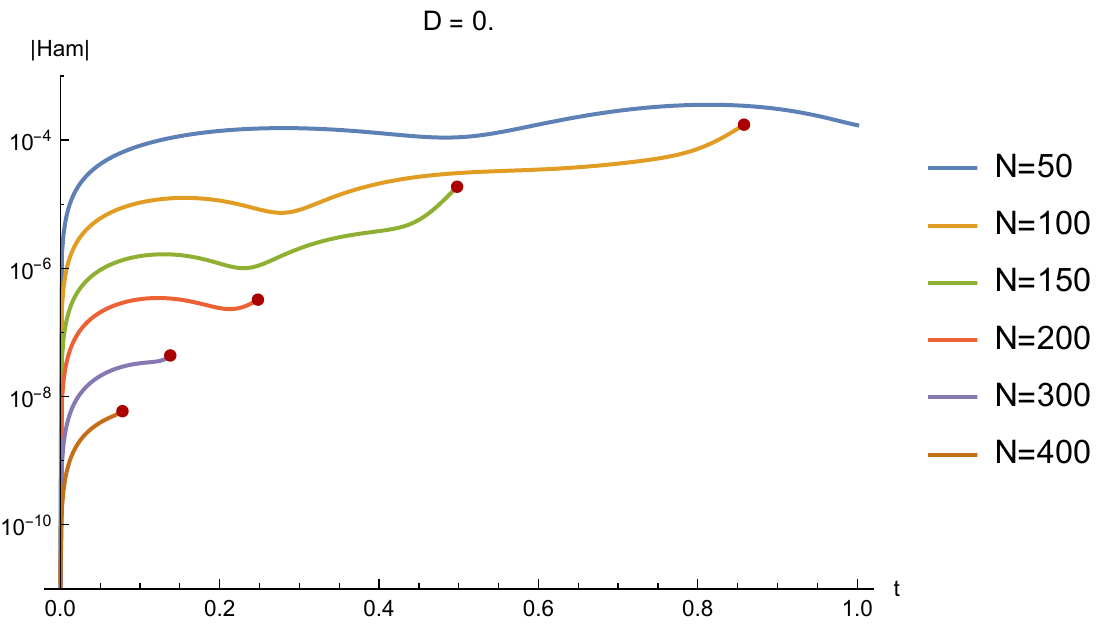} \hspace{0.5cm}    \includegraphics[width=9cm]{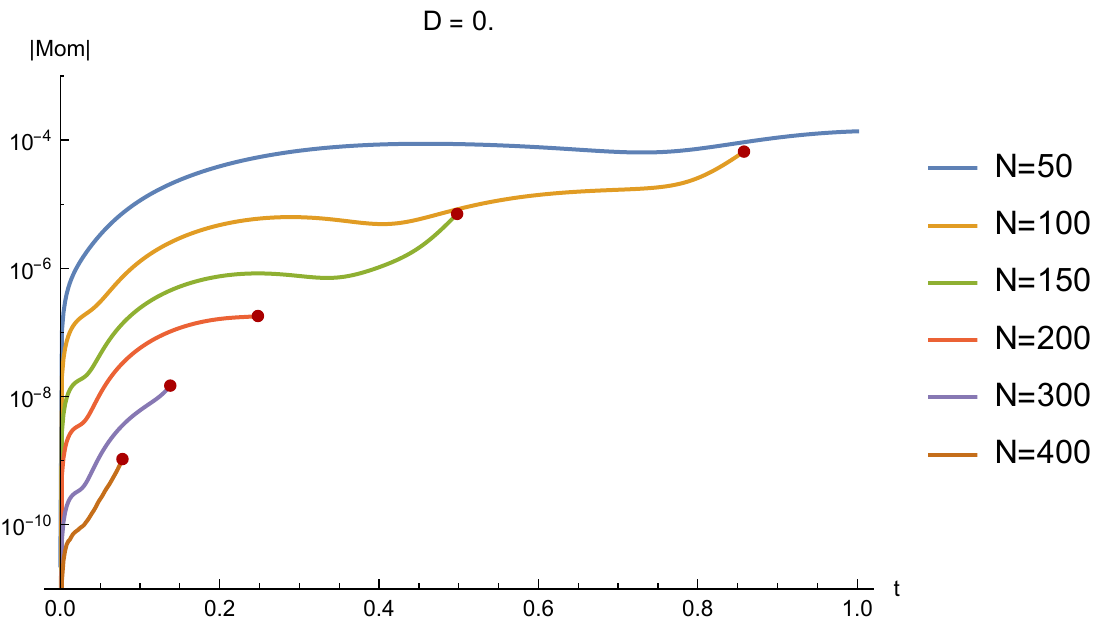}
  }
  \centerline{
  \includegraphics[width=9cm]{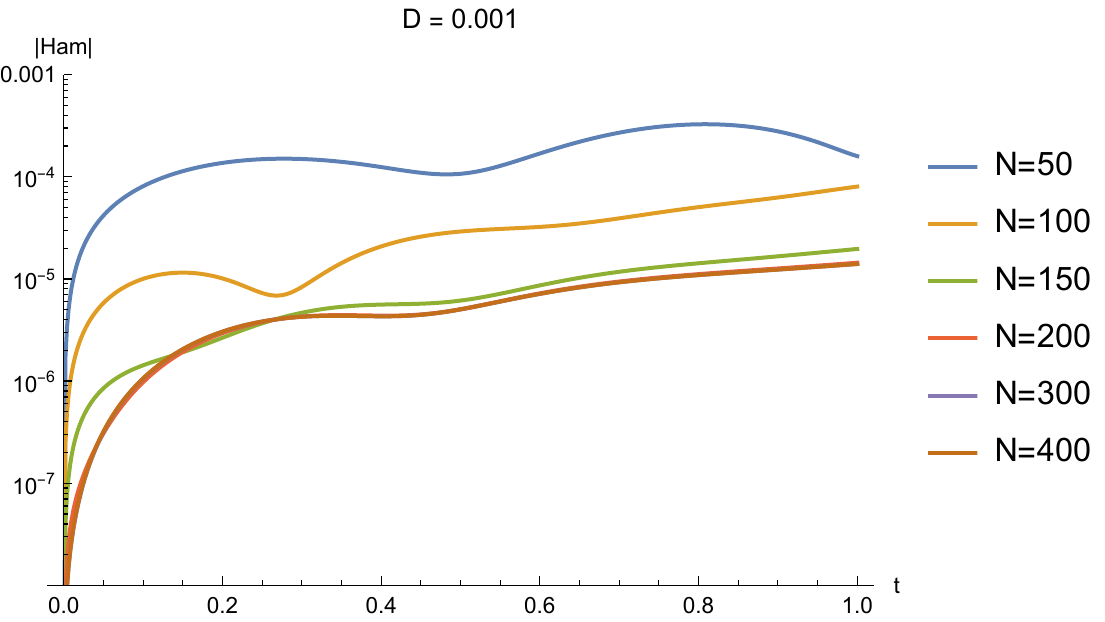} \hspace{0.5cm}    \includegraphics[width=9cm]{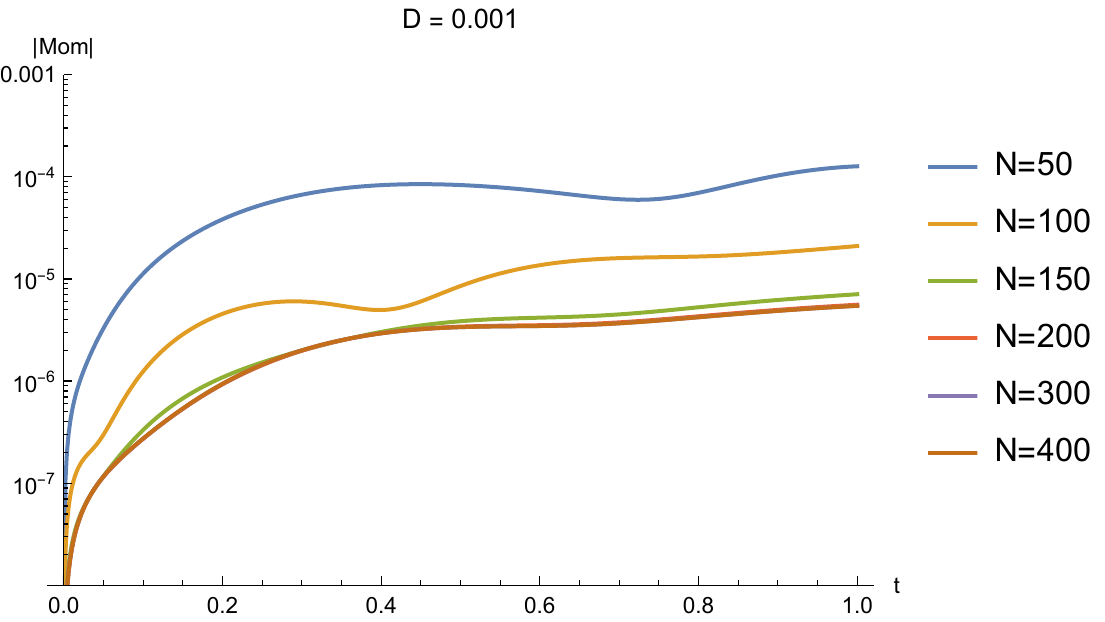}
  }
  \caption{\label{fig:convergence_withN}
    Plots of the norms of the Hamiltonian and momentum constraints in the unstable case $D = 0$ (top) and $D = 0.001$ (bottom), for $A = 0.01$ and varying spatial resolution $N$. With no diffusion, increasing it always improves the conservation of the constraints, at least before the onset of lattice scale instabilities that break the simulations (the times where this occures are indicated by red markers). Once the diffusive effects are included, for a sufficiently large resolution they dominate over the errors introduced by discretization. In particular for $D = 0.001$ this happens below $N = 400$, which was the case for the majority of the plots in this paper.
    }
\end{figure}

Finally in figure~\ref{fig:convergence_withN} we study the effects of varying $N$ between $50$ and $400$ for the unstable case $D = 0$ and the case used throughout this work, $D = 0.001$. Again we have set $A = 0.01$ and evolve until $t = 1$. \\

In the $D = 0$ case, we find that increasing the resolution always leads to an improvement in the values of the constraints, but also pushes forward the onset of lattice scale instabilities which break the simulation, so that for $N = 400$ we can only evolve up to $t \simeq 0.1$. In contrast, if we set $D = 0.001$ then all the resolutions are stable in the plotted time period. However, we see that eventually improving the resolution ceases to improve the results, and all curves obtained for $N \geq 200$ are essentially identical. This is due to the effects of diffusion becoming more important than the discretization error. In other words, we are converging to a solution of the \emph{modified} system of equations, which includes diffusive effects and hence no longer exactly respects the Hamiltonian and momentum constraints. Since there are no appreciable changes to our results above $N = 200$, the choice $N = 400$, which was made for most of the plots in this paper, is justified, as further refining the resolution would not have revealed any new physical effects.

\subsection*{Acknowledgments}
We thank Enrico Barausse, Luis Lehner and Mark Trodden  for helpful discussions.
This work is supported by STFC Consolidated Grant ST/T000791/1. CdR is also supported by a Simons Investigator award 690508 and JK is funded by an STFC studentship.

\clearpage

\addcontentsline{toc}{section}{Bibliography}
\bibliography{refs}

\end{document}